# Janus graphene nanoribbons with a single ferromagnetic zigzag edge

Shaotang Song[1,8], Yu Teng[1,7,8], Weichen Tang[2,4,8], Zhen Xu[3,8], Yuanyuan He[1], Jiawei Ruan[2,4], Takahiro Kojima[3], Wenping Hu[7], Franz J Giessibl[5], Hiroshi Sakaguchi*[3], Steven G Louie*[2,4], Jiong Lu*[1,6]

[1]Department of Chemistry, National University of Singapore, 3 Science Drive 3, Singapore 117543, Singapore.

[2]Department of Physics, University of California, Berkeley, CA94720, USA.

[3]Institute of Advanced Energy, Kyoto University, Uji, Kyoto 611-0011, Japan.

[4]Materials Sciences Division, Lawrence Berkeley National Laboratory, Berkeley, CA 94720, USA.

[5]Institute of Experimental and Applied Physics, University of Regensburg, 93053 Regensburg, Germany.

[6]Institute for Functional Intelligent Materials, National University of Singapore, Singapore, Singapore.

[7]Joint School of National University of Singapore and Tianjin University, International Campus of Tianjin University, Fuzhou 350207, China.

[8]These authors contribute equally: Shaotang Song, Yu Teng, Weichen Tang, Zhen Xu

E-mail: sakaguchi@iae.kyoto-u.ac.jp (H.S.); sglouie@berkeley.edu (S.G.L.); chmluj@nus.edu.sg (J.L)

## Abstract

**Topological design of π-electrons in zigzag-edged graphene nanoribbons (ZGNRs) leads to a wealth of magnetic quantum phenomena and exotic quantum phases.[1-9] Symmetric ZGNRs typically exhibit antiferromagnetically coupled spin-ordered edge states.[1,2] Eliminating cross-edge magnetic coupling in ZGNRs not only enables the realization of a new class of ferromagnetic quantum spin chains,[10] enabling the exploration of quantum spin physics and entanglement of multiple qubits in the 1D limit, [3,11] but also establishes a long-sought carbon-based ferromagnetic transport channel, pivotal for ultimate scaling of GNR-based quantum electronics.[1-3,9,12] Here, we report a general approach for designing and fabricating such ferromagnetic GNRs in the form of Janus GNRs (JGNRs) with two distinct edge configurations. Guided by Lieb's theorem and topological classification theory,[13-15] we devised two JGNRs by asymmetrically introducing a topological defect array of benzene motifs to one zigzag edge, while keeping the opposing zigzag edge unchanged. This breaks structural symmetry and creates a sublattice imbalance within each unit cell, initiating a spin symmetry breaking. Three 'Z-shape' precursors are designed to fabricate one parent ZGNR and two JGNRs with an optimal lattice spacing of the defect array for a complete quench of the magnetic edge states at the "defective" edge. Characterization *via* scanning probe microscopy/spectroscopy and first-principles density functional theory confirms the successful fabrication of JGNRs with ferromagnetic ground state delocalised along the pristine zigzag edge.**

## Introduction

The term "Janus" has been adopted to denote materials that exhibit different properties on two opposing sides or faces.[16,17] In the realm of 2D materials, creating asymmetric Janus materials by breaking the top- and bottom-layer symmetry, provides an attractive route for engineering novel properties and functionalities such as enhanced valley spin splitting, out-of-plane piezoelectricity and second harmonic generation.[18-21] Reducing the dimension of Janus materials from 2D to 1D systems with two different edge structures/topologies (hence properties), especially those with asymmetric zigzag edge in JGNRs, creates new opportunities for the realization of 1D ferromagnetic quantum spin chains and the assembly of multiple qubits. It also enables the creation of ferromagnetic transport channel in the 1D limit,[22,23] which can carry completely spin polarized current without applying in-plane electric field,[24] strain[25] or chemical functionalizations,[26,27] as required for conventional ZGNRs. These features are crucial for advancing the miniaturization of spintronic devices with enhanced energy efficiency.[28]

Although theoretical attempts have been made to break the zigzag structural symmetry by attaching triangular fragments to one zigzag edge of the ZGNR, the resulting ferromagnetic state is more delocalized and extends into the triangular fragments.[29] Moreover, the experimental exploration of such GNRs remain elusive, as the rational design of precursors towards such novel GNRs requires the simultaneous breaking of their structural and spin symmetries. This task is even more challenging than fabricating symmetric ZGNRs, which is notoriously difficult due to the high reactivity of zigzag edges combined with the challenges in intricate design and synthesis of precursors. To date, only two symmetric ZGNRs have been reported, both based on the 'U-shape' precursor design. Moreover, the width of the obtained ZGNRs is limited to having six carbon zigzag chains across the ribbon, namely, 6-ZGNR.[1,2]

## The design principle for JGNRs

Our strategy for creating such JGNRs involves decorating one zigzag edge of ZGNRs with a topological "defect" array (termed as TDZ edge), while keeping the other zigzag edge unchanged. Guided by this principle, we introduce a periodic array of benzene motifs attached to the TDZ edge to break structural symmetry and to create a sublattice imbalance within each unit cell ($\Delta N$ = 1), leaving one unpaired electron in each unit cell (Fig. 2).[13,30] As a result, this introduces a zero-energy band in line with Lieb's theorem, initiating the spin symmetry breaking in this system.[31-33] To effectively quench the spin at the TDZ edge and maintain a single ferromagnetic edge-state band at the zigzag edge, the key is to determine the optimal spacing (denoted as *m, m*=1, 2, 3, etc) between periodic benzene rings at the TDZ edge, which can maximally disrupts the edge states. To achieve this, we applied the topological classification using chiral symmetry developed in Ref. 15. In 1D bipartite lattices such as GNRs, because of the negligible second-nearest-neighbour interactions, chiral symmetry is satisfied, leading to a *Z*-classification of electronic topological phases with a conventional $Z$ index.[15] As the $Z$ index is 0 for vacuum (which is a topological trivial phase), the count of end states can be deduced from the $Z$ index of the GNRs.

For our design principle, we employ an important correspondence here that the *edge* of a JGNR (which can be either a defective zigzag edge or a normal zigzag edge) can be viewed as the *end* of a wide armchair-edged GNR (AGNR) in the large width limit (Fig. S1). This means the existence and characteristics of the *topological edge states* of a defective zigzag edge are similar to the *topological end states* of a large-width AGNR with a similar defective structure at its end. When examining the topological behavior of the end of an AGNR, the *Z* index can be expressed in a straightforward form[15]:

$$Z = W_{notco} - \left\lfloor \frac{W}{3} \right\rfloor \quad (1)$$

where $W$ represents the total number of rows of carbon atoms forming the width of the AGNR, and $W_{notco}$is the number of rows of atoms with unconnected carbon pairs (Fig. S1) within the specific unit cell that is commensurate to the end termination. The topless brackets denote the floor function, which takes the largest integer less than or equal to the value within the brackets.

Fig. 1a shows a family of JGNRs with different spacing of *m* as defined in Fig. 1b. For conceptual understanding of the topological states, we map the edge of a finite-length JGNR to the end of a large-width AGNR with same structure. For a JGNR consisting of *L* identical unit cells along the defect zigzag edge direction, the *associated* AGNR will contain *L* repeat units along the width of AGNR (with *L* be taken to the large limit and equal to a multiple integer of 3). For a specific JGNR with a given *m*, the number of rows of carbon atoms forming the width of the AGNR is given by $W = 2(m+1)L$, where $m$ is the number of missing benzene ring as shown in Fig. 1a. For the TDZ edge of the JGNR, $W_{notco} = mL$, and the corresponding chiral phase index (CPI) (which dictates the number of topological edge states) is $Z_{TDZ} = \frac{m-2}{3}L$ using the Jiang-Louie formula[15] given in Eq. (1). In the case of the non-defective edge (i.e., the normal zigzag edge), the width of the associated AGNR is the same $W = 2(m+1)L$, while $W_{notco} = (m+1)L$, and the corresponding CPI is $Z_{zig} = \frac{m+1}{3}L$ (a detailed analysis can be found in the supplemental material). Moreover, according to Lieb's theorem, in considering the electron spin degree of freedom, the topological end/edge states become spin-polarized, forming a magnetic ground state. As illustrated in Fig. 1b, the sublattice imbalance (difference in numbers of atoms of the A and B sublattice of graphene) for all JGNRs is $\Delta N_L = L$ (since $\Delta N = 1$ is the sublattice imbalance per unit cell as defined in Fig. 1). Thus, the ground state of the system consists of $\frac{m-2}{3}L$ spin-up (by convention) topological states localized at the TDZ edge of JGNRs (or viewed as the end of the large finite-width AGNR) and $\frac{m+1}{3}L$ spin-down topological states localized at the regular zigzag edge at the other side. (Our *ab initio* density functional theory (DFT) calculations that validate our argument numerically can be found in the Supplemental Information, Fig. S1-S3).

With the above topological CPI analysis, we reach the following remarkable prediction without employing any detailed calculations: The sign of the quantity $R$, defined as the negative of the ratio between the CPI of the two edges $R = -\frac{Z_{TDZ}}{Z_{zig}} = -\frac{m-2}{m+1}$, determines the magnetic ordering of the JGNRs (Fig. 1c). For $m = 1$ (with $R > 0$), both edges of JGNR adopt the same spin alignment but with different magnitude, giving rise to a ferromagnetic state. In contrast, for $m > 2$ (where $-1 < R < 0$), the two edges of JGNR exhibit the different spin configurations with varying magnitude, resulting in a ferrimagnetic order. Notably, when $m = 2$ ($R = 0$), no edge state exists at the TDZ edge, leading to the formation of a ferromagnetic ground state localized exclusively at the zigzag edge. The band structure near $E_F$ for the $m = 2$ case (Fig. 1d) is in dramatic contrast to that of the symmetric ZGNRs with antiferromagnetically coupled edge states. In this framework, the conventional symmetric ZGNRs can be viewed as an extreme scenario with $m = \infty$ (with $R = -1$), where the perturbation of the infinitely-spaced benzene rings becomes zero. Therefore, by varying the parameter $m$ associated with the periodicity of topological "defects" on one edge, our theory predicts the emergence of different magnetic ground states with different cross-edge couplings in JGNR, which is unattainable in symmetric ZGNRs (Fig. 1a).

Since (*m* = 2)-JGNR is predicted to feature the unique ferromagnetic ground state with a single ferromagnetic zigzag edge, our study focuses on the synthesis and investigation of this type of JGNRs. Here, we used the notation (*n,m*)-JGNRs to describe the specific JGNRs, where *n* refers to the number of carbon zigzag chain laterally across the ribbon and *m* denotes the benzene ring spacing at the TDZ edge, as defined above. Specifically, we target to fabricate the (*4*,*2*)- and (*5*,*2*)-JGNRs (Fig. 2) with different width and the parent 5-ZGNR (Fig.3) (which can also be viewed as the corresponding (5,∞)-JGNR) to experimentally validate our topological theory predictions and our design principle.

## Precursor design for on-surface synthesis of JGNRs

To achieve the on-surface synthesis of two JGNRs (with *n = 4 and 5*, *m* = 2), we employ a Z-shaped molecular precursor design for the asymmetric fabrication, in contrast with the U-shaped precursor used previously for the on-surface synthesis of the 6-ZGNR.[34-37] The Z-shaped precursor comprises of two independent branch moieties that are interconnected by a single C-C bond. At the termini of the branches, two bromine atoms are strategically positioned as connection sites for thermally induced aryl-aryl coupling. The Z-shaped precursor design also allows for the independent modification of individual branches, facilitating a separate control over the two edge geometries of the resulting JGNRs. Utilizing methylphenanthrene groups for both branches (precursor **3** in Fig. 3) yields two identical zigzag edges, corresponding to the conventional 5-ZGNR. Intriguingly, substituting one of the methylphenanthrene groups to an oligophenyl group results in the modification of one of the zigzag edges to our target TDZ edge with periodically decorated benzene rings (with a spacing of *m* = 2). Additionally, adjusting the length of the oligophenyl groups from biphenyl to triphenyl (precursor **1** and **2** in Fig. 2) enables the modulation of the width of the JGNRs from *n*=4 to *n*=5.

The precursors were successfully obtained *via* multi-step organic synthesis (see Supplementary Information) and then separately deposited onto a clean Au(111) surface under ultrahigh vacuum (UHV) using a Knudsen cell evaporator. Stepwise annealing of the precursor decorated surfaces induces polymerization and cyclodehydrogenation reactions to yield the corresponding (*4*,*2*)- and (*5*,*2*)-JGNRs (Fig. 2) and 5-ZGNR (Fig. 3). The yield of all the GNRs is low under our experimental conditions (Fig. S12 and S13), primarily due to the flexible backbone of the corresponding polymers, which undergo random cyclodehydrogenation reactions during thermal annealing (Fig. S11). In addition, the asymmetric precursors **1** and **2** offer limited control over different polymerization modes (Fig. S14), namely head-to-tail, and head-to-head (or termed as tail-to-tail). If polymerization proceeds in a head-to-head (tail-to-tail) configuration, then the resulting GNRs will feature alternative cove and gulf regions on both edges.[35,36] Here, we focus on the investigation of the targeted JGNRs through the head-to-tail polymerization. The topographic STM images of both (*4*,*2*)-JGNR and (*5*,*2*)-JGNR reveal a characteristic height profile and morphology featuring bright lobes at the intact zigzag edge and larger protrusions at the TDZ edge, corresponding to the position of periodically-spaced benzene rings. The bond-resolved scanning tunnelling microscopy (BR-STM) images of both JGNRs acquired at -10 mV using a CO-functionalized tip exhibit the corrugated hexagon-like rings in the centre and distorted lobe features along the edges, which are presumably attributed to the perturbations induced by the electronic states of the JGNRs in the vicinity of Fermi level ($E_F$).[38,39] In contrast, the BR-STM image of 5-ZGNR acquired at 10 mV using a CO-functionalized tip clearly resolves the characteristic hexagon patterns in the backbone, indicating an absence of electronic states near $E_F$.

To better resolve the backbone framework of both JGNRs, we further conducted the qPlus sensor-based non-contact atomic force microscopy (nc-AFM) images, which relies on monitoring the resonance frequency shift of an oscillating quartz force sensor as the CO functionalized tip scans over the sample in the Pauli repulsive regime.[40,41] The achieved resolution directly confirms the different edge morphologies correspond to the expected JGNRs as defined by the precursor design. It is noted that the outer benzene rings on the TDZ edge appear flattened along the ribbon and larger in size than the ones in the backbone, while the outer rings at the intact zigzag edges appear stretched in the direction perpendicular to the ribbon. Such a different appearance of hexagons at different edges of JGNRs is presumably due to varying degrees of CO-tip bending effect due to different edge topologies. All these results prove the success of the Z-shape precursor design strategy for the synthesis of both JGNRs and 5-ZGNR.

**Scanning tunnelling spectroscopy of the GNRs.**

Previous studies revealed magnetic edge states cannot be directly observed in the wider 6-ZGNR and N-doped 6-ZGNR when sitting on Au(111), because they coupled strongly to the metal substrate.[1,2] Compared to 6-ZGNR, both 5-ZGNR and two JGNRs have a narrow ribbon width, thus leading to a more rigid backbone that could prevent a strong structural downward bending and electronic coupling with substrate.[42] Our differential conductance (d$I$/d$V$) spectroscopic measurements acquired over both 5-ZGNR and JGNRs combined with theoretical calculations confirm that magnetic edge states can survive on Au(111) for these narrower ribbons (Fig. 3). d$I$/d$V$ spectrum obtained at the edge of 5-ZGNR show three distinct peaks located at 760 ± 18 mV (peak A), -300 ± 5 mV (peak B), and -410 ± 7 mV (peak C), yielding an apparent experimental band gap of $\Delta E_{exp} = 1060 \pm 19$ meV (Fig. 3c and Fig. S5). Our experimental gap of 5-ZGNR is smaller than that of the 6-ZGNR (1.5 eV) placed on top of a monolayer NaCl island.[1] Such behaviour is likely attributed to the screening effect of the metal substrate in reducing the quasiparticle energies and band gaps of reduced dimensional systems.[43-46] The d$I$/d$V$ maps recorded at the biases corresponding to these three peak energies reveal the characteristic lobe features predominately located at both zigzag edges, demonstrating reproducible patterns in the DFT-calculated local density of states (LDOS) maps of each of the frontier band states (Figure 3d-f). Note that the slight asymmetry in the d$I$/d$V$ maps of the edge states is presumably arose from tip asymmetry.

As shown in Fig. 4a, d$I$/d$V$ spectrum taken over the zigzag edge of (*4*,*2*)-JGNR also reveals three noticeable features located at 581 ± 19 mV (peak A), 45 ± 5 mV (peak B) and -487 ± 14 mV (peak C), which yields an apparent band gap of $\Delta E_{exp} = 536 \pm 9$ meV for (*4*,*2*)-JGNR. However, peak C is absent in the d$I$/d$V$ curve taken over the additional benzene ring of the TDZ edge. The most pronounced peak B spans across $E_F$, which is tentatively assigned as the partially-filled top VB (spin down and localized mainly along the intact zigzag edge) arising from the charge transfer due to the work function difference between GNR and Au(111). d$I$/d$V$ image (Fig. 4c) acquired at the energy positions of peak B shows a petal-shaped feature on the zigzag edge and flower-shaped features distributed on the entire backbone of the (*4*,*2*)-JGNR. In contrast, the spatial distribution of other two peaks of the (*4*,*2*)-JGNRs reveal similar characteristic zigzag lobe features predominately or exclusively located at the intact zigzag edges for peak A and peak C, respectively (Fig. 4b, 4d, and Fig. S18). Similarly (Fig. 4j), d$I$/d$V$ spectrum of (*5*,*2*)-JGNR also reveals three peak structures centered at 410 ± 11 mV (peak A), 16 ± 2 mV (peak B) and -510 ± 14 mV (peak C), which yields a smaller apparent band gap of $\Delta E_{exp} = 394 \pm 9$ meV. Compared to (*4*,*2*)-JGNR, all the states observed in (*5*,*2*)-JGNR experience a slight downward shift in energy, but their spatial distribution patterns in the d*I*/d*V* maps are rather similar (Fig. 4k-4m, and Fig. S19).

## ***Ab initio* calculation of JGNRs electronic structures**

A direct capture of spectroscopic features and their wave function patterns of electronic states for both 5-ZGNRs and two JGNRs suggest a relatively weak coupling of the ribbons with the Au(111) substrate. We thus conducted *ab initio* density functional theory (DFT) calculations within the local spin density approximation (LSDA) to calculate spin-polarized electronic structures for the free-standing ZGNRs and JGNRs. Our first-principles results offer quantitative evidence that the 5-ZGNR and two JGNRs synthesized on Au(111) all maintain their intrinsic magnetic edge states. Panels in Fig. 3g-3i present the calculated LDOS maps acquired at the energies corresponding to peak A' (conduction band minimum), peak B' (valence band maximum), and peak C' (the valence band states near the Brillouin zone edge X) of the freestanding 5-ZGNR, respectively. The distinctive pattern and relative contrast of protrusions lining the edges of 5-ZGNRs observed in experimental d*I*/d*V* images (Fig. 3d-f) are well reproduced in the corresponding LDOS maps (Fig. 3g-i). The DFT-calculated band structure (Fig. 3l) reveals that the valence band (VB) is more dispersive than the conduction band (CB) in the region of around one-third of Brillouin zone edge near the X point, forming a double-peak structure in the DOS spectrum due to the van Hove singularities at the VB maximum away from X (peak B') and at the X point (peak C'). In contrast, the bottom of the CB is significantly flattened in this region of Brillouin zone, forming only one peak in the DOS spectrum (peak A'). Similar behaviour in electronic band dispersion has been observed in 6-ZGNR in previous study[3]. It is also noted that the experimental gap (980 ± 35 meV) derived from d*I*/d*V* spectra is larger than that of the DFT-calculated one (330 meV). Given DFT's tendency to significantly underestimate quasiparticle bandgaps[47] even when factoring in the screening effects of the underlying Au substrate, it is not surprising that the DFT value is smaller than the experimentally observed gap. Nevertheless, the relative energy positions of the three electronic peaks near $E_F$ along with their corresponding LDOS, are in good agreement with experimental data. In addition, our *ab initio* calculations on 5-ZGNR reveal that an antiferromagnetic alignment of spins across the ribbon width between ferromagnetically ordered edge states is favored as the ground state over both the non-magnetic or a fully ferromagnetic configurations, as illustrated in the spin density plot (the inset of Fig. 3l; note that the absolute spin orientation is arbitrary due to negligible spin-orbit interactions in carbon systems).

In striking contrast, our DFT calculations directly demonstrate that the ground state for both the JGNRs studied is ferromagnetic, with ferromagnetic spin ordered edge states located *only* at the intact zigzag edge (inset of Fig. 4h and 4q). As seen in Fig. 4i and 4r, each one of the originally doubly occupied spin degenerate band splits into a pair of nondegenerate bands with opposite spin orientation. In particular, near zero energy, the splitted two bands (the lowest CB and the highest VB) bracketing the $E_F$ are edge-state bands and show a splitting energy gap of 0.21 eV and 0.19 eV for the (*4*,*2*)-JGNR and (*5*,*2*)-JGNR, respectively. Similar to the 5-ZGNR, the highest VB of both JGNRs is more dispersive than the lowest CB (which both are now nondegenerate), forming a double-peak feature (B' and C') in the DOS spectrum due to the presence of van Hove singularities at the Γ and the X point of the Brillouin zone of the JGNR. Conversely, the first CB is notably flattened, forming a sharp peak (A') in the DOS. Despite variations in their widths, the calculated LDOS patterns of these peak states remain similar, and correspond well with the experimental d$I$/d$V$ maps. Since the physical properties of JGNRs near the $E_F$ are dominated by these edge states, variations in the JGNR's width would only quantitatively change the low-energy physics. According to our prediction on the ($n$,$m$ = 2)-JGNR, if the width of the JGNRs goes to infinity, almost all the states in the lowest CB and the highest VB would be localized at the zigzag edge. However, more bands will appear near the Fermi level.

Previous tight binding calculations on an infinite width *pristine* ZGNRs indicate a *k*-dependent localization length at the zigzag edge, with states being more localized when closer to the X point of the ZGNR Brillouin Zone.[5] Since the edge states of ($n$,$m$ =2)-JGNRs can be viewed as band-folded states of ZGNR, for the peak originating from states near the X point of the highest VB of JGNR (peak C'), they all originate from states near X point of the *pristine* ZGNR. The localization length of those states is small compared to the width of JGNRs, so the theoretical LDOS pattern shows a localized pattern at the zigzag edge (see Fig. 4g and 4p). In contrast, for states more towards to the Γ point (at near 1/3 way from X to Γ or beyond), the localization length of those states in the infinite width prinstine ZGNRs exceeds the width of the JGNRs considered. The corresponding theoretical LDOS pattern (peak B') shows a delocalized pattern throughout the ribbon width (Fig. 4f and 4o). For the peak formed by the CB (peak A'), it comprises states from the entire JGNR Brillouin zone, and the LDOS pattern can be viewed as a mixture of the previous two patterns (Fig. 4e and 4n). We note that in the experiment, peak B located at near $E_F$ for both JGNRs. This is due to the fact that the work function difference between our GNRs (~ 4.5 eV) and Au(111) (~ 5.3 eV) entails a charge transfer from the JGNRs to the substrate, which lifts the valence band maximum upwards to the Fermi level.

## Conclusion

Guided by our topological classification theory, the successful bottom-up synthesis of atomically precise JGNRs with a single ferromagnetic zigzag edge offers a rich platform not only for exploring novel quantum spin physics but also for its technological relevance. For instance, such JGNRs could serve as a model system for realizing ferromagnetic Heisenberg spin-1/2 chains. Achieving 1D ferromagnetic states in JGNRs also enables the development of fully spin-polarized current channels. Additionally, the ability to precisely tune the bandgaps of JGNRs by adjusting their width and edge structure allows for fine control over band alignment and spin-polarized transport barriers. These features are crucial for advancing the miniaturization of spintronic devices with enhanced energy efficiency. Furthermore, the ability to assemble multiple spin centers in 1D JGNRs opens new avenues for creating robust qubit platforms, leveraging individual π-spins or their collective spin states to achieve enhanced noise resilience.[48]

## Method

**In-solution and on-surface synthesis.** For the synthesis of the precursor of 5-ZGNR, both branches of the Z-shape precursor are concurrently generated through the Suzuki coupling of 5,5'-dibromo-2,2'-diiodo-1,1'-biphenyl with methylphenyl boronic acid, then generating an ethynyl group *via* the cleavage of the triisopropylsilyl. Subsequent $PtCl_2$ catalysed cyclization of the ethynyl moiety produces two methylphenanthrene motifs connected by a C-C bond (precursor **3**), serving as branches to construct the zigzag edges. In contrast, for the precursor of JGNRs, one of the two methylphenanthrene branches is substituted by biphenyl and triphenyl groups by an additional step of Suzuki coupling. The synthetic procedures of the precursors are detailed in Supplementary Information. Au(111) single crystal (MaTeck GmbH) was cleaned by multiple cycles of $Ar^+$ sputtering and annealing. Knudsen cell (MBE-Komponenten GmbH) was used for the deposition of precursor molecules onto clean Au(111) surfaces under ultrahigh vacuum conditions (base pressure, $<2 \times 10^{-10}$ mbar) for on-surface synthesis of product **1**. The precursor was sublimated at varying temperature for 1 minute to achieve a moderate coverage on Au(111). After the deposition of precursors, the sample was stepwise annealed at elevated temperatures as stated in the main text for 20 min to induce intramolecular dehydrogenation (Fig. S12).

**STM/BR-STM/nc-AFM and d*I*/d*V* characterization.** The experiments were conducted in Scienta Omicron LT-STM/AFM systems operated under ultrahigh vacuum (base pressure, p $< 2 \times 10^{-11}$ mbar) at a temperature of T = 4.5 K. All the BR-STM and nc-AFM images were taken in constant height mode with a CO functionalized tip. The microscope of Scienta Omicron was equipped with qPlus sensors S0.8 (see table I in ref. 38) with a resonance frequency of $f_0$ = 39.646K Hz, a stiffness of 3600 N/m, and a quality factor of 119115. nc-AFM images were collected at a constant-height frequency modulation mode using an oscillation amplitude of $A$ = 20 pm. The tip-sample distance with respect to an STM set point is indicated in the figure caption for the corresponding AFM image.The images were analyzed and processed with Gwyddion software. The d*I*/d*V* spectra were collected using a standard lock-in technique with a modulated frequency of 479 Hz. The modulation voltages for individual measurements are provided in the corresponding figure captions. The STM tip was calibrated spectroscopically against the surface state of Au(111) substrate.

**Computational Method.** First-principles DFT calculations in the LSDA were performed using the Quantum Espresso packages.[49,50] A supercell geometry was employed, with a 10 Å vacuum spacing placed in all non-periodic directions to prevent interaction between replicas. The atomic geometry was fully relaxed until all components of the forces on each atom were smaller than 0.01 eV/Å. A 100 Ry wavefunction energy cut-off, along with scalar relativistic and norm-conserving pseudopotentials for C and H, were used.[51,52]

**Acknowledgement**

J. Lu acknowledges the support from MOE grants (MOE2019-T2-2-044, MOE T2EP50121-0008, MOE-T2EP10221-0005) and Agency for Science, Technology and Research (A*STAR) under its AME IRG Grant (Project715 No. M21K2c0113). S.G. Louie acknowledges the support from the U.S. National Science Foundation under grant number DMR-2325410, which provided all the theoretical topological formulation and analyses as well as the density functional calculations. H. Sakaguchi acknowledges the support from KAKENHI program No. 22H01891. T. Kojima acknowledges the support from KAKENHI program No. 23K04521. S. Song acknowledges the support from A*STAR under its AME YIRG Grant (M22K3c0094).

**Competing interests**

F.J.G. holds patents on the qPlus sensor. The other authors declare no competing interest.

**Contributions**

J. Lu supervised the project and organized the collaboration. S.S. and J. Lu conceived and designed the experiments. S.S. and Y.T. carried out the STM/nc-AFM measurements. Z.X., S.S., T.K. and H.S. synthesized the organic precursors. S.G.L. conceived the topological equivalence of defect edge of JGNR with end of AGNR, and supervised the theoretical analyses and calculations. W.T., J.R. and Y. H. performed the theoretical studies and DFT calculations. F.J.G. assisted in the optimization of the nc-AFM measurements and instrumentation. S.S., W.T., S.G.L. and J. Lu wrote the manuscript with inputs from all authors.

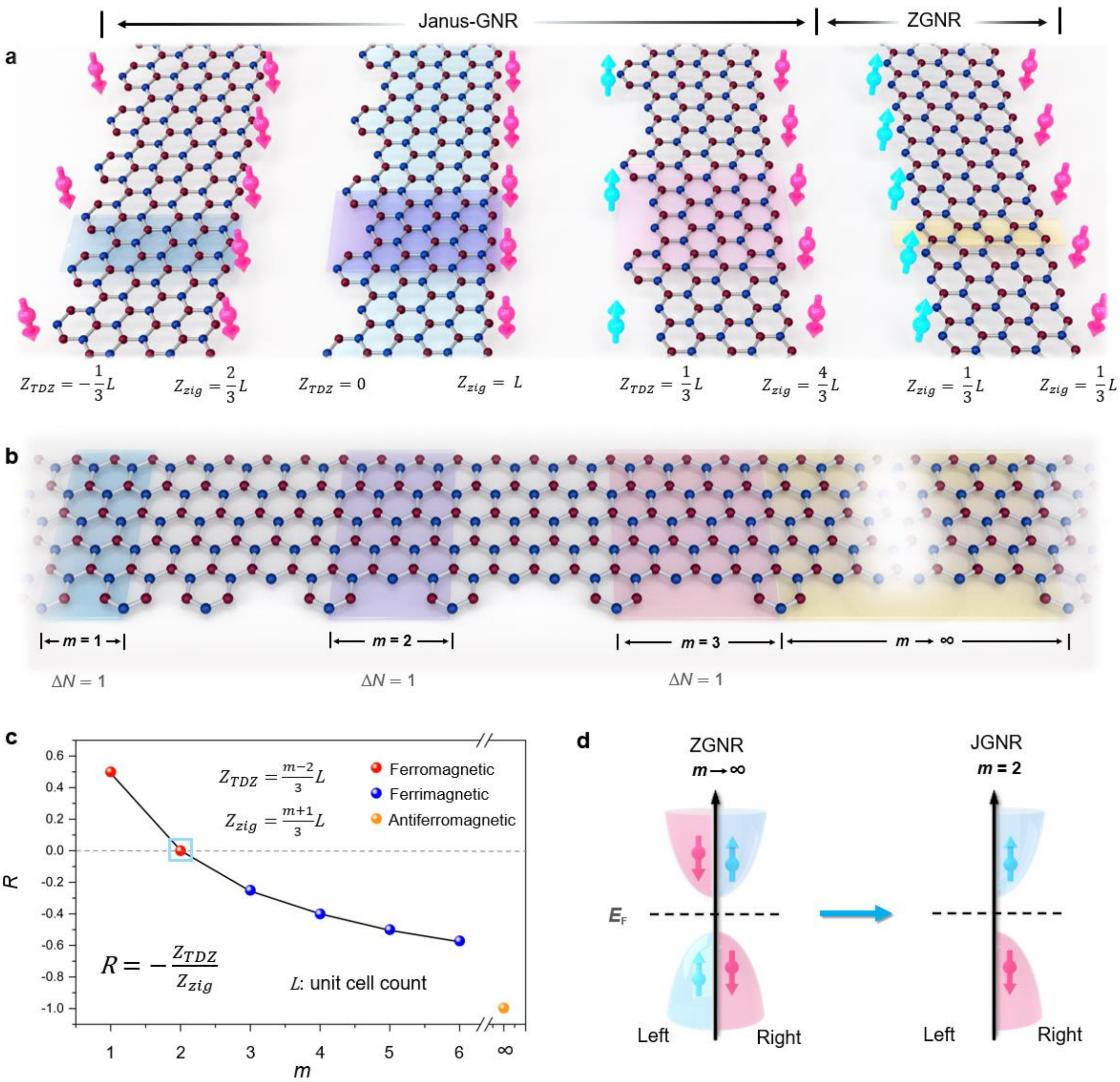

**Figure 1| Design principle of JGNRs. a.** Schematic illustration of JGNRs with various TDZ edges of different $m$ values. Red and blue arrows represent spin-up and spin-down of edge states (by convention), respectively. The numbers at the bottom represent the CPI index of the corresponding TDZ ($Z_{TDZ}$) and intact zigzag ($Z_{zig}$) edges, where $L$ is the number of unit cells for a finite-length ribbon along the zigzag direction (the unit cells are shaded by colors). **b.** The schematic diagram of the repeated unit of JGNRs, illustrating a gradual increase in "defective" site spacing between adjacent benzene rings, progressing from left to right: 1, 2, 3, and towards infinity (∞). The sublattice imbalance per unit cell remains as $\Delta N = 1$ for all the cases except for $m = \infty$. **c.** The graph illustrates $m$-dependent magnetic orderings between the two edges, which can be described using the negative sign of the ratio of the CPI index for the two edges of JGNRs (defined as $R = -\frac{Z_{TDZ}}{Z_{zig}}$). As $m$ increases, $R$ transforms from a positive to negative value, crossing 0 at $m = 2$. From Lieb's theorem and CPI theory, JGNR

undergoes a transition from ferromagnetic to antiferromagnetic through a region of ferromagnetic coupling. **d**. Schematic diagram of the spin-polarized band structure near the fundamental bandgap going from ZGNR ($m = \infty$, antiferromagnetic coupling between the two symmetric zigzag edges that are ferromagnetically ordered) to JGNR ($m = 2$, ferromagnetic order only at the single zigzag edge).

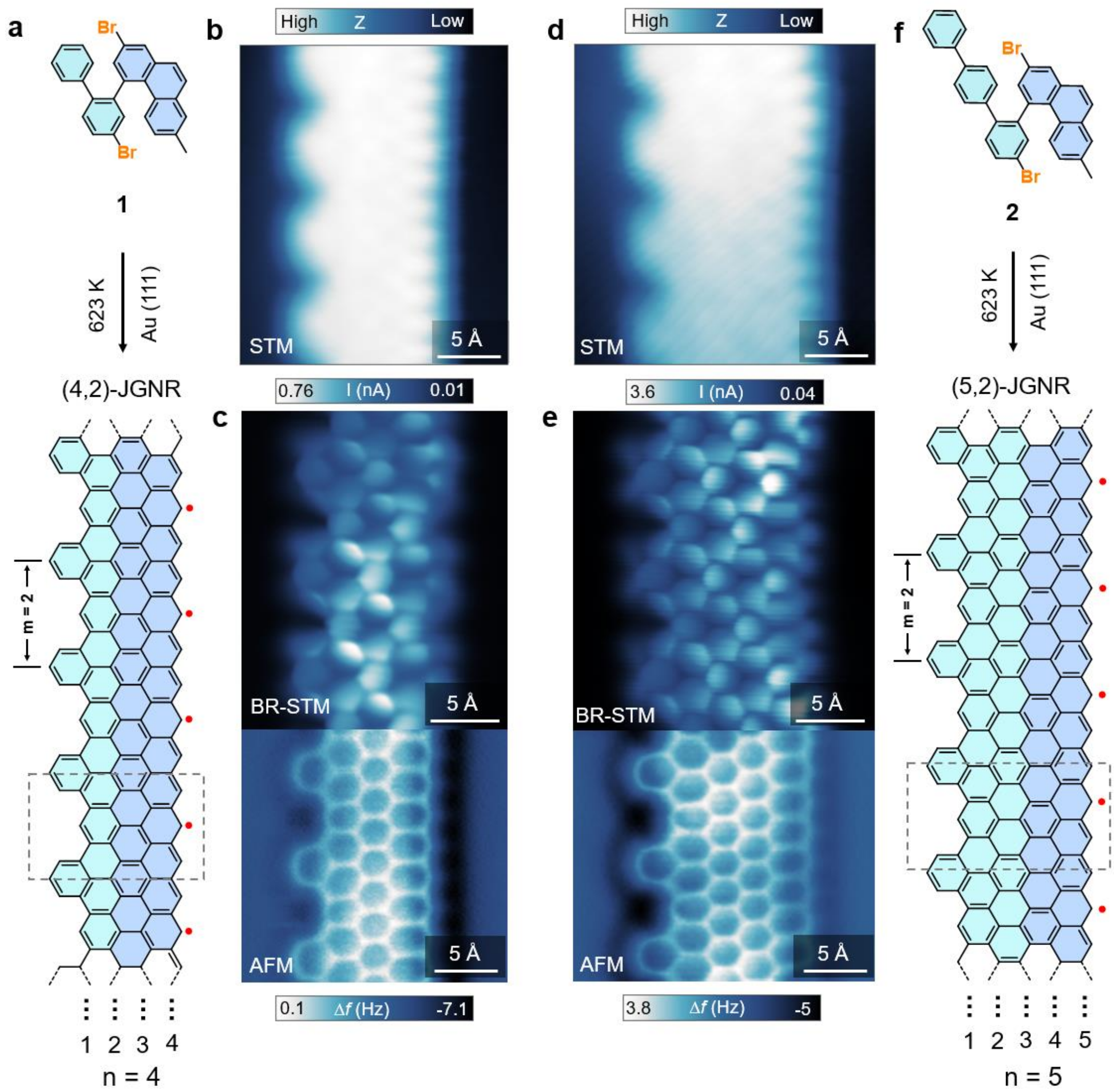

**Figure 2 | On-surface synthesis and structural characterization of the JGNRs. a**, On-surface synthetic strategy of (*4,2*)-JGNR. **b.** Constant current STM image of (*4,2*)-JGNR ($V_s$ = -800mV, $I$=100pA) **c.** BR-STM image ($V_s$ = -10 mV) and nc-AFM image ($V_s$ = 10 mV) of (*4,2*)-JGNR acquired at constant height mode using a qPlus sensor with a CO functionalized tip. **d**. Constant current STM image of (*5,2*)-JGNR ($V_s$ = -800mV, $I$=100pA). **e.** BR-STM image ($V_s$ = -10 mV) and nc-AFM image ($V_s$ = 10 mV) of the (*5,2*)-JGNR

acquired at constant height mode using a qPlus sensor with a CO functionalized tip. **f.** On-surface synthetic strategy of (*5,2*)-JGNR.

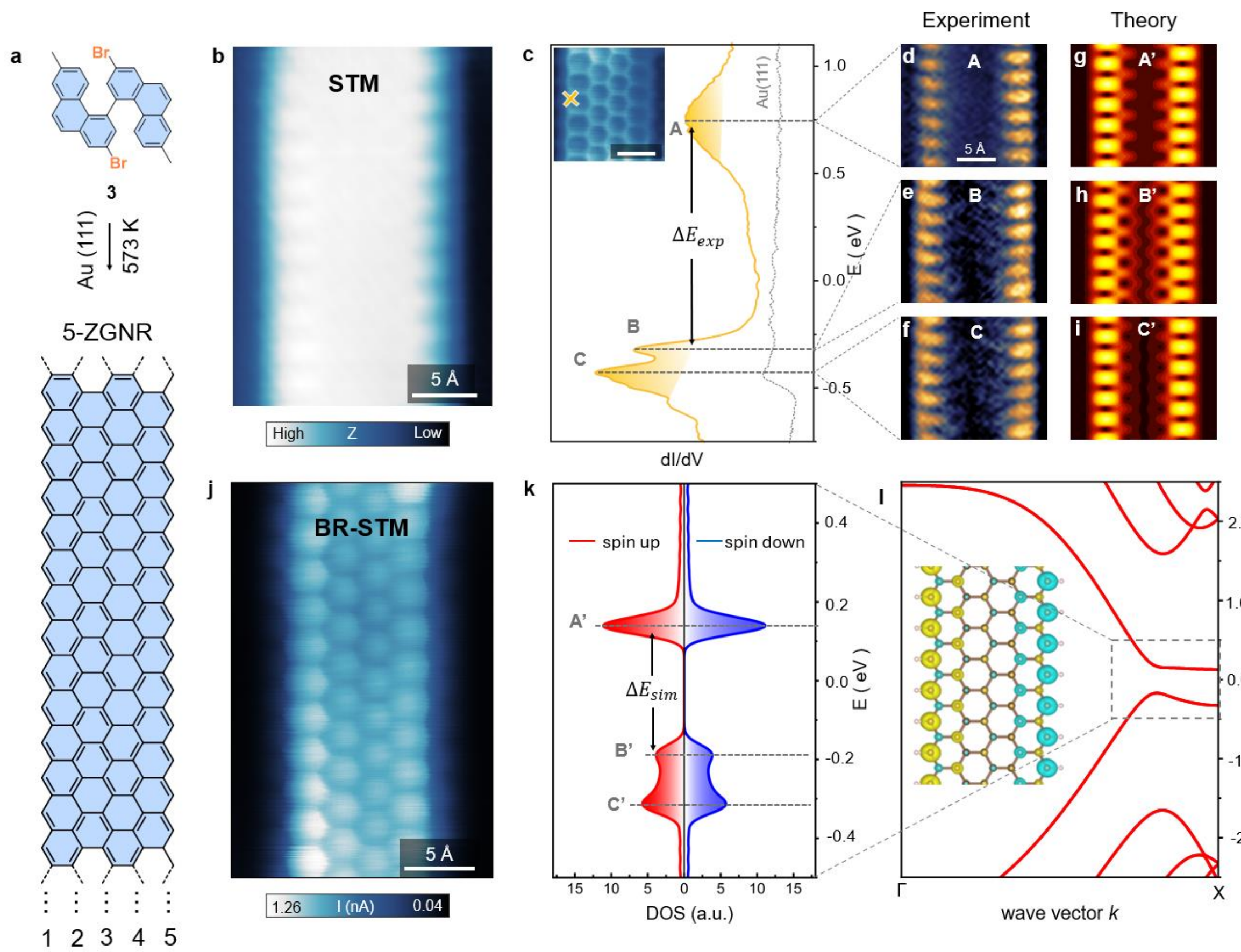

**Figure 3| Fabrication and characterization of 5-ZGNR. a**. On-surface synthetic strategy of 5-ZGNRs. **b**. Constant current STM image of the 5-ZGNR ($V_s$ = 10 mV) **c**. d$I$/d$V$ point spectroscopy of 5-ZGNR on Au(111) at the position marked in the inset panel (red cross). Dashed line shows the Au(111) reference spectrum ($V_{ac}$ = 20 mV). Inset is the nc-AFM image acquired at constant height mode using a qPlus sensor with a CO functionalized tip ($V_s$ = 10 mV). Scale bar is 5Å. **d**, **e**, **f**. Constant-current d$I$/d$V$ map recorded at a voltage bias of +750, -225 and -470 mV respectively. ($V_{ac}$ = 10 mV). **g**, **h**, **f**. DFT-calculated LDOS at the energetic positions of A', B', and C' of the DOS spectrum as marked in panel **k**, respectively. LDOS are computed at a height of 4 Å above the atomic plane of the 5-ZGNR. **j**. BR-STM image of 5-ZGNR ($V_s$ = 10 mV). **k**. DFT-calculated DOS of spin-up (red) and spin-down (blue) of 5-ZGNR (broadened by 27 meV Gaussian). **l.** The DFT-calculated band structure of 5-ZGNR. Inset shows the calculated spin density distribution of a free-standing 5-ZGNR.

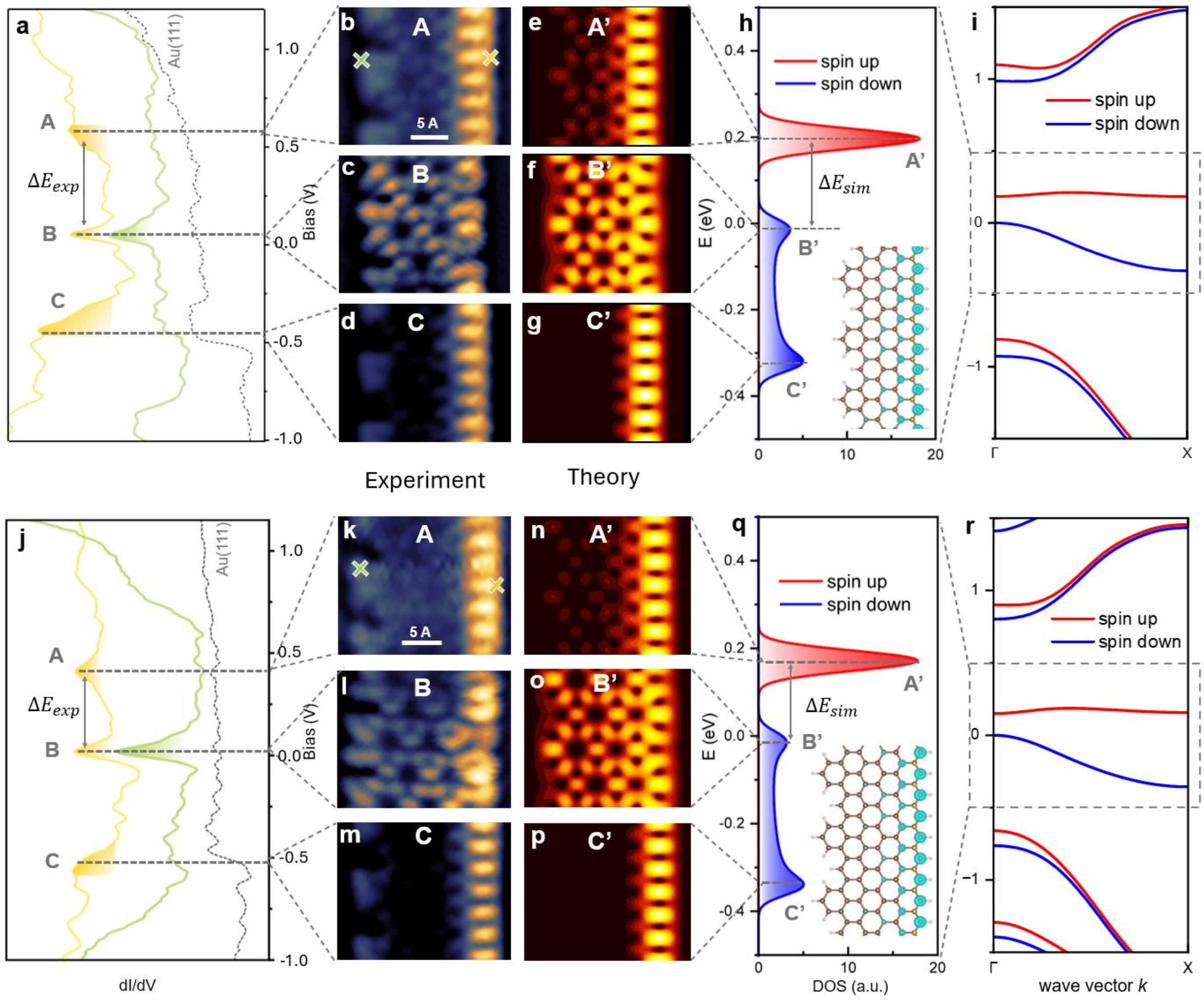

**Figure 4| Electronic structure of JGNRs. a**. d$I$/d$V$ point spectroscopy of (*4,2*)-JGNR on Au(111) at the corresponding colored position marked in panel **b** (yellow and green cross). Dashed line shows the Au(111) reference spectrum ($V_{ac}$ = 20 mV). **b, c, d.** Constant-current d$I$/d$V$ maps recorded at the energy positions of A, B, and C in panel **a**, respectively ($V_{ac}$ = 10 mV). **e, f, g.** DFT- calculated LDOS at the energetic positions of A', B', and C' marked in panel **h**, respectively. LDOS is computed at a height of 4 Å above the atomic plane of the (*4,2*)-JGNR. **h.** DFT-calculated DOS of spin-up (red) and spin-down (blue) of (*4,2*)-JGNR (broadened by 27 meV Gaussian). Inset is the calculated spin density distribution of a free-standing (*4,2*)-JGNR. **i.** The DFT calculated band structure for (*4,2*)-JGNR. **j.** d$I$/d$V$ point spectroscopy of (*5,2*)-JGNR on Au(111) at the corresponding colored position marked in panel **k** (red and blue cross). Dashed line shows the Au(111) reference spectrum ($V_{ac}$ = 20 mV). **k, l, m.** Constant-current d$I$/d$V$ maps recorded at the energy positions of A, B, and C in panel **j**, respectively ($V_{ac}$ = 10 mV). **n, o, p.** Calculated DFT LDOS at the energetic positions of A', B', and C' marked in panel **q**, respectively. LDOS is computed at a height of 4 Å

above the atomic plane of the (*5,2*)-JGNR. **q.** Calculated DFT DOS of spin-up (red) and spin-down (blue) of (*5,2*)-JGNR. Inset is the calculated spin density distribution of a free-standing (*5,2*)-JGNR. **r.** The DFT-calculated band structure of (*5,2*)-JGNR.

Supplementary information for

# Janus graphene nanoribbons with a single ferromagnetic zigzag edge

Shaotang Song[1,9], Yu Teng[1,7,9], Weichen Tang[2,4,9], Zhen Xu[3,9], Yuanyuan He[1], Jiawei Ruan[2,4], Takahiro Kojima[3], Wenping Hu[7,8], Franz J Giessibl[5], Hiroshi Sakaguchi*[3], Steven G Louie*[2,4], Jiong Lu*[1,6]

[1]Department of Chemistry, National University of Singapore, 3 Science Drive 3, Singapore 117543, Singapore.

[2]Department of Physics, University of California, Berkeley, CA94720, USA.

[3]Institute of Advanced Energy, Kyoto University, Uji, Kyoto 611-0011, Japan.

[4]Materials Sciences Division, Lawrence Berkeley National Laboratory, Berkeley, CA 94720, USA.

[5]Institute of Experimental and Applied Physics, University of Regensburg, 93053 Regensburg, Germany.

[6]Institute for Functional Intelligent Materials, National University of Singapore, Singapore, Singapore.

[7]Joint School of National University of Singapore and Tianjin University, International Campus of Tianjin University, Fuzhou 350207, China.

[8]Tianjin Key Laboratory of Molecular Optoelectronics, Department of Chemistry, School of Science, Tianjin University, Tianjin, 300072 China.

[9]These authors contribute equally: Shaotang Song, Yu Teng, Weichen Tang, Zhen Xu

E-mail: sakaguchi@iae.kyoto-u.ac.jp (H.S.); sglouie@berkeley.edu (S.G.L.); chmluj@nus.edu.sg (J.L)

## 1. Theoretical calculations

### 1.1 $Z_2$ index and Z index in GNR systems

The $Z_2$ index is a topological invariant that characterizes the electronic structure of a material. The $Z_2$ index (0 or 1) distinguishes between trivial and non-trivial topological phases. In the trivial phase ($Z_2$=0), there are no protected edge states at the interface with vaccum . In contrast, the non-trivial phase ($Z_2$=1) has robust edge states (or end states in the case of 1D systems) at the interface with vaccum that are protected against perturbations [1].

However, the $Z_2$ theory requires the system to have time-reversal and spatial symmetry and does not precisely predict the exact number of protected edge states. To address this, a more robust topological classification based on chiral symmetry was developed, as chiral symmetry is a good approximate symmetry for GNR structures [2].

This classification led to the chiral phase index (Z index) for GNRs, where different topological classes are given by an invariant that is an integer Z. Different Z (which can take on any integer values) indices correspond to different topological phases. The trivial phase (Z=0) does not have protected end states at the vacuum/material interface, while the non-trivial phases (Z≠0) have robust end states. The number of end states is given by the Z index [2].

**1.2 The Z index for the end of AGNR and the edge of JGNR**

The Z index of the AGNRs, with specific widths and end terminations, has been shown to have an explicit form [2]. This involves simply counting the number of unconnected carbon pairs and the total number of carbon pairs in the unit cell of the AGNR that is commensurate with its end (as shown by the area enclosed by the orange dashed line in Fig. S1) [2]:

$$Z = W_{notco} - \lfloor \frac{W}{3} \rfloor \qquad (1).$$

Here, $W$ represents the total number of rows of carbon atoms forming the width of the AGNR, and $W_{\text{notco}}$ denotes the number of rows of atoms with carbon pairs not connected by σ-bonds within the specific unit cell that is commensurate to an anticipated termination. The floor function, indicated by the topless brackets, takes the largest integer less than or equal to the value within the brackets.

When considering the edges of a JGNR (which includes a TDZ edge and a regular zigzag edge), we can always find an associated wide AGNR whose end matches the edge of the JGNR (as illustrated in Fig. S1). This mapping enables us to calculate the number of edge states at either edge of the JGNR by calculating the number of end states of the corresponding AGNR.

In Fig. S1, each unit cell of the JGNR is enclosed by two grey dot lines and the unit cells are labelled 1,2,3 … $L$, with a total of $L$ unit cells of a finite-length ribbon (note that the unit cell of JGNR is different from that of AGNR). In mapping the edge of JGNR to the end of a wide AGNR, the corresponding AGNR contains $L$ repeat units along its width direction. This means that $W$, the width of the wide AGNR, equals $L$ times the width of each repeat unit, and $W_{\text{notco}}$ equals $L$ times the number of not connected carbon pair in each repeat unit. This relationship will be used in detailed calculations in section S1.3.

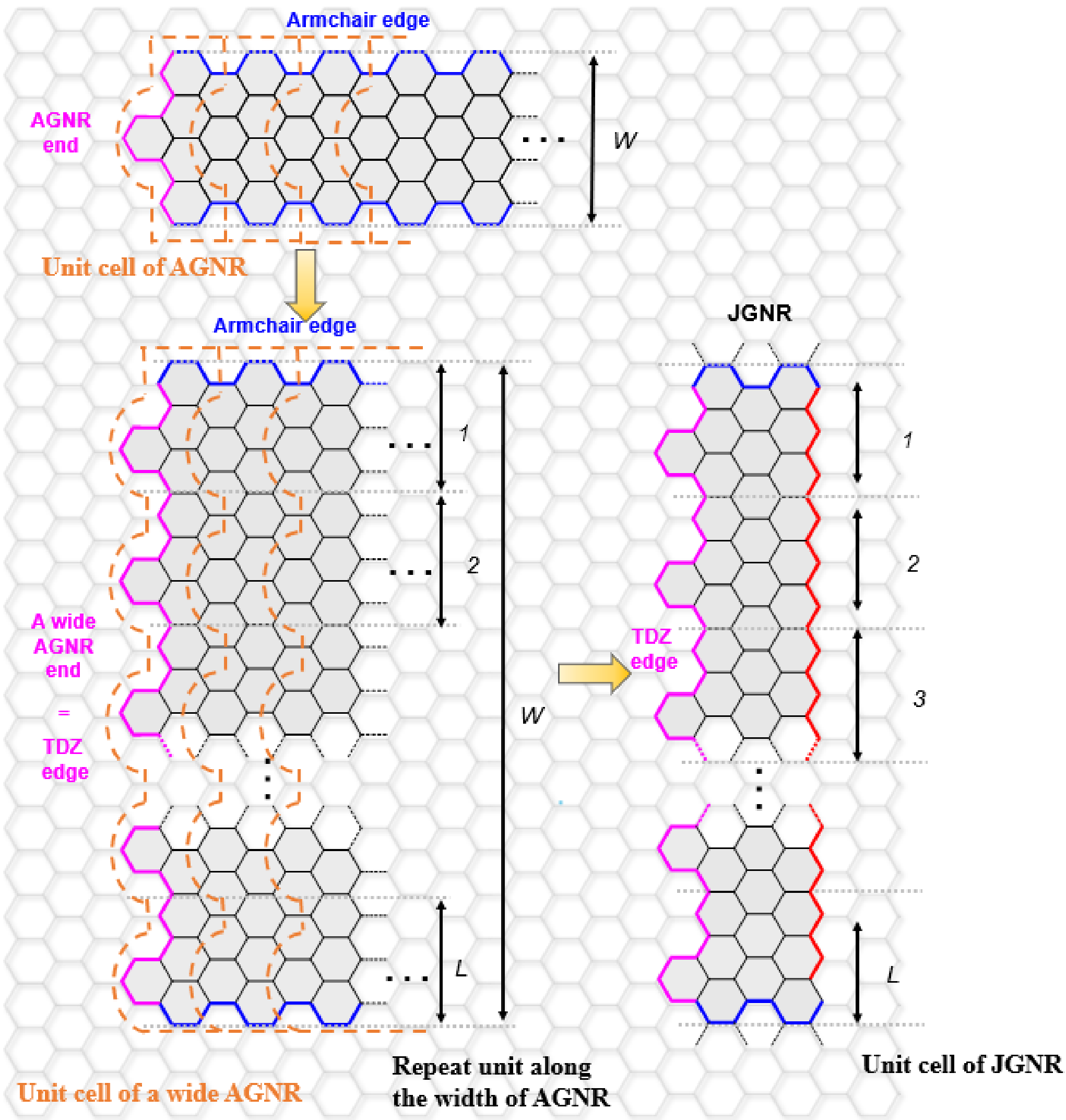

**Figure S1.** The correspondence between the end of a wide AGNR and the edge of JGNR.

**1.3 The calculation of number of edge states for JGNRs with different edges**

In the main text and section S1.2, we made the argument that the topological edge states of a ZGNR/JGNR can be deduced from the topology of an associated wide AGNR with different structural ends, and present the results of the $Z$ index of the AGNR that determine the edge states for the different JGNRs. Here, we provide the calculations of different edges in more detail.

First, we map the edge of the ZGNR/JGNR to the end of a wide AGNR. Fig. S2(a), (b), (c), and (d) show the corresponding wide AGNR for the $m$ = 1 TDZ edge, $m$ = 2 TDZ edge, $m$ = 3 TDZ edge, and a regular zigzag edge, respectively. Next, we applied Eq. (1) to compute the Z index. The detailed calculations of Z for the TDZ and pristine edge of certain finite-length JGNRs, when considered as the ends of wide AGNRs, are as follows:

(a) For the case where the distance between additional benzene rings on the TDZ edge is $m = 1$, as shown in Fig. S2(a) and there are $L$ additional benzene rings in the JNGR, the number of unconnected carbon pair is $L$ and the number of connected carbon pair is $3L$ in the unit cell of the corresponding wide AGNR. The total number of carbon pairs is $4L$, and $Z = L - \frac{4L}{3} = -\frac{1}{3}L$ for the unit cell.

(b) For the case where the distance between additional benzene rings on the TDZ edge is $m = 2$, as shown in Fig. S2(b), the number of unconnected carbon pair is $2L$ and the number of connected carbon pair is $4L$. The total number of carbon pair is $6L$ and $Z = 2L - \frac{6L}{3} = 0$ for the unit cell. That means no edge states for the TDZ edge in this case.

(c) For the case where the distance between additional benzene rings on the TDZ edge is $m = 3$, as shown in Fig. S2(c), the number of unconnected carbon pair is $3L$ and the number of connected carbon pair is $5L$ . The total number of carbon pair is $8L$ and $Z = 3L - \frac{8L}{3} = \frac{1}{3}L$ in the unit cell.

(d) For a regular zigzag edge, as shown in Fig. S2(d), the number of unconnected carbon pair is 1/benzene ring and the number of connected carbon pair is 1/benzene ring. The total number of carbon pair is 2/ benzene ring and $Z = 1 - \frac{2}{3} = \frac{1}{3}$/ benzene ring. Notice that when a unit cell contains $L$ benzene rings at the regular zigzag edge, the corresponding $Z = \frac{1}{3}L$ in the unit cell.

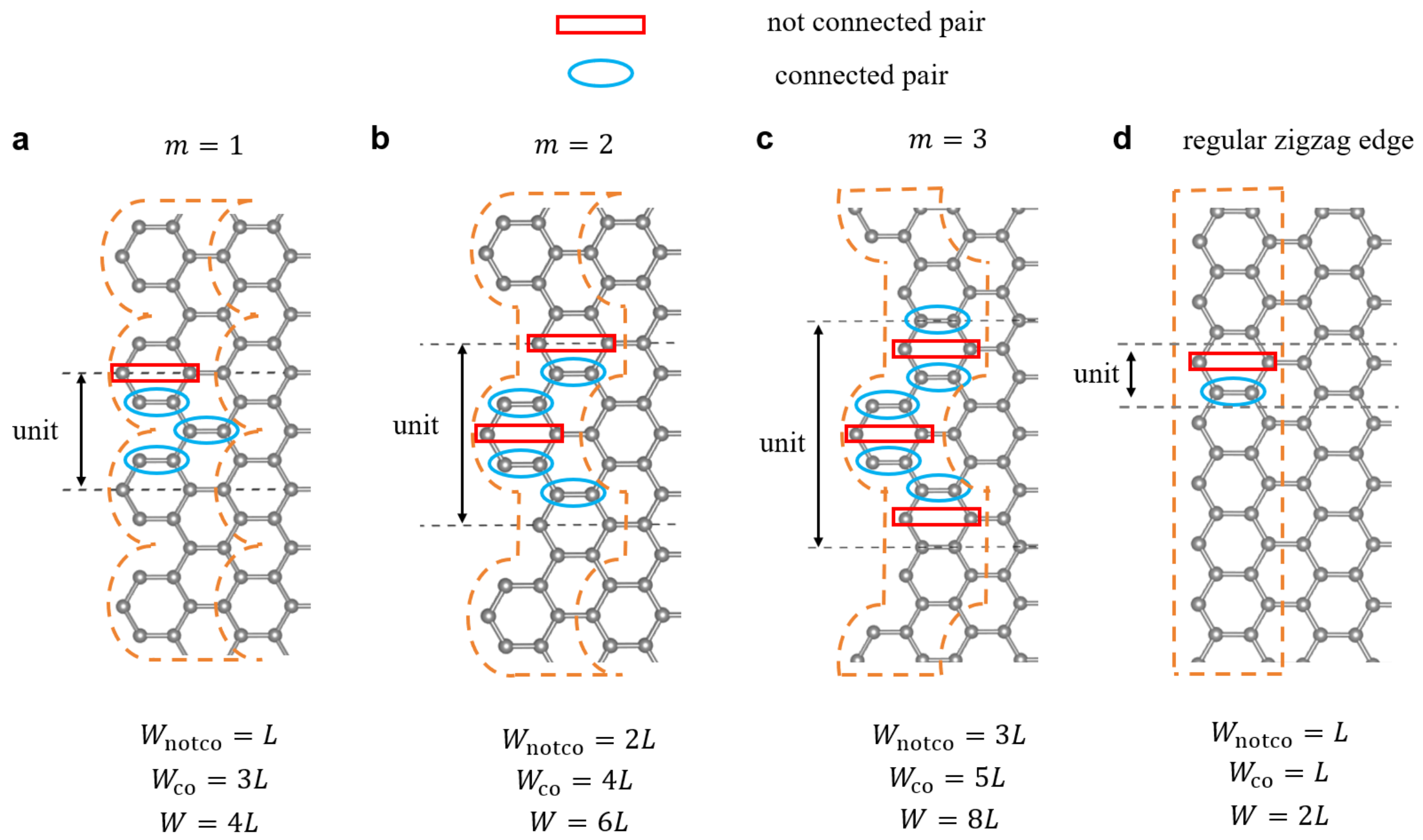

**Figure S2.** Sketch of the defective edge of JGNR with different $m$ value, which is mapped to the end of a large width AGNR with unit cell enclosed by the orange dashed curve that is commensurate with its termination on the left (the topological defect edge of interest). $W$ is the number of carbon rows forming the length of the finite-size JGNR (which is equivalent to the number of carbon rows along the width of the associated AGNR). $W_{co}$ and $W_{notco}$ are the number of connected and unconnected carbon pairs, respectively, in the unit cell of the wide AGNR. $L$ is the number of repeat units along the width of the associated AGNR direction (i.e., $L$ is number of unit cells forming the length of the finite-size ZGNR of physical interest), and the repeat unit is denoted by the area between the two grey dashed lines. The relationships between $L$ and $W$, $W_{co}$ *and* $W_{notco}$ are shown. **a**. A defect edge with $m = 1$, **b**. A defect edge with $m = 2$, **c**. A defect edge with $m = 3$, **d**. A regular zigzag edge, the repeat unit along the width of AGNR direction is a benzene ring. These quantities are important in the calculation of $Z$ index.

### 1.4 Electronic structure of JGNRs with different edges

In the main text, we present predictions on the ground state spin configurations for the JGNRs based on the chiral phase index (CPI) Z and Lieb's theorem. For the case in which the distance between additional benzene rings on the TDZ edge is $m = 2$, we validate our results through direct verification using first-principles DFT calculations, as detailed in the main text. Here, we extend our analysis by providing a direct verification of our results for other $m$ values through first-principles DFT calculations. To effectively illustrate the localization of the edge states, we utilize wider samples to showcase our results. Since the physical properties near the Fermi energy are dominated by the edge states, the width of samples only quantitatively changes the low-energy physics.

(1) The distance between additional benzene rings on the TDZ edge is $m = 1$.

In this case, the $Z$ index for the TDZ edge equals to $-\frac{1}{3}$ per additional benzene ring, and the $Z$ index for the zigzag edge equals to $\frac{2}{3}$ per benzene ring. It is essential to highlight that the positive and negative signs in this context do not directly correspond to the spin configuration; instead, they indicate which sublattice the edge states will localize. However, as given in Ref. 1, a negative sign for $Z$ on the left end gives the same sublattice localization as a positive sign on the right end, and the same sublattice sites have same spin polarization. The sublattice imbalance is equal to 1 per unit cell. Consequently, according to Lieb's theorem, the spin configuration for the system comprises $\frac{1}{3}$ (along the length of the JGNR) spin-down edge states per unit cell at the TDZ edge and $\frac{2}{3}$ spin down edge states per unit cell at the zigzag edge. This configuration is confirmed by the spin density calculated by DFT, as illustrated in Fig. S3 (a). Further examination of the band structure and wave function characteristics in Fig. S3 (b)-(d), enables the identification of the distribution of these edge states in the Brillouin zone. In the highest valence band, states around two-thirds of the Brillouin zone near the Γ point are localized at the zigzag edge, while states around one-third of the Brillouin zone near the X point are localized at the TDZ edge.

(2) The distance between additional benzene rings on the TDZ edge is $m = 3$.

In this case, the $Z$ index for the TDZ edge equals to $\frac{1}{3}$ per additional benzene ring (or unit cell along the JGNR), and the $Z$ index for the zigzag edge equals to $\frac{4}{3}$ per such unit cell. Thus, the edge states of the two sides are on opposite sublattice sites. The sublattice imbalance is equal to 1 per unit cell. Therefore, to satisfy Lieb's theorem (and also from the sign of the $Z$ values), the spin configuration for the system should be $\frac{1}{3}$ spin-up edge states per unit cell at the TDZ edge and $\frac{4}{3}$ spin down edge states per unit cell at the zigzag edge. This configuration is confirmed by the spin density distribution calculated by DFT, as shown in Fig. S4 (a). A detailed analysis of the band structure and wave function characteristics shown in Fig. S4 (b)-(d) enables the identification of the distribution of these edge states within the Brillouin zone. The states in the highest spin-down valence band and those at around one-third of the Brillouin zone near the X point in the second highest spin-down valence band are localized at the zigzag edge. The states at around one-third of the Brillouin zone near the X point in the highest spin-up valence band are localized at the defect edge.

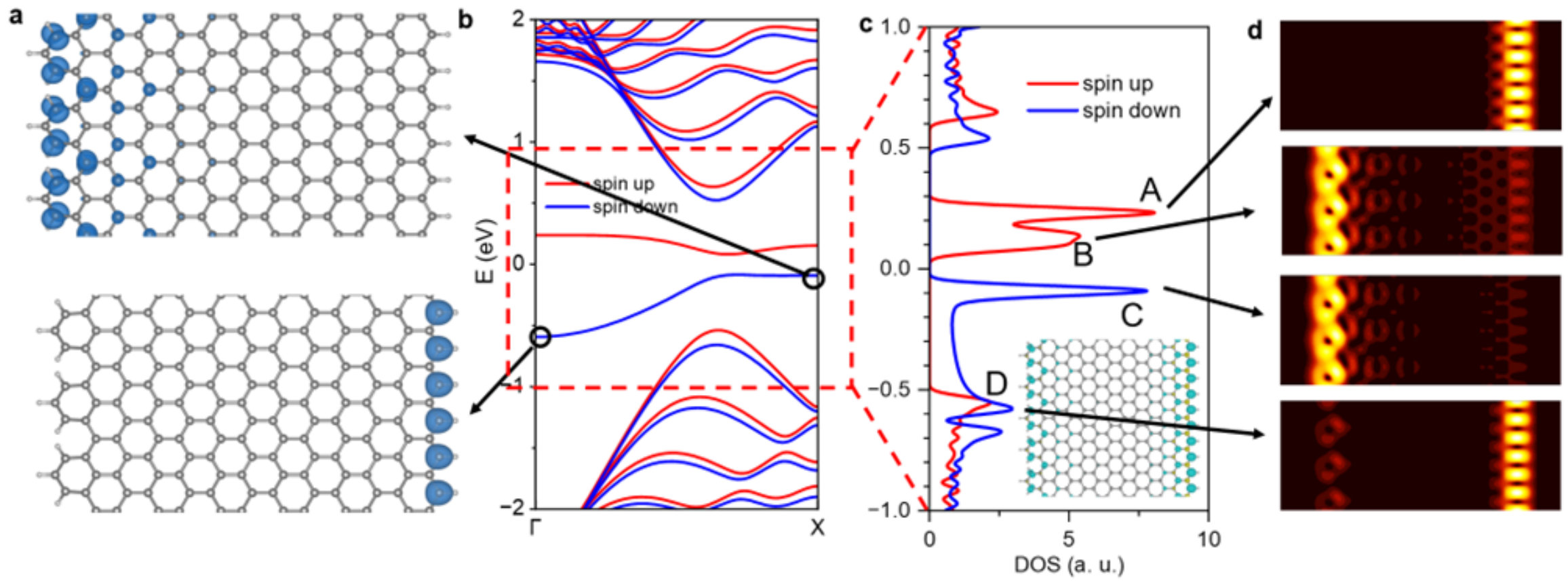

**Figure S3.** The DFT calculations of (12,1)-JGNR. The distance between additional benzene rings on the TDZ edge being $m = 1$. **a**. Calculated wavefunction character for the first valence band at the X point (upper panel) and Γ point (lower panel), respectively. **b**. DFT calculated band structure. **c**. DFT calculated DOS of spin-up (red) and spin-down (blue) of (12,1)-JGNR. Inset is calculated spin density distribution. **d**. DFT calculated LDOS at the energetic positions of A, B, C and D marked in panel **c**, respectively. LDOS is computed at a height of 4 Å above the atomic plane of the (12,1)-JGNR.

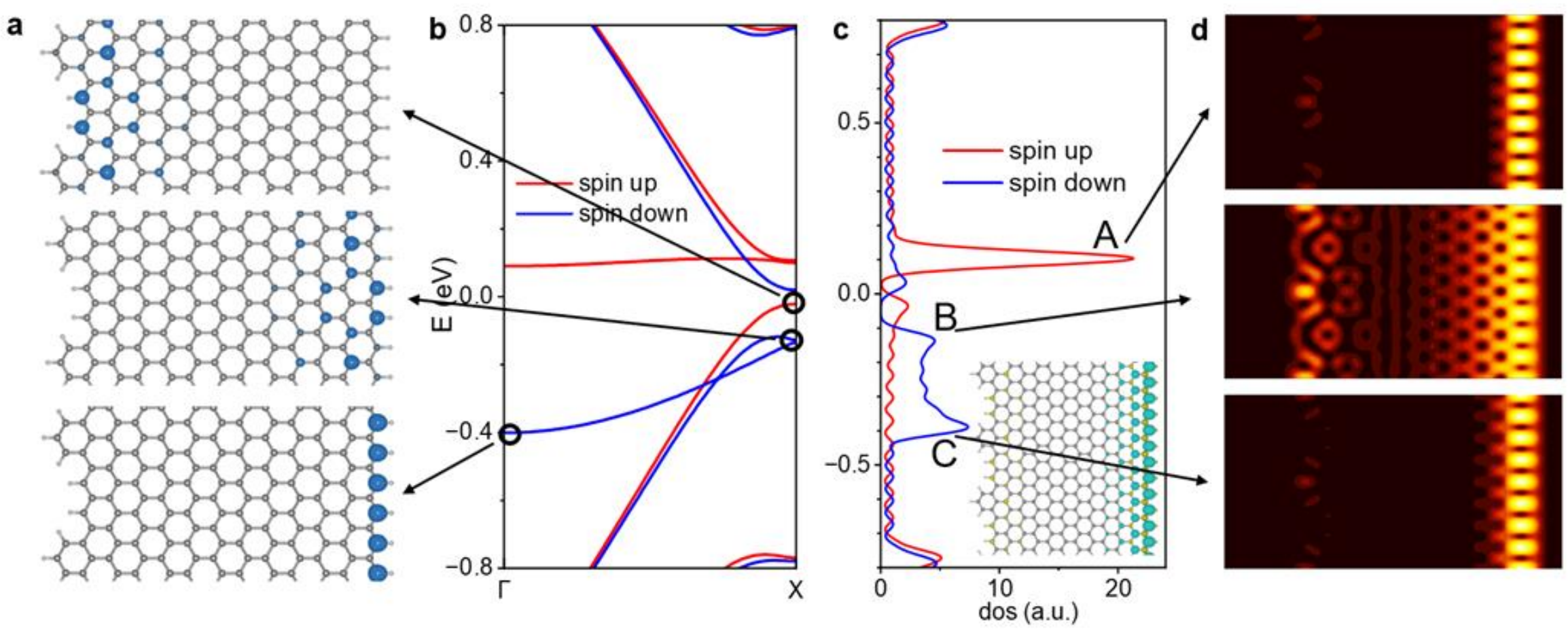

**Figure S4.** The DFT calculations of (*12*,*3*)-JGNR. The distance between additional benzene rings on the defect edge being $m = 3$. **a**. Wavefunction character for the first spin up valence band at the X point (top panel), the second spin down valence band at the X point (middle panel) and the first spin down valence band at the Γ point (bottom panel), respectively. **b**. DFT calculated band structure. **c**. DFT calculated DOS of spin-up (red) and spin-down (blue) of (*12*,*3*)-JGNR. Inset is the calculated spin density distribution. **d**. Calculated DFT LDOS at the energetic positions of A, B, C marked in panel **c**, respectively. LDOS is computed at a height of 4 Å above the atomic plane of the (*12*,*3*)-JGNR.

**1.5 Stability of the JGNRs' electronic structure and magnetism upon charge doping**

In the main text, our DFT simulations were done for the JGNRs in the vacuum. However, in the experiment, the properties of JGNRs were measured on a gold substrate. It is known that doping and hybridization/screening effects of gold tend to reduce the magnetic order in the GNRs [3,4]. Therefore, we discuss the effects of charge doping here, demonstrating that the ferromagnetism and localized spin density of JGNRs remain stable against hole doping and can survive on a gold substrate.

Since the electronic properties of (4,2)-JGNR and (5,2)-JGNR are similar near the Fermi level, and charge doping primarily affects the states near the Fermi level, we use (4,2)-JGNR to illustrate the effect of charge doping. The work function of JGNRs, calculated by DFT, is around 4.2 eV; therefore, hole doping is expected. Since the work function of a gold surface can vary depending on the environment, we calculated the electronic structures of JGNRs under several different hole doping levels and determined the experimental value of hole doping by matching the theoretical results with experimental results. Fig. S5, S6, and S7 show the band structures, spin densities, and the simulated $dI/dV$ mapping of (4,2)-JGNR under 0.1, 0.2, and 0.4 holes doping per unit cell. Note that in the experiment, peak B is consistently around 0.0 eV (with the exact position varying from sample to sample); therefore, we can estimate that the hole doping level is around 0.1 holes per unit cell.

From the results, we can see that the electronic structure of the system is stable up to 0.2 holes per unit cell. The localized spin density is even more robust against hole doping up to 0.4 holes per unit cell. This is because the states that contribute to the localized spin density at the zigzag edge are from the first valence band near the X point, which are deeper in the Fermi Sea and hardly impacted by the doping effect.

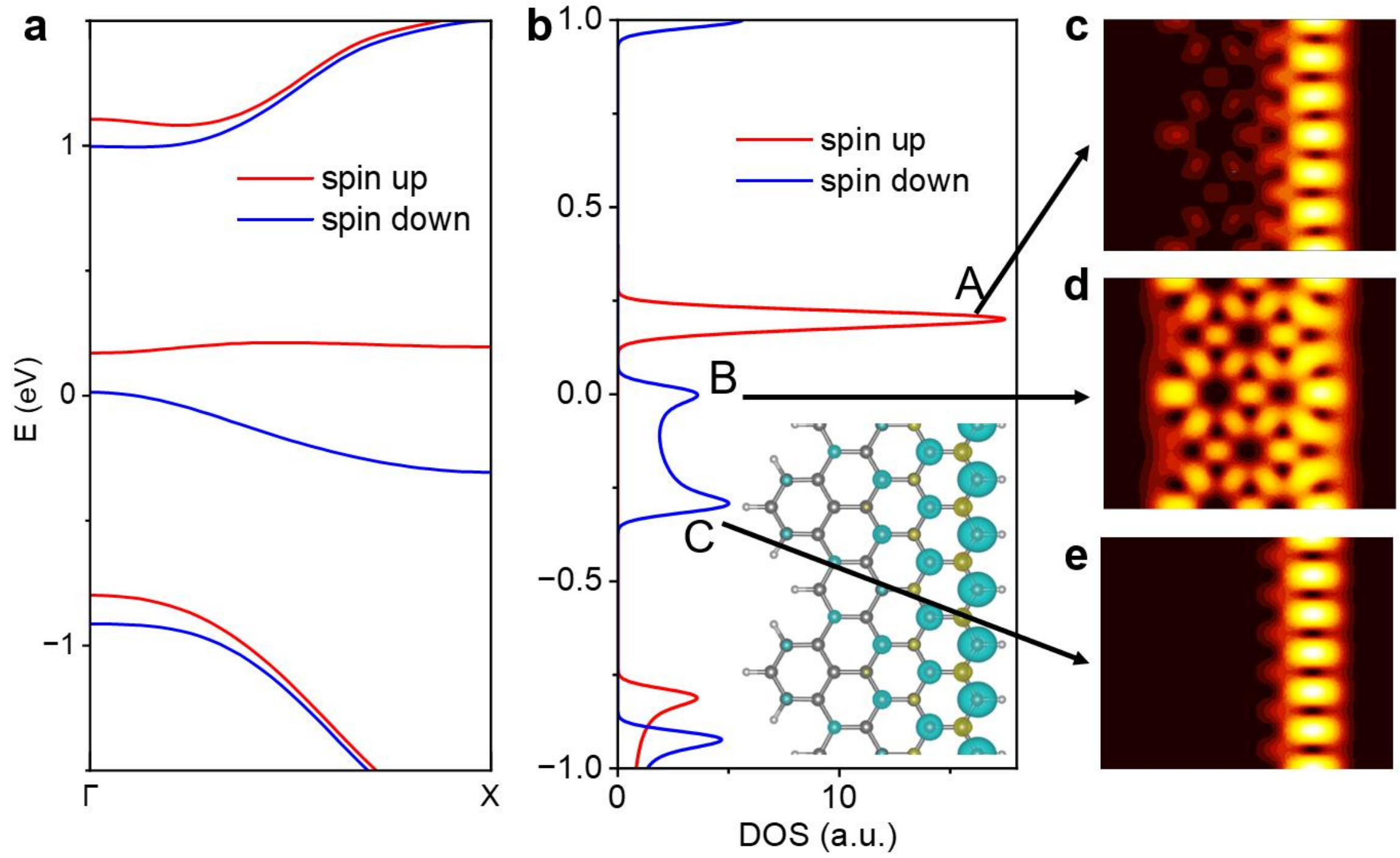

**Figure S5.** The DFT calculations of (*4*,*2*)-JGNR under 0.1 holes doping per unit cell. **a**. DFT calculated band structure. **b**. Calculated DFT DOS of spin-up (red) and spin-down (blue) of (*4*,*2*)-JGNR, inset is calculated spin density distribution. **c**, **d**, **e**. Calculated DFT LDOS at the energetic positions of A, B, C marked in panel **b**, respectively. LDOS is computed at a height of 4 Å above the atomic plane of the (*4*,*2*)-JGNR.

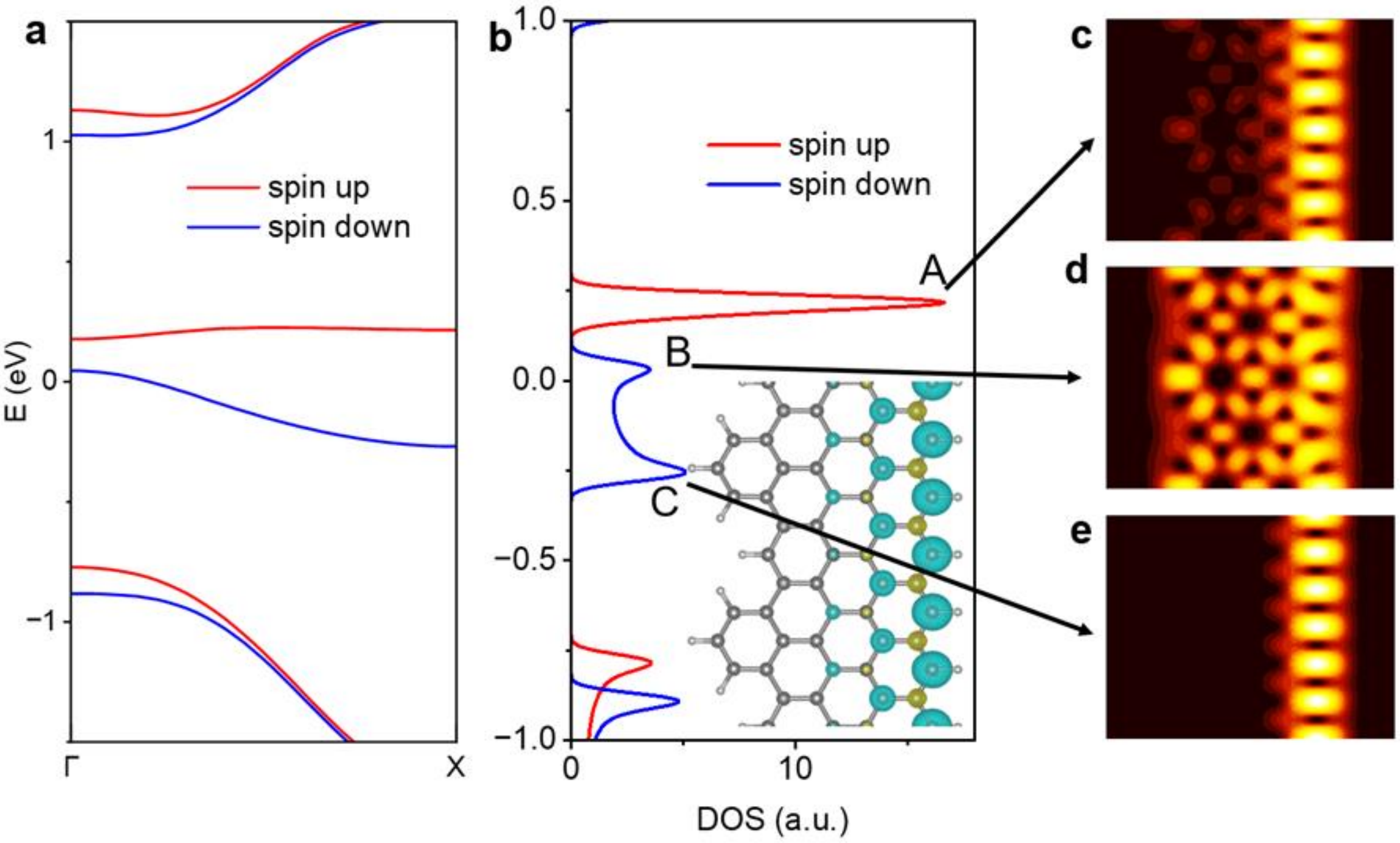

**Figure S6.** The DFT calculations of (4,2)-JGNR under 0.2 holes doping per unit cell. **a**. DFT calculated band structure. **b**. Calculated DFT DOS of spin-up (red) and spin-down (blue) of (4,2)-JGNR, inset is calculated spin density distribution. **c**, **d**, **e**. Calculated DFT LDOS at the energetic positions of A, B, C marked in panel **b**, respectively. LDOS is computed at a height of 4 Å above the atomic plane of the (4,2)-JGNR.

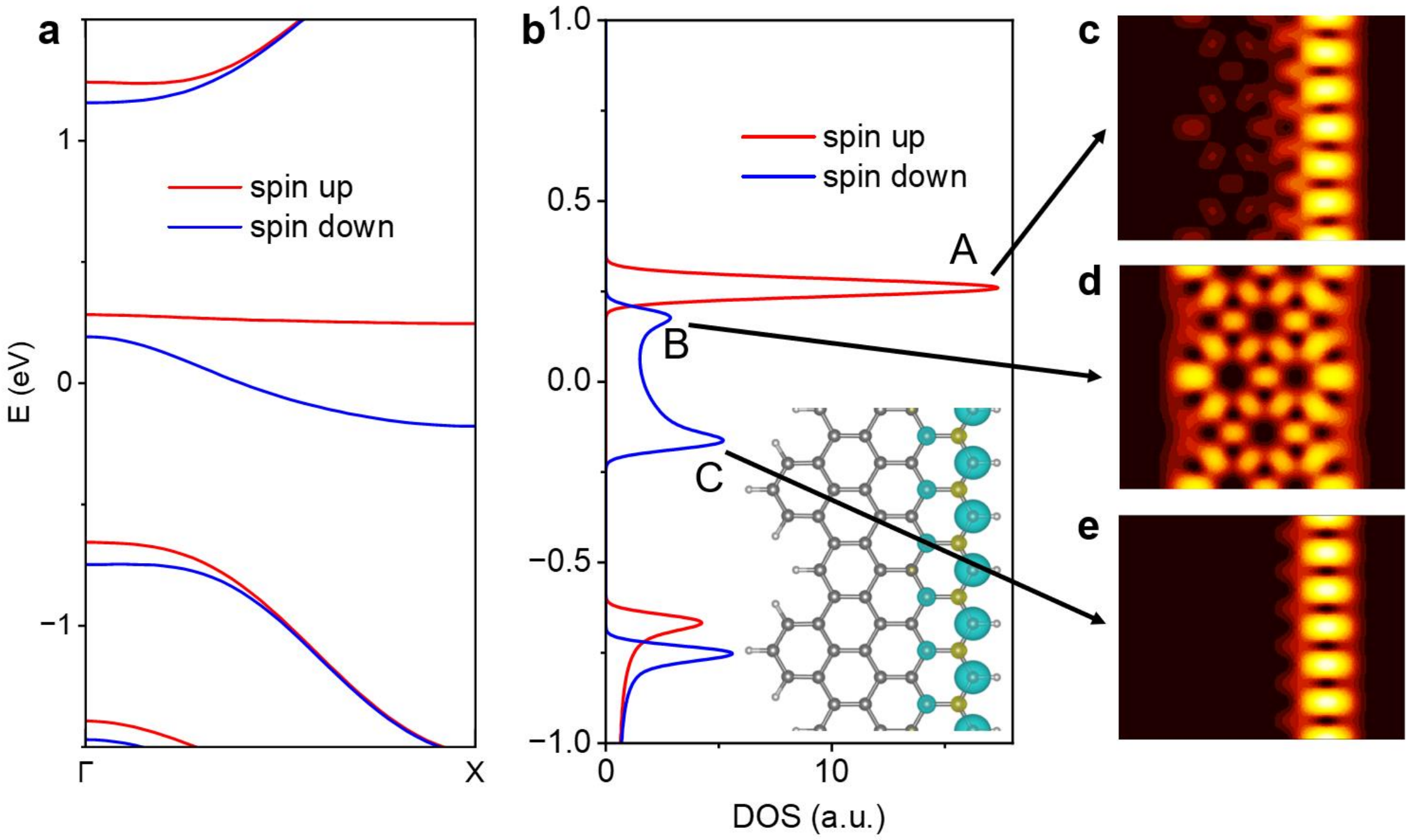

**Figure S7.** The DFT calculations of (*4*,*2*)-JGNR under 0.4 holes doping per unit cell. **a**. DFT calculated band structure. **b**. Calculated DFT DOS of spin-up (red) and spin-down (blue) of (*4*,*2*)-JGNR, inset is calculated spin density distribution. **c**, **d**, **e**. Calculated DFT LDOS at the energetic positions of A, B, C marked in panel **b**, respectively. LDOS is computed at a height of 4 Å above the atomic plane of the (*4*,*2*)-JGNR.

### 1.6 The electronic structure and the simulated d*I*/d*V* mapping of non-magnetic phase

In our DFT simulation, we can force the system to be in a non-magnetic phase and calculate the electronic structure of the system. The electronic band structures and the simulated d*I*/d*V* mappings of the non-magnetic phase for (*4*,*2*)-JGNR under different doping levels are presented in Figures S8, S9, and S10. Unlike the ferromagnetic phase, the DOS spectrum shows only one or two peaks near the Fermi surface. Note that in the DFT calculation, the ferromagnetic phase of the system always has lower energy compared to the non-magnetic phase and is theoretically more stable.

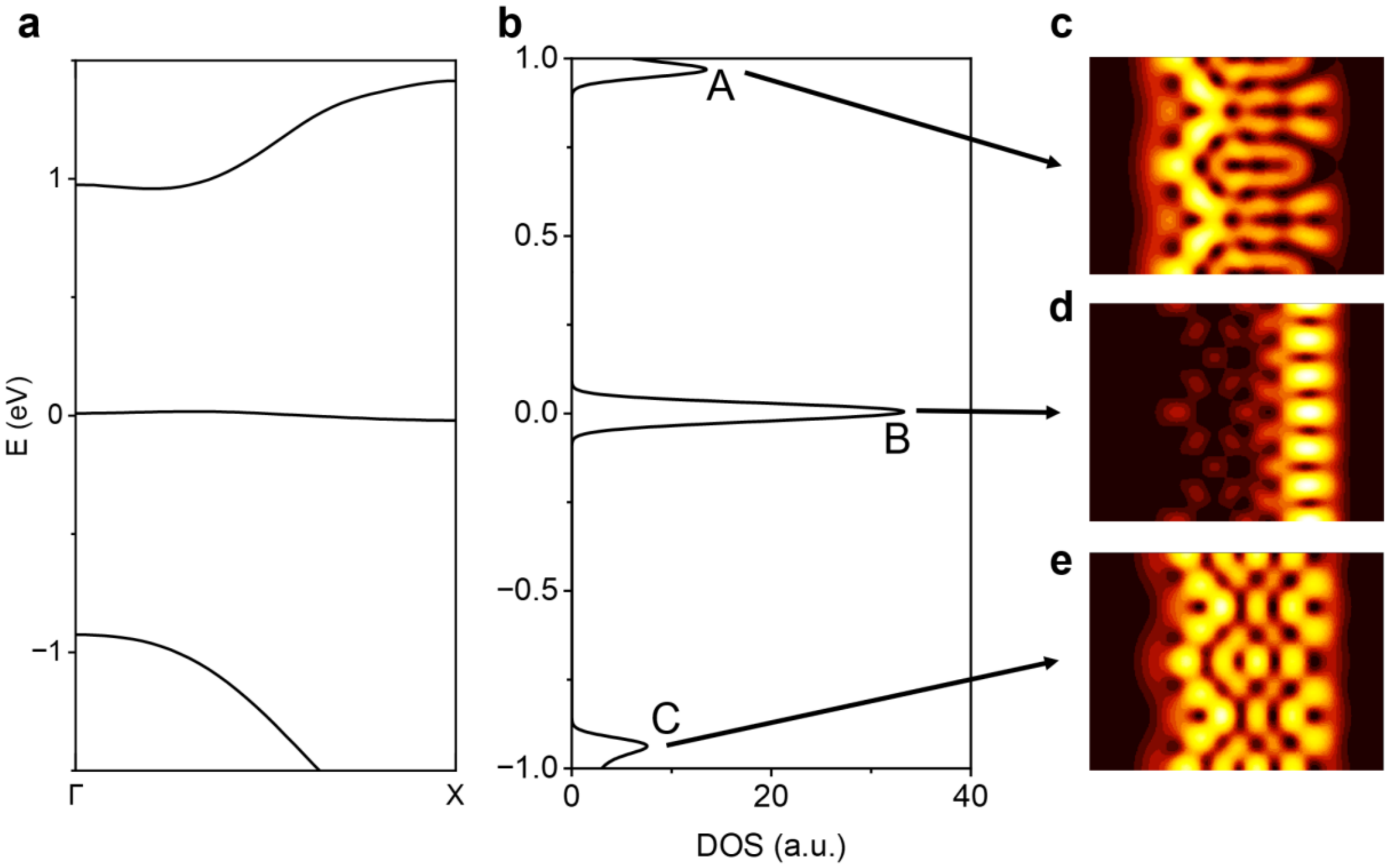

**Figure S8.** The DFT calculations of the non-magnetic phase of (*4*,*2*)-JGNR under 0.1 holes doping per unit cell. **a**. DFT calculated band structure. **b**. Calculated DFT DOS of (*4*,*2*)-JGNR. **c**, **d**, **e**. Calculated DFT LDOS at the energetic positions of A, B, C marked in panel b, respectively. LDOS is computed at a height of 4 Å above the atomic plane of the (*4*,*2*)-JGNR.

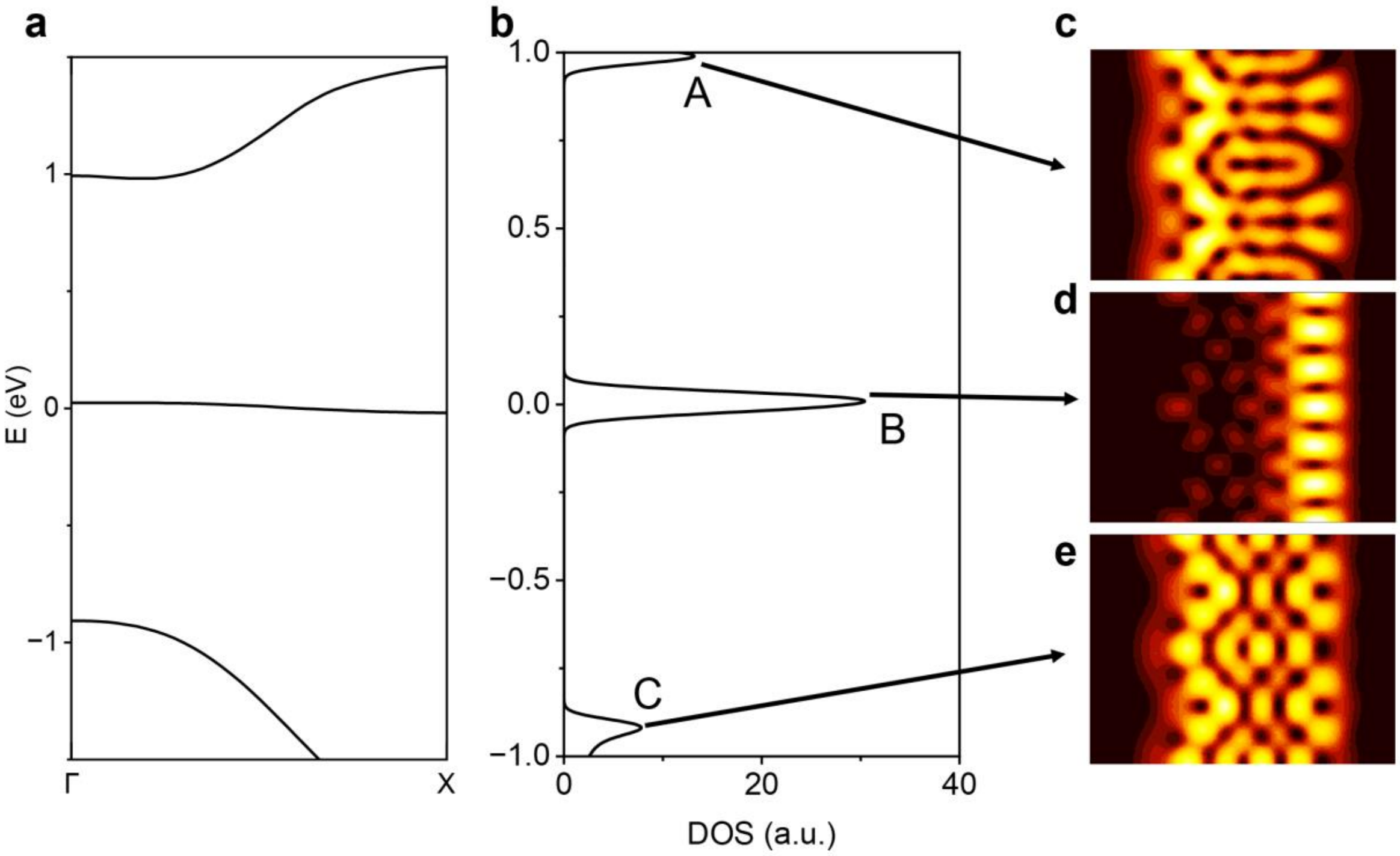

**Figure S9.** The DFT calculations of the non-magnetic phase of (*4*,*2*)-JGNR under 0.2 holes doping per unit cell. **a**. DFT calculated band structure. **b**. Calculated DFT DOS of (*4*,*2*)-JGNR. **c**, **d**, **e**. Calculated DFT LDOS at the energetic positions of A, B, C marked in panel **b**, respectively. LDOS is computed at a height of 4 Å above the atomic plane of the (*4*,*2*)-JGNR.

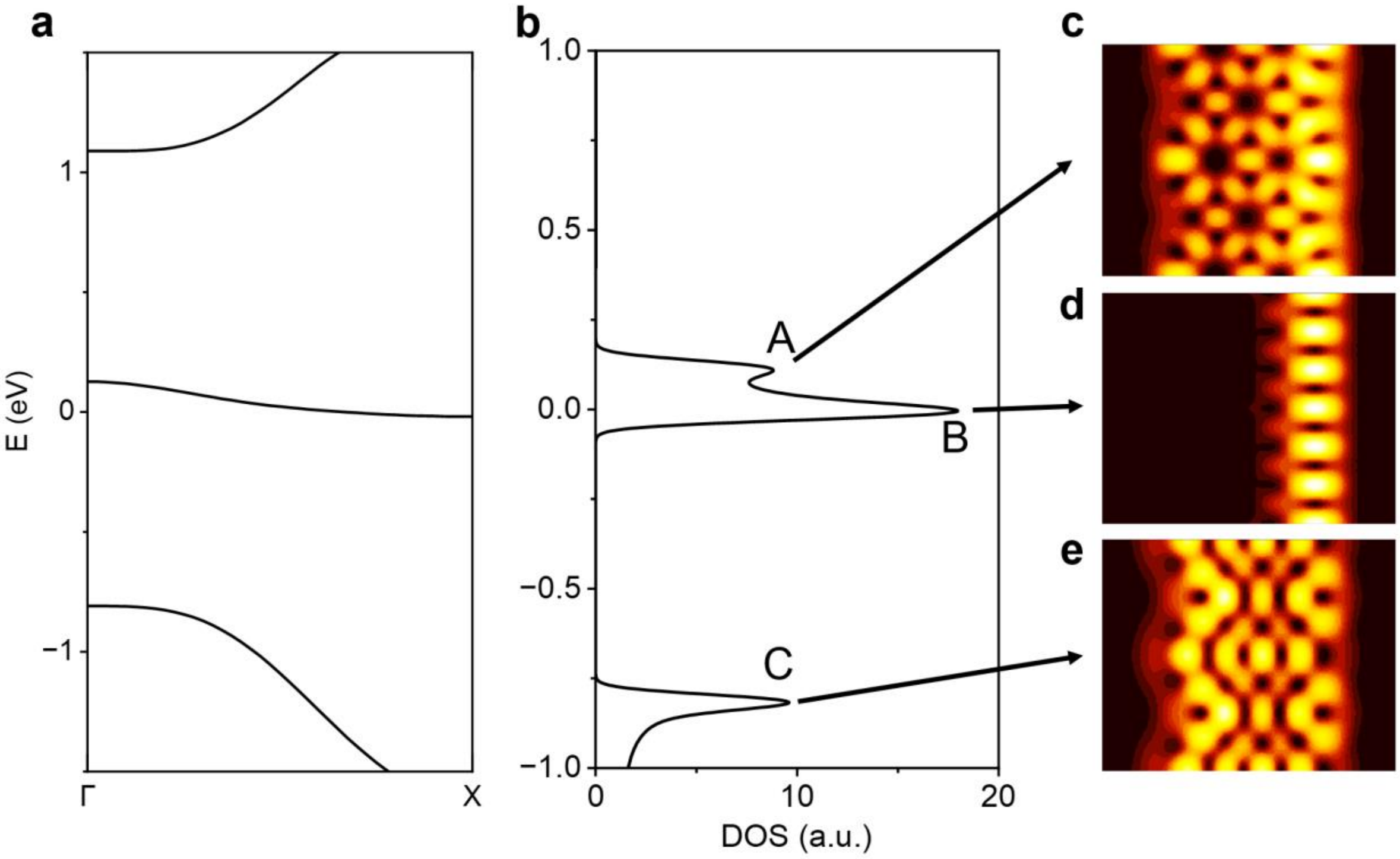

**Figure S10.** The DFT calculations of the non-magnetic phase of (*4*,*2*)-JGNR under 0.4 holes doping per unit cell. **a**. DFT calculated band structure. **b**. Calculated DFT DOS of (*4*,*2*)-JGNR. **c**, **d**, **e**. Calculated DFT LDOS at the energetic positions of A, B, C marked in panel **b**, respectively. LDOS is computed at a height of 4 Å above the atomic plane of the (*4*,*2*)-JGNR.

## 2. On-surface synthesis and electronic structure characterization of JGNRs and ZGNR

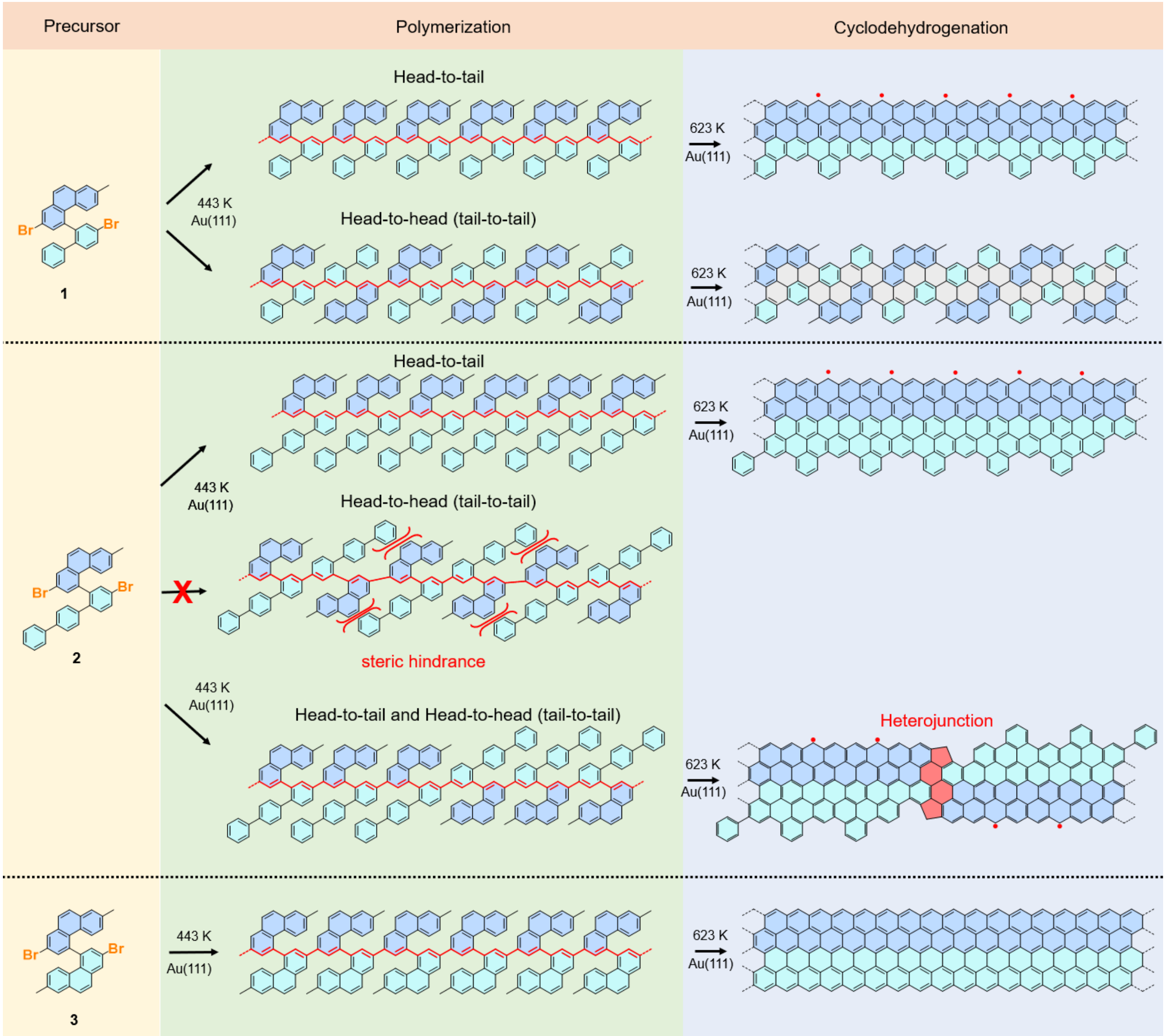

**Figure S11.** On-surface reaction schemes of the three precursors. Head-to-tail polymerization and subsequential cyclodehydrogenation of precursor **1** and **2** leads to the formation of target JGNRs. Head-to-head (tail-to-tail) polymerization of precursor **1** lead to GNRs with alternating cove and gulf regions on the edges. Steric hindrance prevents the head-to-head (tail-to-tail) polymerization of precursor **2**. Heterojunction structure derived from precursor **2** is formed by head-to-head (tail-to-tail) polymerization of two head-to-tail polymerized polymer fragments. Note, one possible approach for improving the yield is to fuse the two branches of the precursors to form rigid and planar backbone polymers.

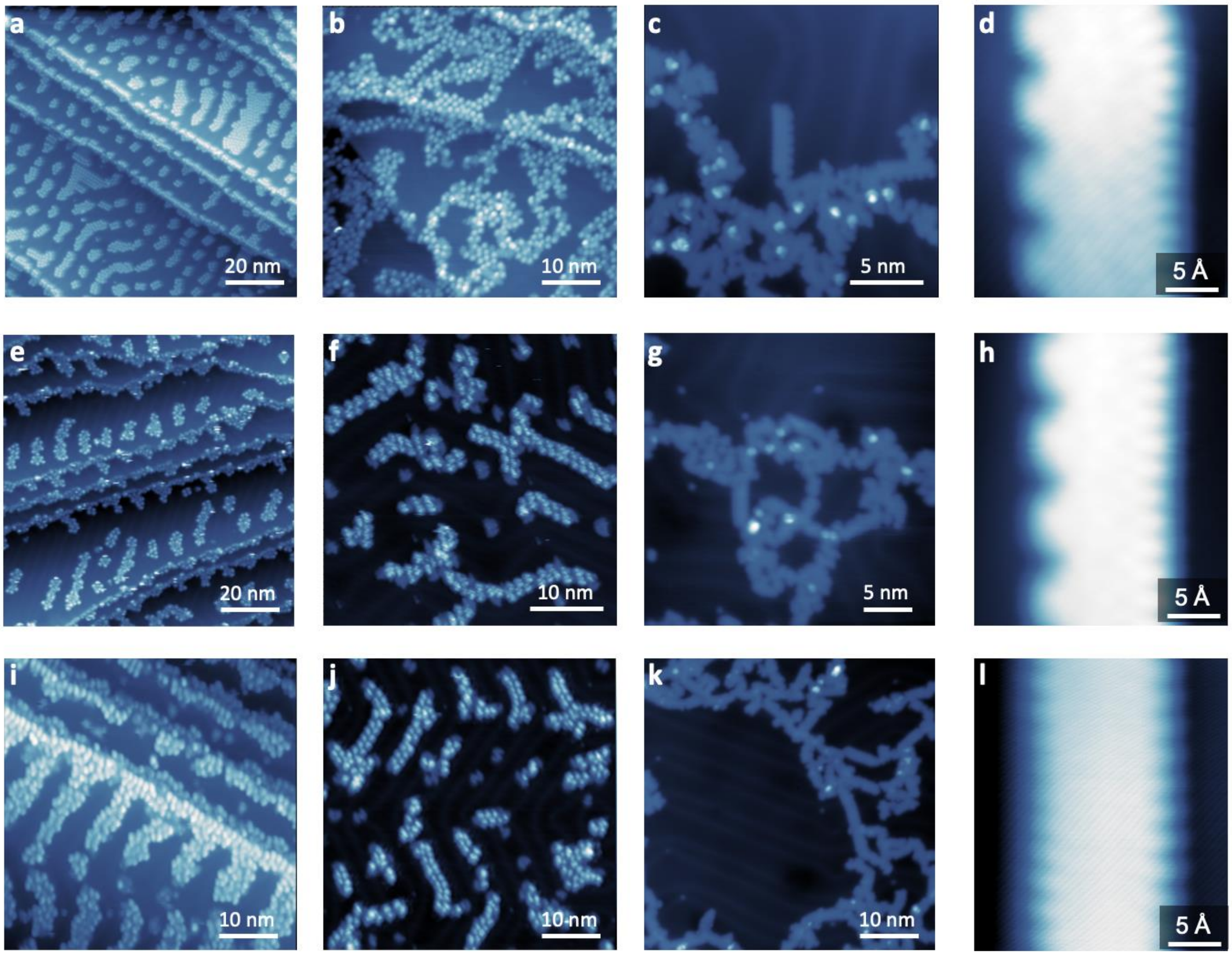

**Figure S12. a, e, f.** STM topographic image of molecular precursor **1, 2,** and **3** as deposited on Au(111), respectively. **b, f, j.** The STM topographic images of polymer chains obtained by annealing the sample of **a, e, f** up to 473, 473, and 423 K, respectively. **c, g, k.** STM topographic images of fully cyclized (*5,2*)-JGNR, (*4,2*)-JGNR and 5-ZGNR through a subsequent annealing at 623, 623, and 573 K ($V_s$ = 1 V, $I_t$ = 100 pA), respectively. **d, h, l.** Constant current zoom-in STM images of an individual (*5,2*)-JGNR, (*4,2*)-JGNR and 5-ZGNR, respectively ($V_s$ = 800 mV, $I_t$ = 200 pA, using a CO-functionalized tip).

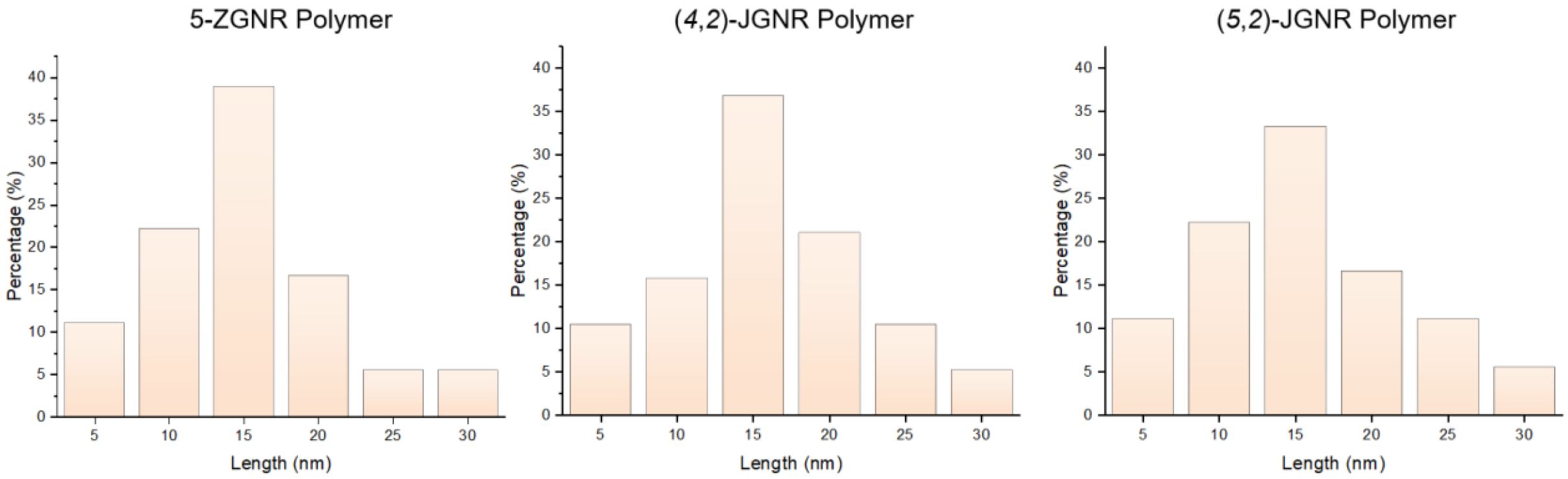

**Figure S13**. The statistical analysis of polymer-length distributions shows that all three types of polymers exhibit a normal distribution, ranging from approximately 5 to 30 nm, with the largest portion centered around 15 nm.

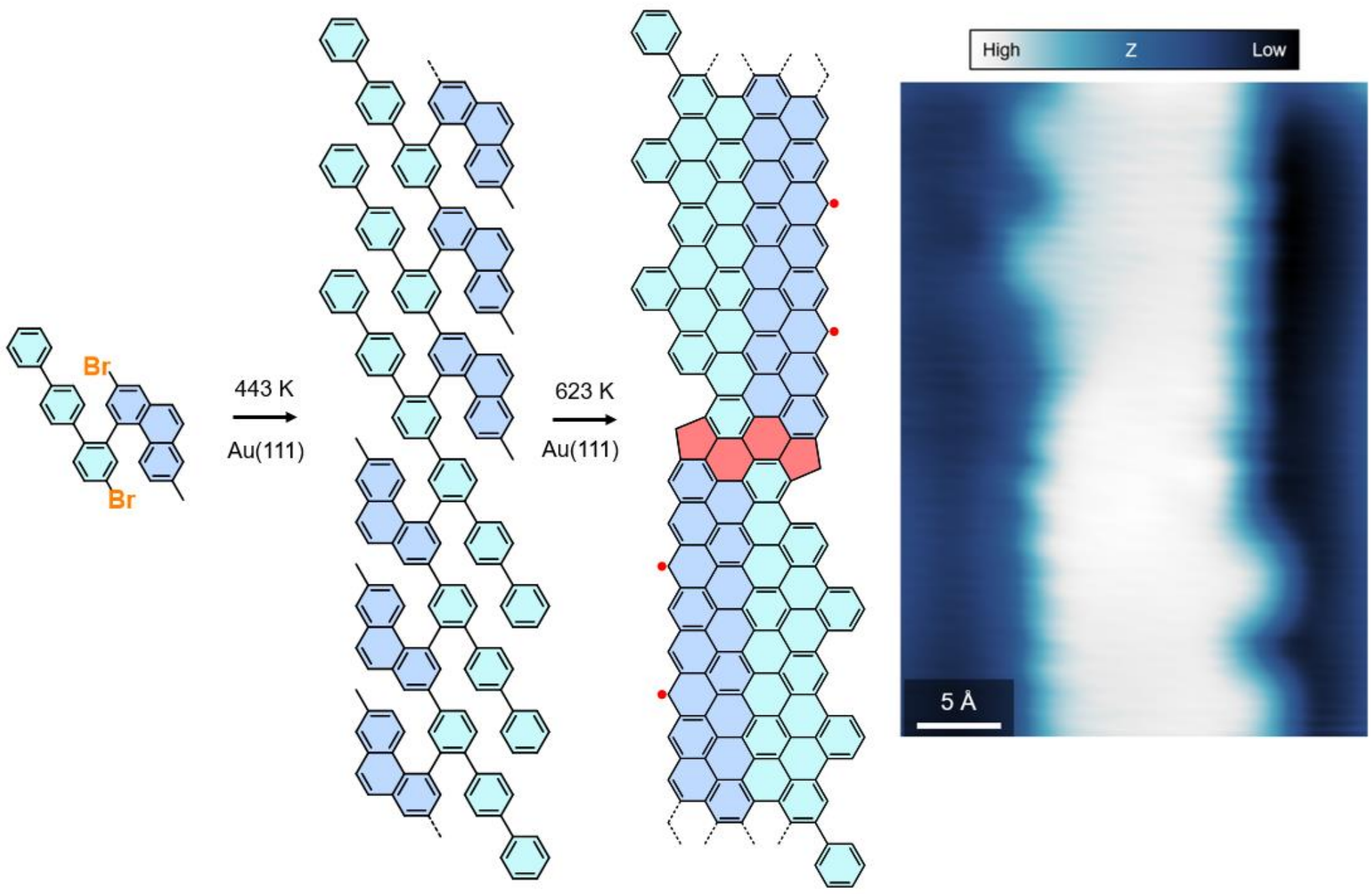

**Figure S14**. Heterojunction structure derived from head-to-head (tail-to-tail) polymerization and subsequent cyclodehydrogenation.

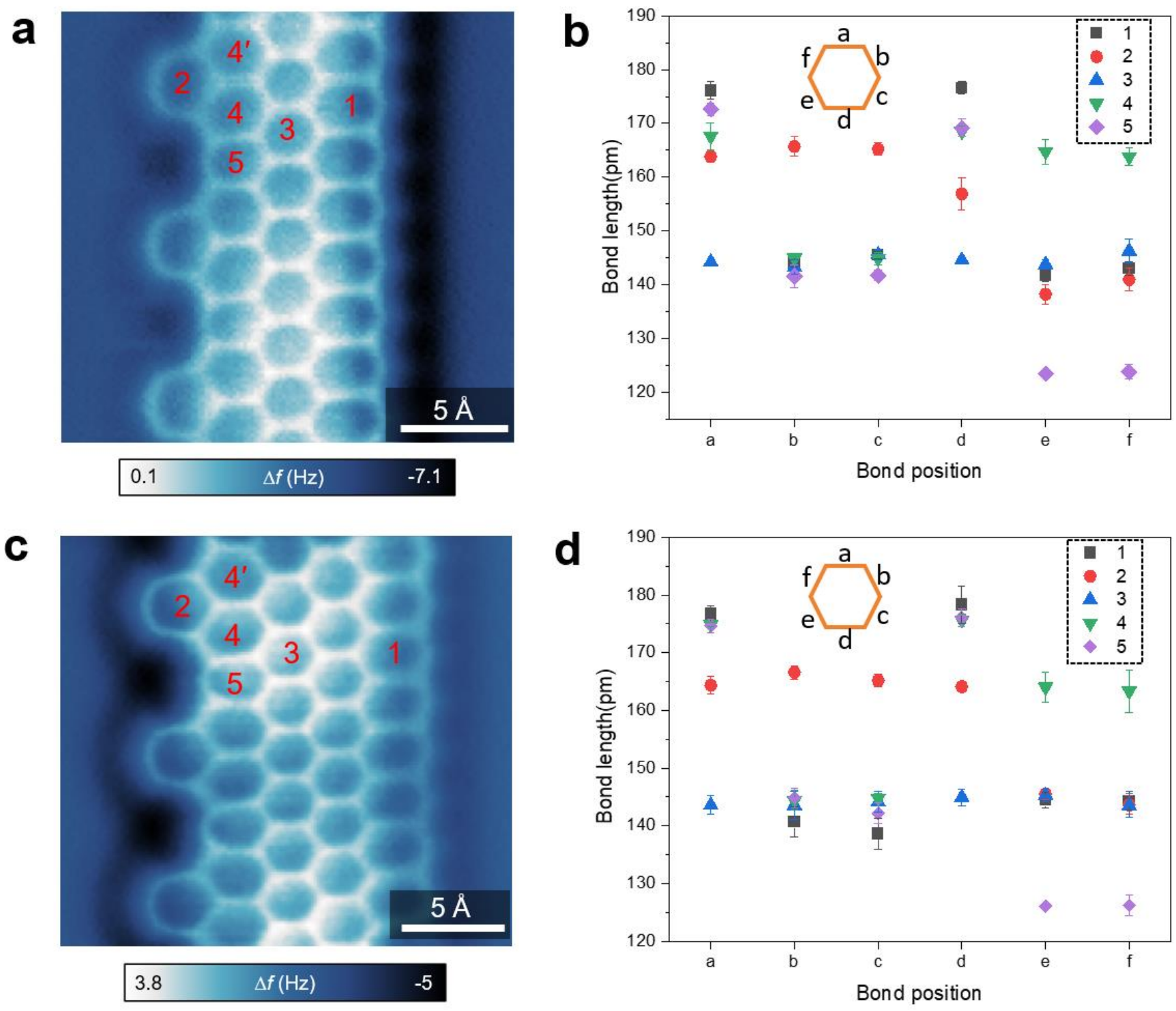

**Figure S15.** Bond-length analysis of the (*4*,*2*)-JGNR and (*5*,*2*)-JGNR. There are five different kinds of hexagons in both JGNRs (labelled as hexagon **1**-**5**). Hexagon **4** and **4′** are mirror symmetrical. The six C-C bond length of the hexagons in the middle of the JGNRs (hexagon **3**) are almost identical (~ 144 pm), while the hexagons at the edges are distorted in different manners. The C-C bonds of *a* and *d* in hexagon **1** on the right edges of both JGNRs are apparently longer (~176 pm), while the bonds of *b* and *c* are slightly shorter (~142 pm for (*4*,*2*)-JGNR, and ~139 pm for (*5*,*2*)-JGNR). The topological defective benzene motif (hexagon **2**) on the left edges of both JGNRs appears a large distortion, where the length of the four bonds *a*, *b*, *c*, and *d* spans from ~164 to ~167 pm. As a result, its neighbouring two rings (hexagon **4** and **4′**) undergoes an in-plane "stretch" towards outside of the ribbon. Finally, this leads to a squeeze of hexagon **5**, which makes the bonds *e* and *f* of hexagon **5** shorter (~125 pm).

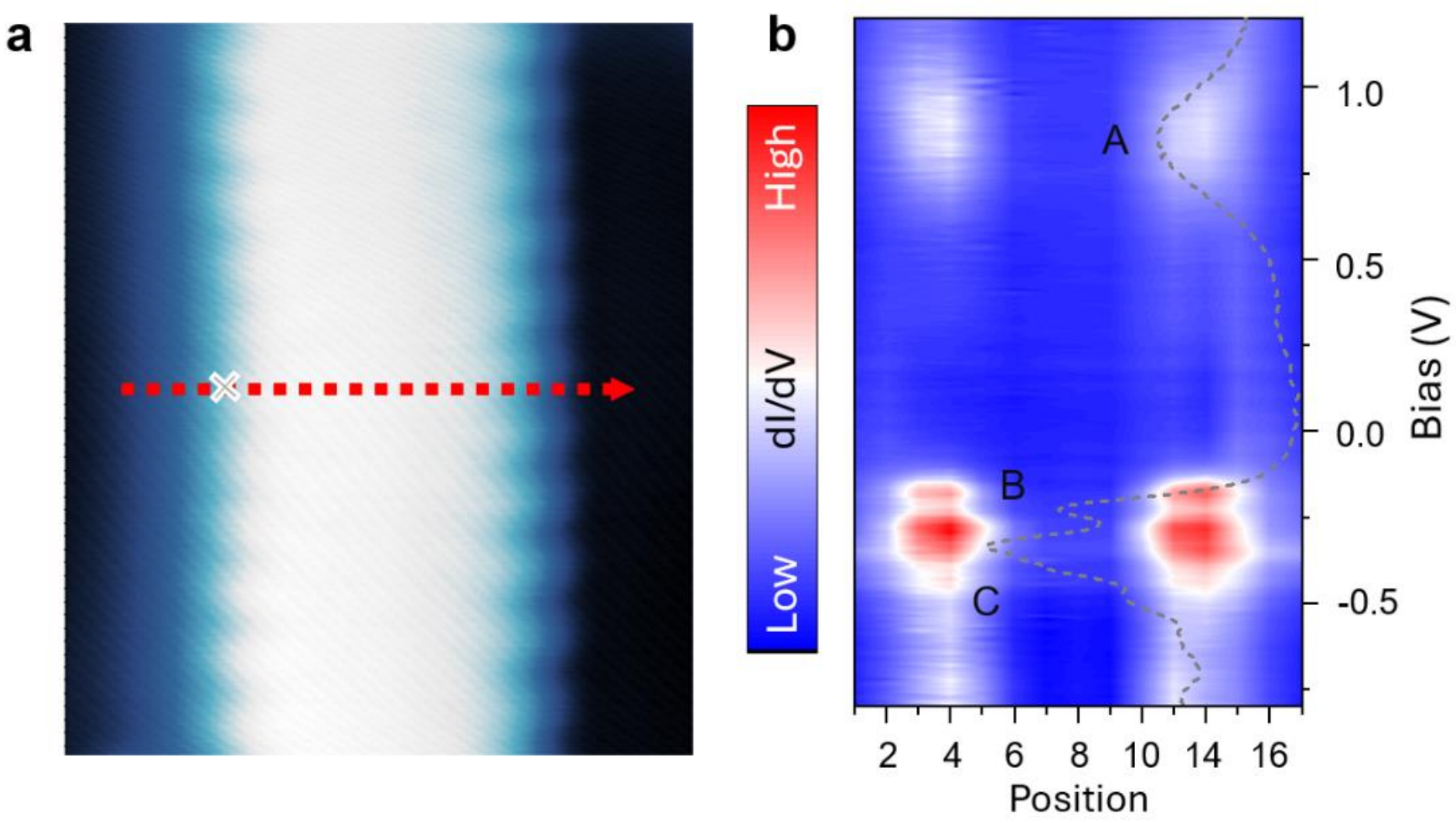

**Figure S16. a.** Constant current STM images of 5-ZGNR, ($V_s$ = 800 mV, $I_t$ = 200 pA, CO-functionalized tip). **b.** Color-coded d$I$/d$V$ spectra collected as a function of position along the red dotted line (from left to right) in **a** ($V_{ac}$ = 20 mV, $V_s$ = 1.5 V). The dashed line overlayed on **b** is the point d$I$/d$V$ spectrum acquired at the position labelled with a cross in **a**.

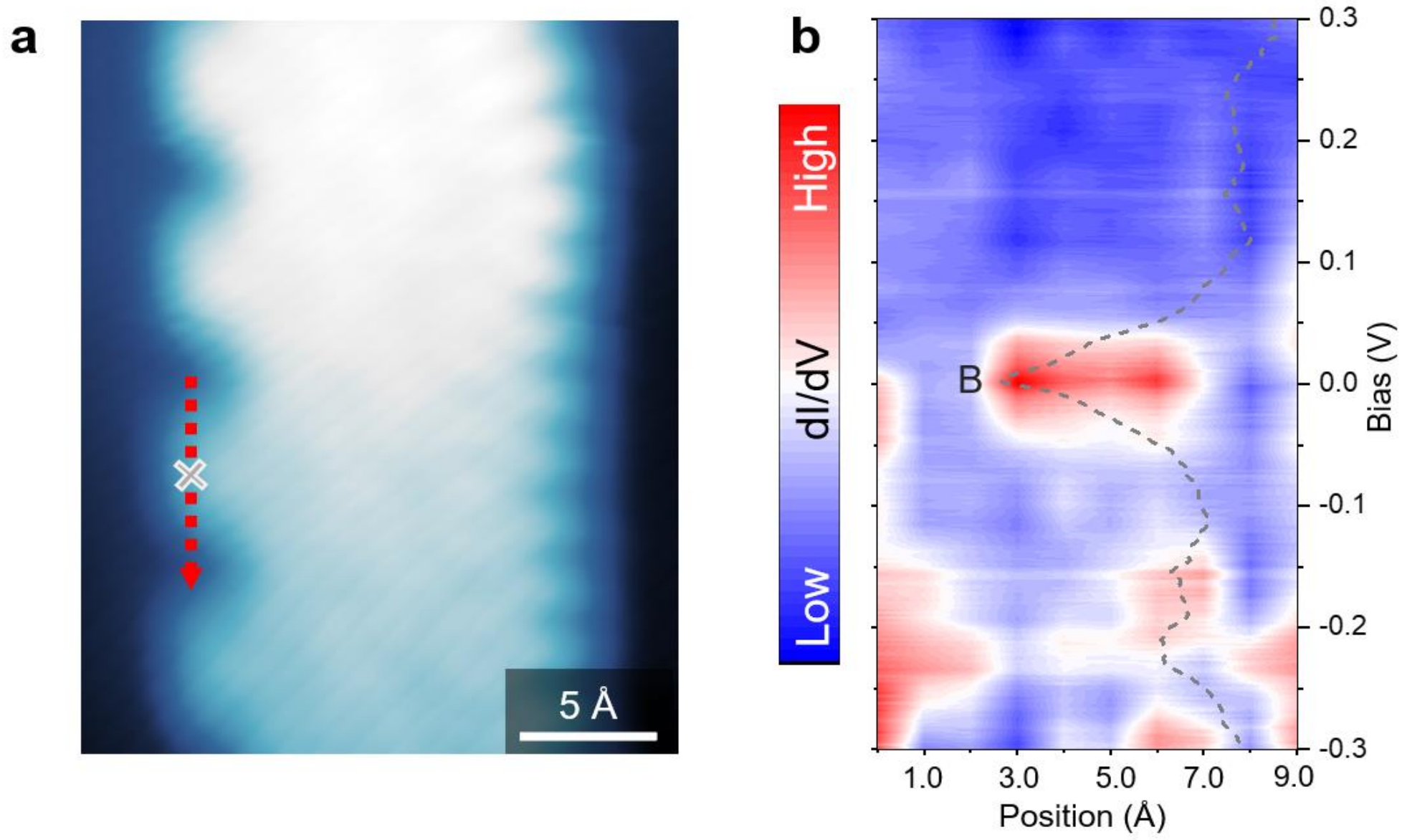

**Figure S17. a.** Constant current STM images of (*5*,*2*)-JGNR, ($V_s$ = 800 mV, $I_t$ = 200 pA, CO-functionalized tip). **b.** Color-coded d$I$/d$V$ spectra collected as a function of position along the red dotted line (from top to bottom) in **a** ($V_{ac}$ = 20 mV, $V_s$ = 1.5 V). The dashed line overlayed on **b** is the point d$I$/d$V$ spectrum acquired at the position labelled with a cross in **a**.

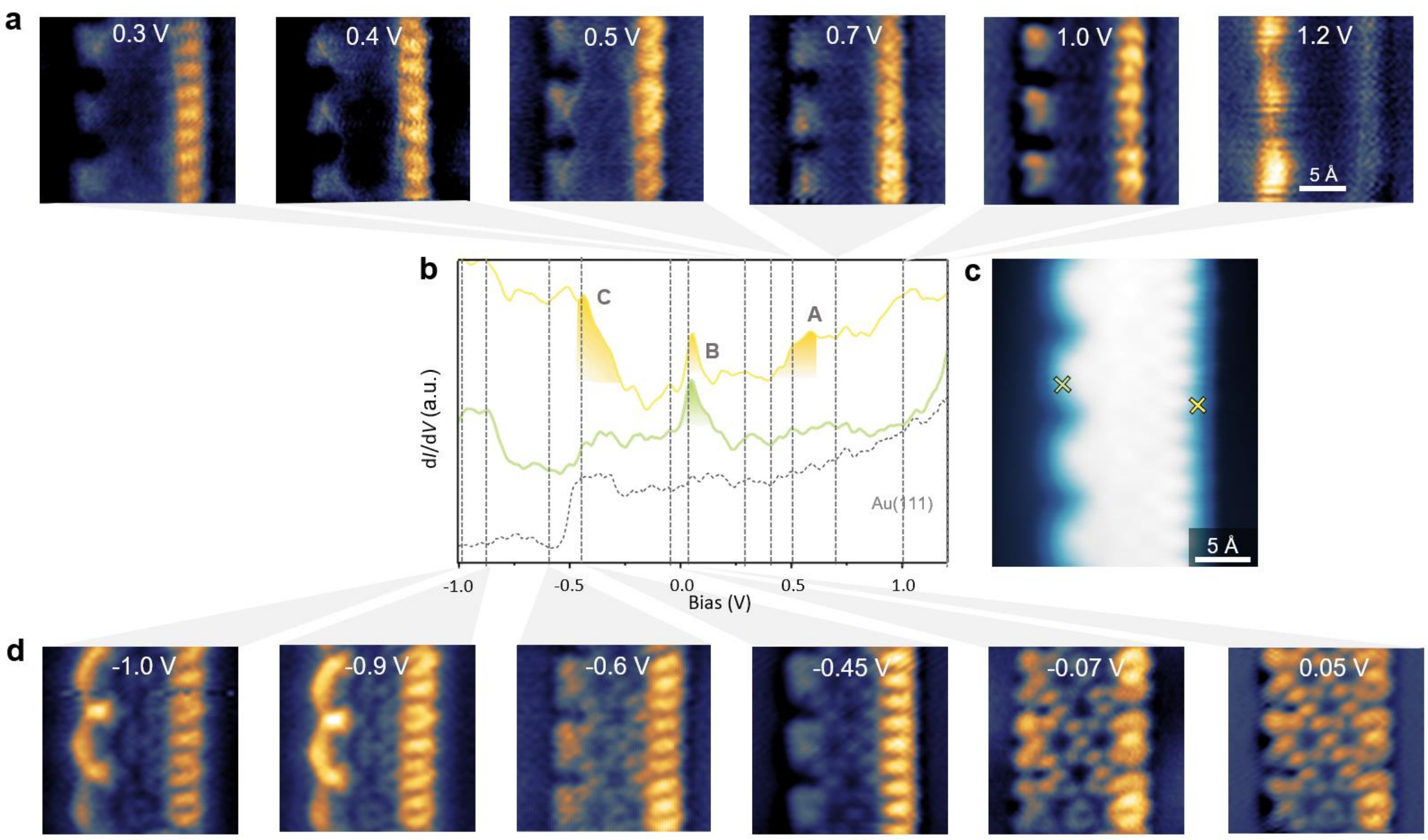

**Figure S18. a, d.** Constant-current d$I$/d$V$ maps of (*4,2*)-JGNR recorded at various sample biases marked in **b** by dashed line ($V_{ac}$ = 10 mV). **b.** d$I$/d$V$ point spectrum of (*4,2*)-JGNR on Au(111) acquired at the position marked in the panel **c** (yellow and green cross). Grey dashed line spectrum represents the Au(111) reference spectrum ($V_{ac}$ = 20 mV). **c.** Constant current STM images of (*4,2*)-JGNR, ($V_s$ = 800 mV, $I_t$ = 200 pA, CO-functionalized tip).

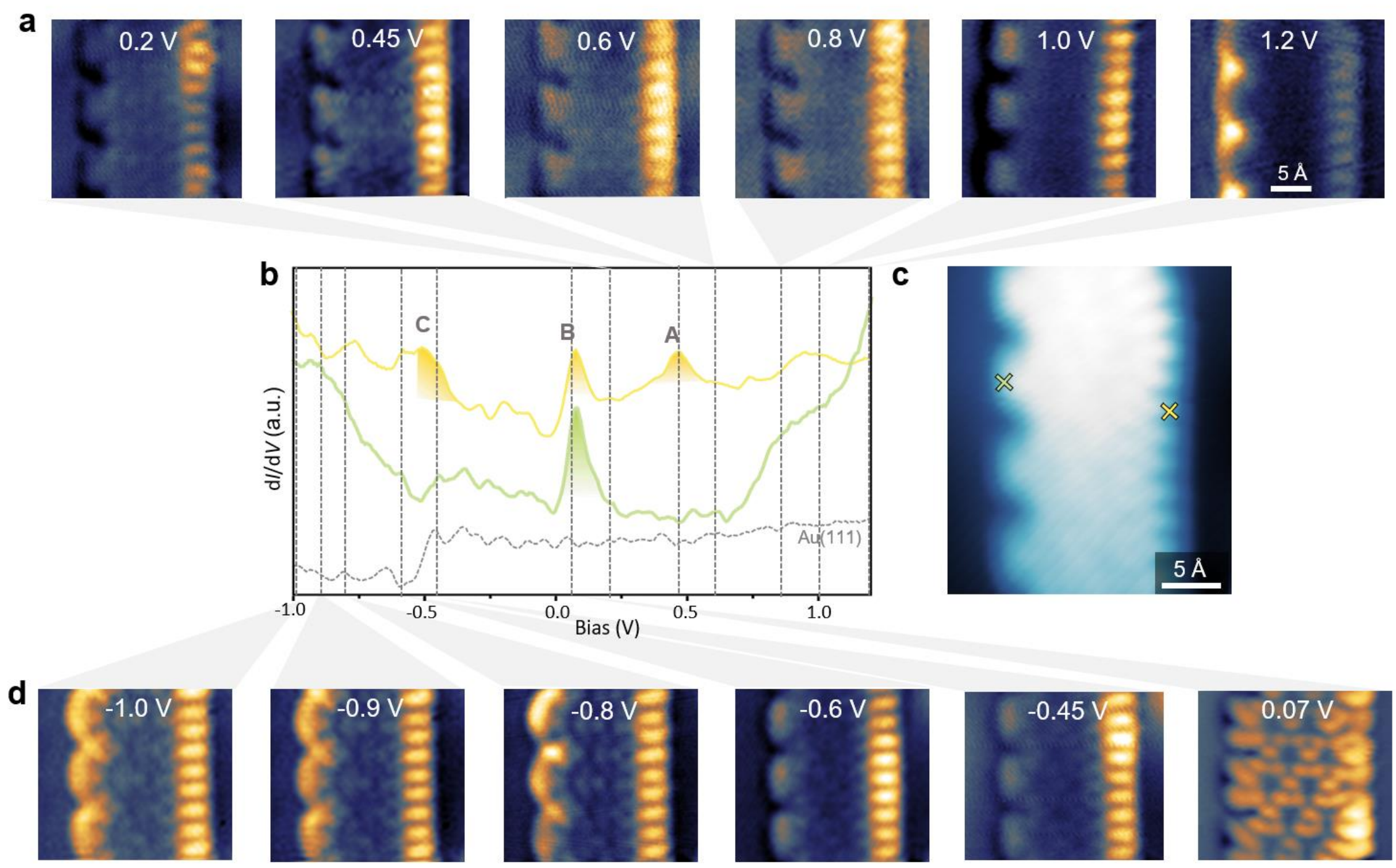

**Figure S19. a, d.** Constant-current d$I$/d$V$ maps of (*5,2*)-JGNR recorded at various sample biases marked in **b** by dashed line ($V_{ac}$ = 10 mV). **b.** d$I$/d$V$ point spectrum of (*5,2*)-JGNR on Au(111) acquired at the position marked in the panel **c** (yellow and green cross). Grey dashed line spectrum represents the Au(111) reference spectrum ($V_{ac}$ = 20 mV). **c.** Constant current STM images of (*5,2*)-JGNR, ($V_s$ = 800 mV, $I_t$ = 200 pA, CO-functionalized tip).

## 3. Organic synthesis of the precursors

All reagents and starting materials were obtained from commercial suppliers and used without further purification. All reactions were done under nitrogen atmosphere. Gel permeation chromatography (GPC) was performed on an LC-9210NEXT or LC-9201 system (Japan Analytical Industry) with JAIGEL 1H and 2H polystyrene columns (eluent: $CHCl_3$, flow: 9.5 mL/min). The $^1$H NMR and $^{13}$C NMR spectra were recorded in solution of $CDCl_3$ and $CD_2Cl_2$:$CS_2$ (1:4) on JEOL JNM-ECP300 at 300 MHz, Bruker AMX500 at 500 MHz and ASCENDTM400 at 400 MHz with tetramethylsilane (TMS) as internal standard. The chemical shifts were recorded in ppm and the multiplicity of each signal is given by the abbreviation: s=singlet, d=doublet, dd=doublet of doublet, t=triplet and m=multiplet. High-resolution (HR) APCI mass spectra were recorded on a MicrOTOF-QII instrument. Compound **6** was synthesized according to literature. [8]

The single crystals of the precursor molecules were obtained by vapor diffusion of *n*-hexane into the $CH_2Cl_2$ solutions of the precursors. The single-crystal X-ray diffraction (SXRD) data was measured at low temperature (T = 100 K), using a four circles goniometer Kappa geometry, Bruker AXS D8 Venture, equipped with a Photon 100 CMOS active pixel sensor detector. A Molybdenum monochromatized ($\lambda$ = 0.71073 Å) X-Ray radiation was used for the measurement. The structure was solved and refined using the SHELXT, VERSION 2014/5 Software.

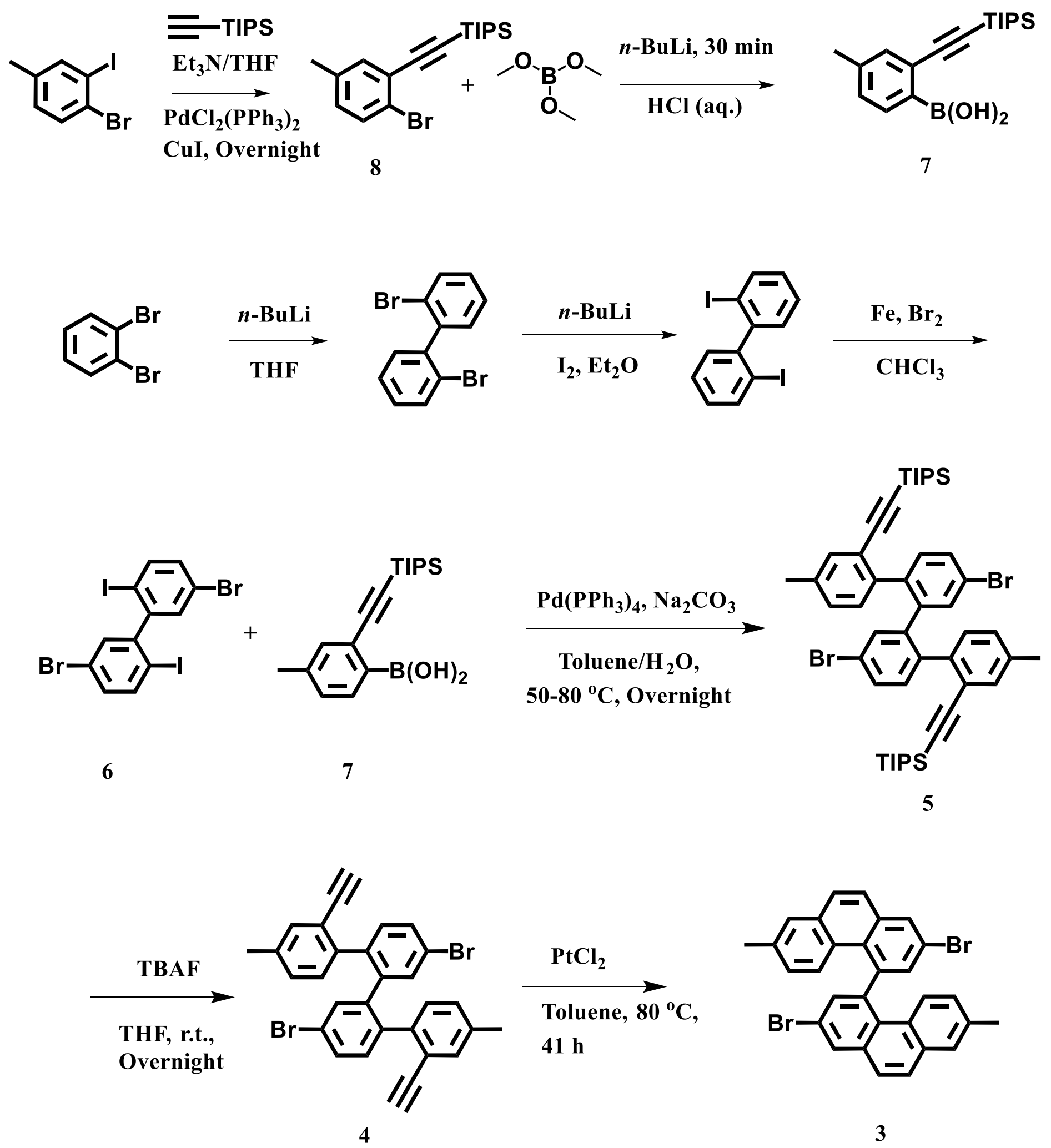

**Scheme 1.** The synthetic route of precursor **3**.

To a 25 mL Schlenk tube was added 4-Bromo-3-iodotoluene (152.4 mg, 0.513 mmol), $PdCl_2(PPh_3)_2$ (18 mg, 0.026 mmol), and CuI (10.2 mg, 0.0513 mmol). The flask was evacuated and backfilled with $N_2$ gas for 3 times. A degassed mixture of 3.8 mL THF and 2 mL $Et_3N$ was added to the Schlenk tube. Finally, (triisopropylsilyl)acetylene (0.13 mL, 0.565 mmol) was added to the above mixture. The reaction mixture was stirred at r.t. for overnight. Saturated aqueous $NH_4Cl$ solution was added to quench the reaction. The organic phase was extracted with DCM, dried with $Na_2SO_4$. After removing the solvent by vacuum, the crude product was purified by silica gel column (hexane) and GPC. **8** was obtained as yellow oil 175 mg, yield 97%. $^1$H NMR (400 MHz, Chloroform-*d*) δ 7.42 (d, *J* = 8.2 Hz, 1H), 7.31 (d, *J* = 2.3 Hz, 1H), 6.95 (dd, *J* = 8.3, 2.2 Hz, 1H), 2.26 (s, 3H), 1.14 (s, 21H). $^{13}$C NMR (101 MHz, Chloroform-*d*) δ 136.81, 134.36, 132.07, 130.43, 125.27, 122.40, 104.98, 95.58, 18.70, 18.58, 11.33.

**8** (1000 mg, 2.846 mmol) in 25 mL THF was treated with *n*-BuLi (2 mL, 1.59 M in hexane) at – 78 °C and stirred for 30 min. Trimethyl borate (0.63 mL, 0.932 g/mL) was added dropwise, then the solution kept 30 min before warmed up to r.t. for overnight. HCl (aq.) (20 mL, 1N) was added and stirred for 30 min. The organic phase was extracted with EtOAc three times. The crude white powder was purified by washing with hexane. **7** was obtained as white powder 707.9 mg, yield 78.6%. $^1$H NMR (300 MHz, Chloroform-*d*) δ 7.88 (d, *J* = 7.7 Hz, 1H), 7.35 (d, *J* = 1.7 Hz, 1H), 7.20 (dd, *J* = 7.7, 1.7 Hz, 1H), 5.99 (s, 2H), 2.36 (s, 3H), 1.16 (s, 21H). $^{13}$C NMR (75 MHz, Chloroform-*d*) δ 140.94, 135.62, 133.62, 129.47, 126.82, 108.66, 95.22, 21.25, 18.58, 11.26.

**6** (340.6 mg, 0.6 mmol), **7** (344 mg, 1.088 mmol), and $Na_2CO_3$ (636 mg, 6 mmol) were mixed with toluene (6 mL) and water (3 mL). The mixture was degassed three times with freeze-pump-thaw technique. Then $Pd(PPh_3)_4$ (70 mg, 0.06 mmol) was added under $N_2$ atmosphere, and heated under 50 °C for 5 h, 65 °C for 5 h, and 80 °C overnight. The organic phase was extracted with DCM after cooling down, and dried with $Na_2SO_4$. The product **5** was purified by silica gel column (hexane : DCM = 3 : 1) and GPC, obtained 31.3 mg as white powder, yield 6.07%. After purification, some impurities still remained. The product was used for next step without further purification.

**5** (336 mg, 0.395 mmol) was dissolved in THF (30 mL) and treated with TABF (6 mL, 1 M in THF) at 0 °C, then slowly warmed up to r.t. and stirred for overnight. The reaction was quenched with $H_2O$. The organic phase was diluted with $CH_2Cl_2$, washed with $H_2O$, and dried with $Na_2SO_4$. The solvent was removed by vacuum to obtain crude product. The crude product was further purified by GPC. **4** was obtained as white powder 165 mg, yield 77.3%. $^1$H NMR (400 MHz, Chloroform-*d*) δ = 7.59 (s, 2H), 7.40 (dd, *J* = 8.3, 2.1 Hz, 2H), 7.31 - 7.00 (m, 4H), 6.84 (d, *J* = 7.6 Hz, 2H), 6.41 (s, 2H), 2.72 (s, 2H), 2.26 (s, 6H). $^{13}$C NMR (101 MHz, Chloroform-*d*) δ = 141.10, 139.13, 137.89, 136.74, 134.73, 133.92, 132.65, 130.31, 130.07, 129.13, 129.08, 128.28, 125.35, 121.31, 121.00, 83.02, 79.49, 20.85.

**4** (160 mg, 0.296 mmol) and $PtCl_2$ (34 mg, 0.128 mmol) were charged to an oven-dried Schlenk tube. The mixture was degassed by vacuum for 30 min and dry toluene (36 mL) was added under nitrogen atmosphere. The reaction mixture was heated at 80 °C for 41 h. The solvent was removed by vacuum after cooling down. The product was purified by by silica gel column (hexane : DCM = 3: 1) and GPC to obtain **3** as white powder 53.1 mg, yield 33.2%.

**Precursor 3**. $^1$H NMR (400 MHz, Chloroform-*d*) δ = 8.13 (d, *J* = 2.2 Hz, 2H), 7.81 - 7.70 (m, 6H), 7.63 (s, 2H), 7.37 (d, *J* = 2.3 Hz, 2H), 6.84 (dd, *J* = 8.8, 2.0 Hz, 2H), 2.36 (s, 6H). $^{13}$C NMR (126 MHz, Chloroform-*d*) δ = 142.95, 136.46, 135.13, 133.47, 132.24, 130.81, 129.00, 128.38, 128.33, 128.22, 127.81, 127.25, 126.74, 126.69, 126.46, 119.45, 21.06. HRMS (APCI, positive) m/z calcd for $C_{30}H_{20}Br_2$ $[M+H]^+$ 539.0004, found 539.0003.

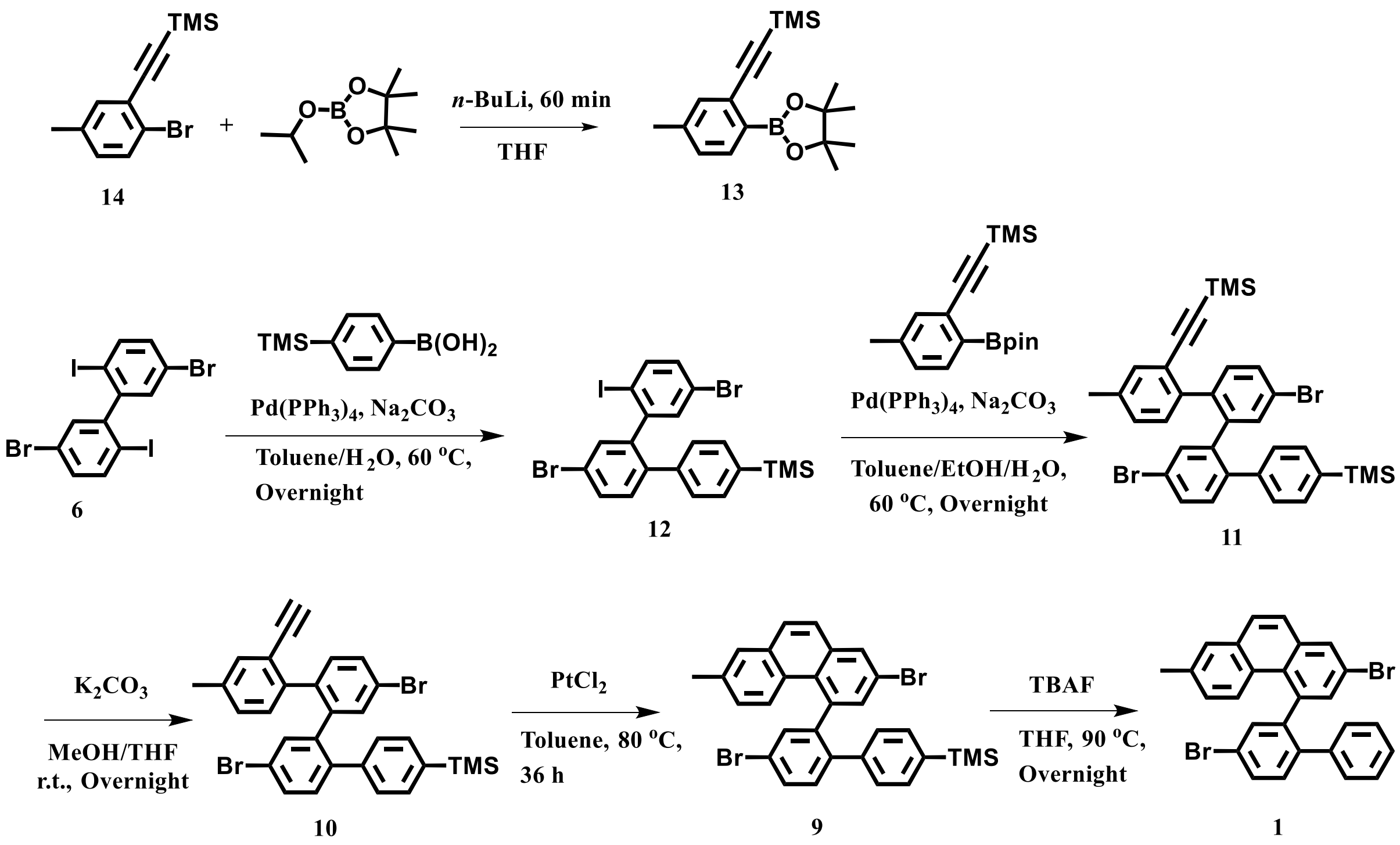

**Scheme 2.** The synthetic route of precursor **1**.

**14** (1000 mg, 3.742 mmol) was treated with *n*-BuLi (2.6 mL, 1.59 M in hexane) at – 78 °C and stirred for 60 min. 2-isopropoxy-4,4,5,5-tetramethyl-1,3,2-dioxaborolane (1.14 mL, 5.613 mmol) was added dropwise, then the solution was slowly warmed up to r.t. and stirred overnight. The reaction was quenched with $H_2O$. The organic phase was extracted with chloroform for three times, washed with brine, dried with $Na_2SO_4$. The crude white powder was purified by washing with hexane. **13** was obtained as white powder 934 mg, yield 79.4%. $^1$H NMR (300 MHz, Chloroform-*d*) δ 7.60 (d, *J* = 7.7 Hz, 1H), 7.33 (s, 1H), 7.09 (d, *J* = 7.6 Hz, 1H), 2.30 (s, 3H), 1.35 (s, 12H), 1.24 (s, 9H). $^{13}$C NMR (75 MHz, Chloroform-*d*) δ 140.38, 135.17, 133.65, 129.36, 128.54, 127.69, 105.82, 83.71, 75.03, 24.90, 24.84, 24.80, 21.26, -0.01.

**6** (1669 mg, 2.96 mmol), (4-(trimethylsilyl)phenyl)boronic acid (603 mg, 3.11 mmol), and $Na_2CO_3$ (1569 mg, 14.8 mmol) were mixed with toluene (20 mL) and water (10 mL). The mixture was degassed three times with freeze-pump-thaw technique. Then $Pd(PPh_3)_4$ (171 mg, 0.148 mmol) was added under $N_2$ atmosphere, and heated under 60 °C overnight. The organic phase was extracted with chloroform after cooling down. The crude product was purified by silica gel column (hexane : DCM = 2 : 1) and GPC to obtain **12** as white powder 513.1 mg, yield 29.6%. $^1$H NMR (500 MHz, Methylene Chloride-$d_2$) δ 7.64 (d, $J$ = 8.5 Hz, 1H), 7.61 (dd, $J$ = 8.4, 2.1 Hz, 1H), 7.40 – 7.35 (m, 3H), 7.30 (d, $J$ = 8.3 Hz, 1H), 7.28 (d, $J$ = 2.4 Hz, 1H), 7.14 – 7.10 (m, 2H), 7.09 (dd, $J$ = 8.5, 2.4 Hz, 1H), 0.22 (s, 9H). $^{13}$C NMR (126 MHz, Methylene Chloride-$d_2$) δ 147.93, 144.10, 141.35, 140.73, 140.64, 140.33, 134.67, 134.27, 133.95, 133.01, 132.61, 129.68, 122.90, 121.67, 98.98, -0.53.

**12** (200 mg, 0.341 mmol), **13** (140 mg, 0.444 mmol), and $Na_2CO_3$ (216 mg, 2.05 mmol) were mixed with toluene (6 mL), EtOH (2 mL) and water (2 mL). The mixture was degassed three times with freeze-pump-thaw technique. Then $Pd(PPh_3)_4$ (40 mg, 0.034 mmol) was added under $N_2$ atmosphere, and heated under 60 °C overnight. The organic phase was extracted with chloroform after cooling down. The crude product was purified by silica gel column (hexane : DCM = 2 : 1) and GPC to obtain **11** as white powder 127.6 mg, yield 57.8%. $^1$H NMR (500 MHz, Methylene Chloride-$d_2$) δ 7.53 (d, $J$ = 2.3 Hz, 1H), 7.51 – 7.39 (m, 4H), 7.29 (d, $J$ = 7.7 Hz, 2H), 7.06 – 6.99 (m, 3H), 6.78 – 6.69 (m, 3H), 2.21 (s, 3H), 0.25 (s, 9H), 0.10 (s, 9H). $^{13}$C NMR (126 MHz, Methylene Chloride-$d_2$) δ 135.22, 134.13, 133.23, 131.77, 131.13, 130.42, 129.59, 128.91, 20.85, -0.10, -1.11.

**11** (127.6 mg, 0.2 mmol) was dissolved in THF/MeOH (18 mL/36 mL), and treated with $K_2CO_3$ (177 mg, 1.28 mmol) for overnight at room temperature. The organic phase was extracted with $CHCl_3$, dried with $Na_2SO_4$, then the solvent was removed by vacuum. The crude product was purified by GPC to obtain **10** as white powder (118 mg, quantitatively converted) which was directly used for next step without further purification.

**10** (553.3 mg, 0.963 mmol) and $PtCl_2$ (68 mg, 0.256 mmol) were charged to an oven-dried Schlenk tube. The mixture was degassed by vacuum for 30 min and dry toluene (72 mL) was added under nitrogen atmosphere. The reaction was heated at 80 °C for 36 h. The solvent was removed by vacuum after cooling down. The crude product was purified by silica gel column (hexane : DCM = 2 : 1) and GPC to obtain **9** as white powder 394 mg, yield 71.2%. $^{1}$H NMR (500 MHz, Methylene Chloride-$d_2$) δ 7.92 (d, $J$ = 2.3 Hz, 1H), 7.68 (dd, $J$ = 8.3, 2.2 Hz, 1H), 7.65 (d, $J$ = 8.7 Hz, 1H), 7.62 – 7.53 (m, 3H), 7.50 (d, $J$ = 2.1 Hz, 1H), 7.40 (d, $J$ = 8.3 Hz, 1H), 7.35 (d, $J$ = 2.2 Hz, 1H), 7.06 (dt, $J$ = 8.1, 1.8 Hz, 3H), 6.87 (dt, $J$ = 8.0, 1.6 Hz, 2H), 2.45 (s, 3H), 0.10 (s, 9H). $^{13}$C NMR (126 MHz, Methylene Chloride-$d_2$) δ 144.93, 140.82, 140.77, 140.28, 139.88, 137.56, 135.51, 134.43, 134.26, 133.77, 133.54, 133.50, 132.79, 132.20, 131.48, 129.52, 129.26, 128.92, 128.75, 128.65, 127.69, 127.09, 122.73, 119.44, 21.89, -0.68.

**9** (394 mg, 0.686 mmol) was dissolved in THF (22 mL) and treated with TABF (22 mL, 1 M in THF) for overnight at 90 °C. Then the solvent was removed by vacuum. The crude product was purified by silica gel column (hexane : DCM = 2 : 1) and GPC to obtain **1** as white powder 240.4 mg, yield 69.78%.

**Precursor 1**. $^{1}$H NMR (500 MHz, Methylene Chloride-$d_2$) δ 7.89 (d, $J$ = 2.2 Hz, 1H), 7.71 – 7.67 (m, 2H), 7.62 (d, $J$ = 8.8 Hz, 1H), 7.59 – 7.53 (m, 2H), 7.48 (d, $J$ = 2.1 Hz, 1H), 7.43 (d, $J$ = 8.3 Hz, 1H), 7.30 (dd, $J$ = 2.3, 0.9 Hz, 1H), 7.08 (dd, $J$ = 8.8, 2.0 Hz, 1H), 7.03 – 6.98 (m, 1H), 6.96 – 6.93 (m, 4H), 2.51 (s, 3H). $^{13}$C NMR (126 MHz, Methylene Chloride-$d_2$) δ 144.35, 140.04, 139.79, 139.63, 136.68, 134.92, 133.86, 133.79, 133.28, 132.86, 131.59, 131.00, 129.19, 129.11, 129.09, 129.00, 128.29, 128.23, 128.18, 127.35, 127.17, 126.75, 122.63, 119.41, 21.74. HRMS (APCI, positive) m/z calcd for $C_{27}H_{18}Br_2$ $[M+H]^+$ 500.9848, found 500.9841.

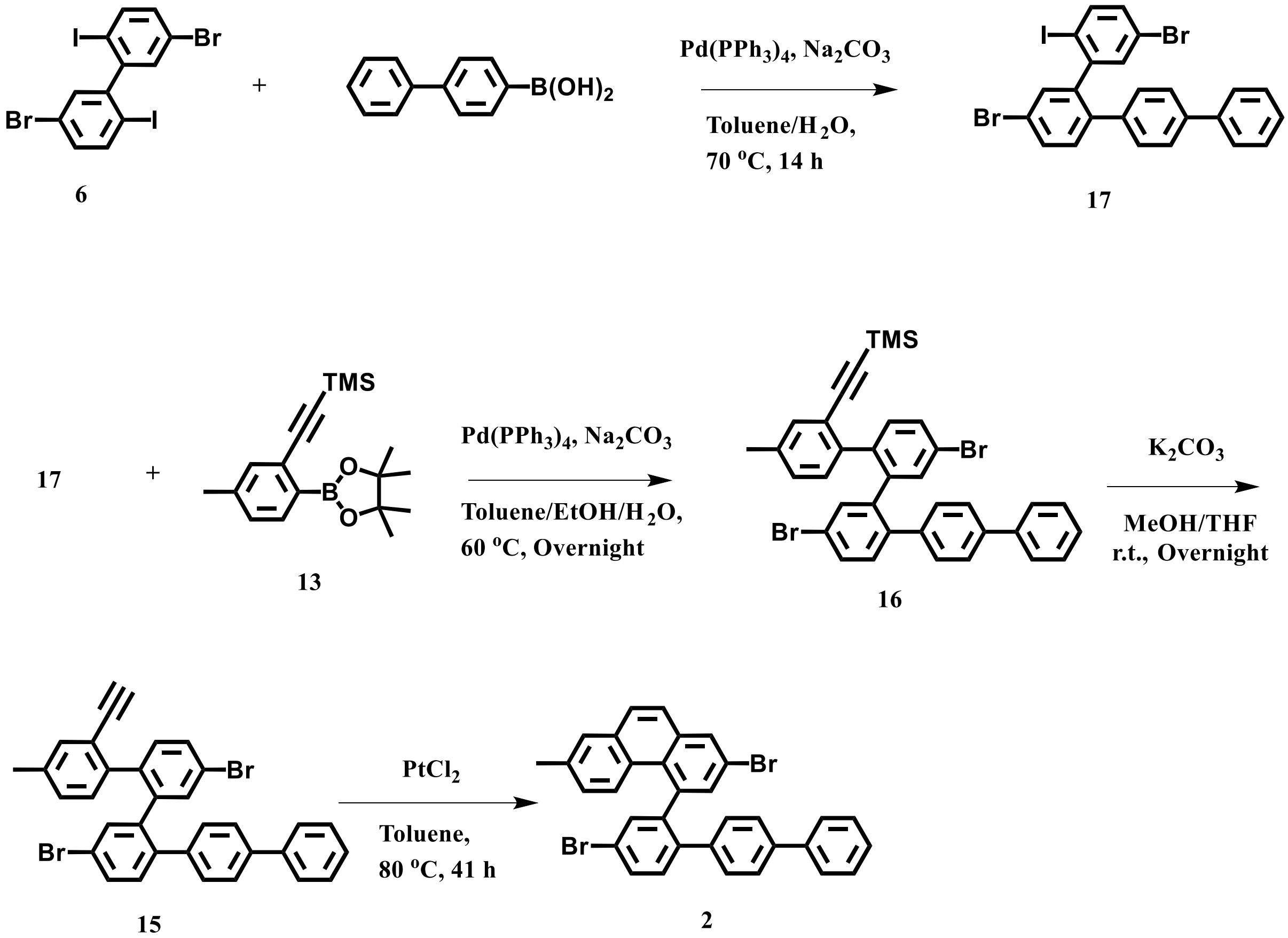

**Scheme 3.** The synthetic route of precursor **2**.

**6** (2000 mg, 3.55 mmol), biphenyl boronic acid (844 mg, 4.26 mmol), and $Na_2CO_3$ (1881 mg, 17.75 mmol) were mixed with toluene (30 mL) and water (15 mL). The mixture was degassed three times with freeze-pump-thaw technique. Then $Pd(PPh_3)_4$ (205 mg, 0.178 mmol) was added under $N_2$ atmosphere, and heated under 70 °C for 14 h. The organic phase was extracted with chloroform after cooling down. The crude product was purified by silica gel column (hexane : DCM = 5: 1) and GPC to obtain **17** as white powder 580.9 mg, yield 27.8%. $^1H$ NMR (500 MHz, Chloroform-*d*) δ 7.68 – 7.61 (m, 2H), 7.59 – 7.53 (m, 2H), 7.50 – 7.44 (m, 2H), 7.43 – 7.38 (m, 3H), 7.36 (dd, $J$ = 8.2, 1.8 Hz, 1H), 7.34 – 7.29 (m, 2H), 7.23 – 7.18 (m, 2H), 7.12 – 7.07 (m, 1H). $^{13}C$ NMR (126 MHz, Chloroform-*d*) δ 141.22, 134.50, 134.05, 132.85, 132.70, 132.46, 130.75, 129.57, 128.22, 127.60, 127.26.

**17** (382.5 mg, 0.648 mmol), **13** (305.6 mg, 0.972 mmol), and $Na_2CO_3$ (412.3 mg, 3.89 mmol) were mixed with toluene (6 mL), EtOH (2 mL) and water (2 mL). The mixture was degassed three times with freeze-pump-thaw technique. Then $Pd(PPh_3)_4$ (74.9 mg, 0.0648 mol) was added under $N_2$ atmosphere, and heated under 60 °C overnight. The organic phase was extracted with chloroform after cooling down. The crude product was purified by silica gel column (hexane : DCM = 2 : 1) and GPC to obtain **16** as white powder 233.4 mg, yield 55.4%. $^1$H NMR (500 MHz, Methylene Chloride-$d_2$) δ 7.61 – 7.55 (m, 4H), 7.49 – 7.26 (m, 9H), 7.04 (d, $J$ = 9.6 Hz, 2H), 6.74 (d, $J$ = 8.4 Hz, 3H), 2.22 (s, 3H), -0.01 (s, 9H). $^{13}$C NMR (126 MHz, Methylene Chloride-$d_2$) δ 135.14, 134.09, 133.01, 131.76, 131.21, 130.36, 129.89, 129.07, 128.86, 127.63, 127.38, 126.96, 121.45, 97.70, 20.86, -0.31.

**16** (233.4 mg, 0.359 mmol) was dissolved in THF/MeOH (32 mL/64 mL) and treated with $K_2CO_3$ (318 mg, 2.3 mmol) for overnight at room temperature. The organic phase was extracted with $CHCl_3$, dried with $Na_2SO_4$. Then the solvent was removed by vacuum. The crude product was purified by GPC to obtain **15** as white powder (202.2 mg, yield 97.4%) which was directly used for next step without further purification.

**15** (202.2 mg, 0.35 mmol) and $PtCl_2$ (34 mg, 0.128 mmol) were charged to an oven-dried Schlenk tube. The mixture was degassed by vacuum for 30 min and dry toluene (48 mL) was added under nitrogen atmosphere. The reaction was heated at 80 °C for 41 h. The solvent was removed by vacuum after cooling down. The crude product was purified by silica gel column (hexane : DCM = 2 : 1) and GPC to obtain **2** as white powder 129.7 mg, yield 64%. **Precursor 2**. $^1$H NMR (300 MHz, Methylene Chloride-$d_2$) δ 7.93 (d, $J$ = 2.3 Hz, 1H), 7.72 – 7.65 (m, 2H), 7.62 – 7.52 (m, 4H), 7.44 (d, $J$ = 8.3 Hz, 1H), 7.40 – 7.29 (m, 5H), 7.28 – 7.22 (m, 1H), 7.15 (dt, $J$ = 8.4, 2.1 Hz, 2H), 7.08 (dd, $J$ = 8.8, 2.1 Hz, 1H), 6.94 (dt, $J$ = 8.4, 2.0 Hz, 2H), 2.45 (s, 3H). $^{13}$C NMR (75 MHz, Methylene Chloride-$d_2$) δ 143.89, 140.13, 139.64, 139.20, 138.88, 138.41, 136.48, 134.43, 133.30, 133.22, 132.68, 132.30, 131.15, 130.43, 129.02, 128.55, 128.37, 128.16, 127.84, 127.68, 127.56, 127.16, 126.59, 126.56, 126.00, 121.72, 118.40, 20.80. HRMS (APCI, positive) m/z calcd for $C_{33}H_{22}Br_2$ $[M+H]^+$ 577.0161, found 577.0145.

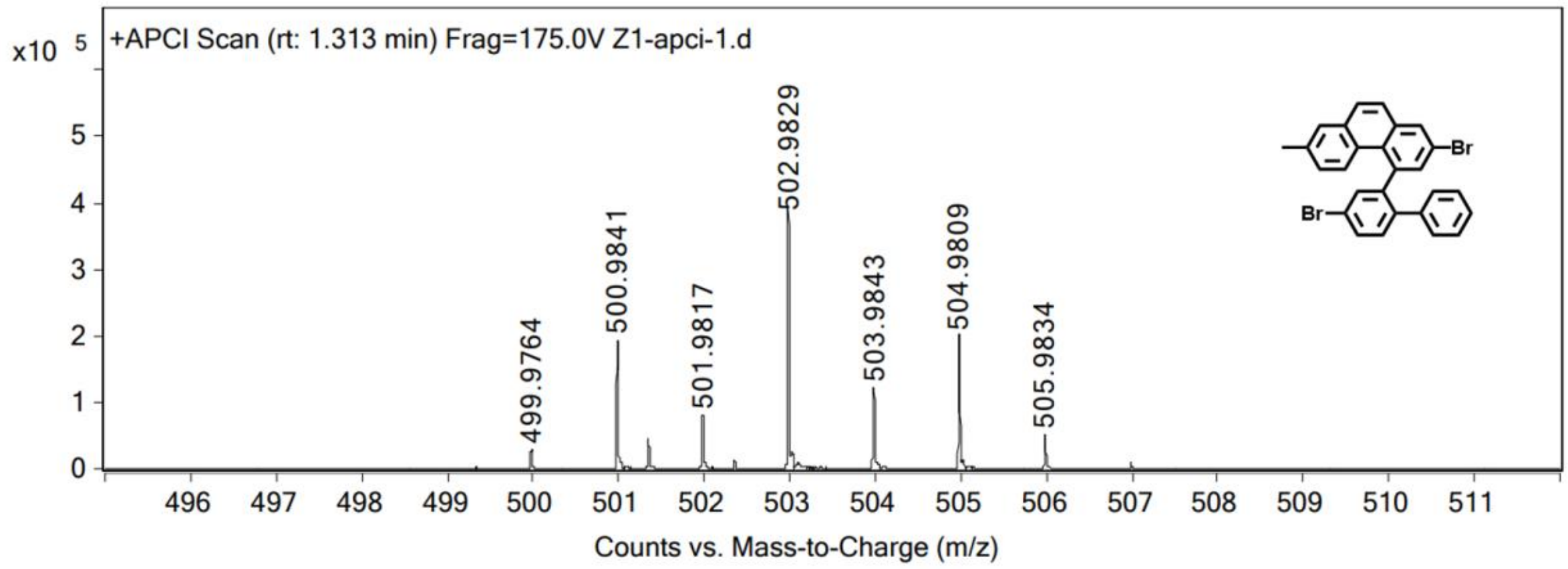

**Figure S20.** High resolution mass spectrum (APCI) of the precursor **1**.

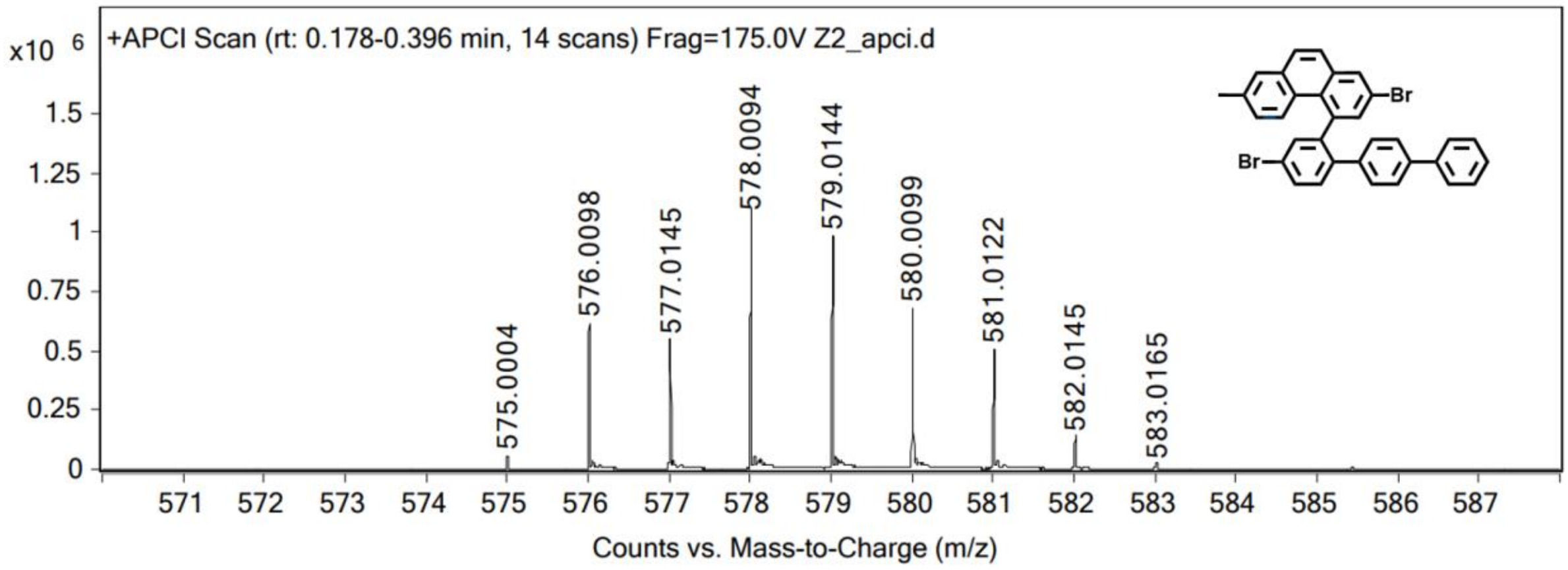

**Figure S21.** High resolution mass spectrum (APCI) of the precursor **2**.

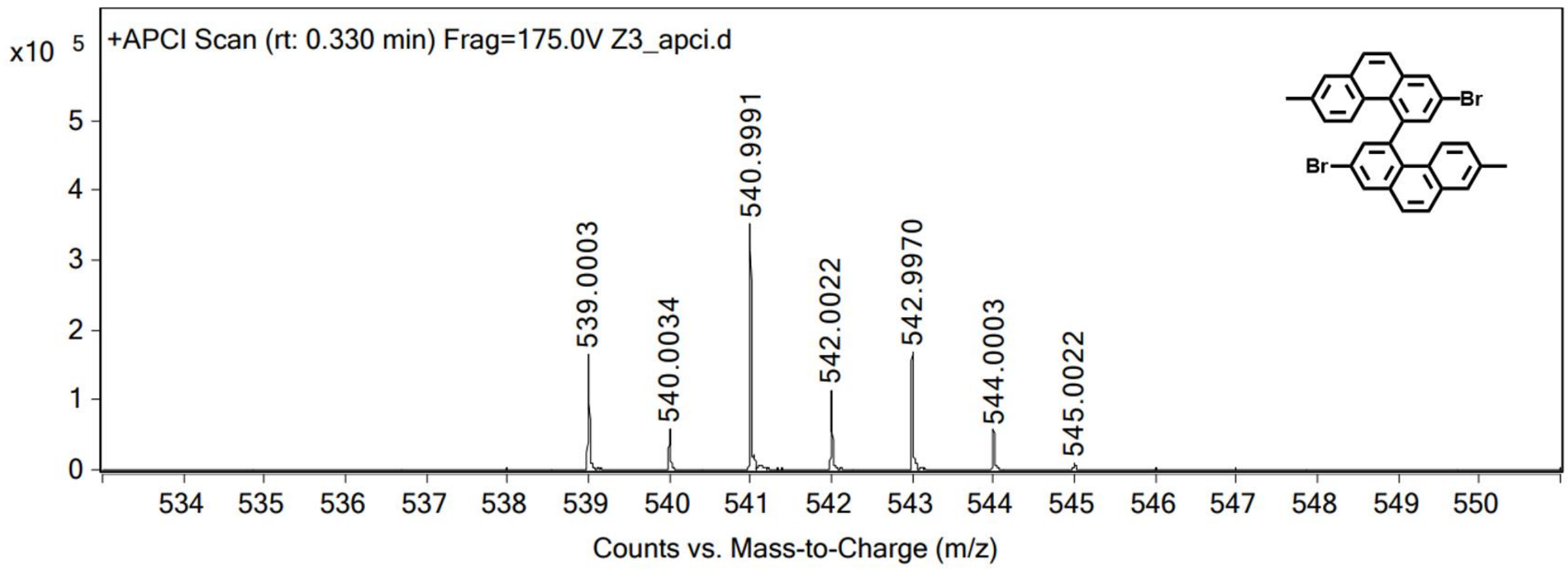

**Figure S22.** High resolution mass spectrum (APCI) of the precursor **3**.

**Single crystal data of the precursors**

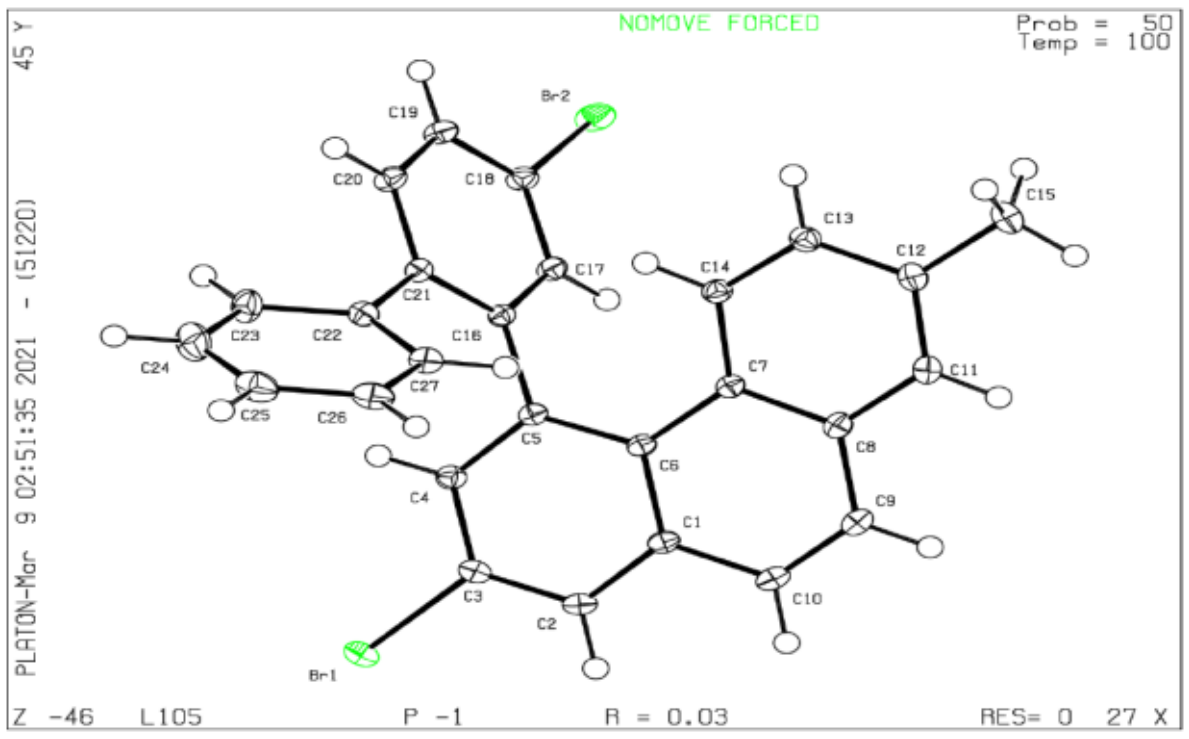

Table S1. Crystal data and structure refinement for precursor **1**.

| | | |
|---|---|---|
| Identification code | L105 | |
| Empirical formula | $C_{27} H_{18} Br_2$ | |
| Formula weight | 502.23 | |
| Temperature | 100(2) K | |
| Wavelength | 0.71073 Å | |
| Crystal system | Triclinic | |
| Space group | P-1 | |
| Unit cell dimensions | a = 8.2463(5) Å | a= 116.134(2)°. |
| | b = 11.2688(6) Å | b= 94.646(2)°. |
| | c = 12.2738(7) Å | g = 94.160(2)°. |
| Volume | 1013.09(10) $Å^3$ | |
| Z | 2 | |
| Density (calculated) | 1.646 $Mg/m^3$ | |
| Absorption coefficient | 4.012 $mm^{-1}$ | |
| F(000) | 500 | |
| Crystal size | 0.576 x 0.566 x 0.376 $mm^3$ | |
| Theta range for data collection | 2.028 to 40.401°. | |
| Index ranges | -15<=h<=15, -20<=k<=20, -22<=l<=22 | |
| Reflections collected | 94886 | |
| Independent reflections | 12824 [R(int) = 0.0628] | |
| Completeness to theta = 25.242° | 99.8 % | |
| Absorption correction | Semi-empirical from equivalents | |
| Max. and min. transmission | 0.7479 and 0.3139 | |
| Refinement method | Full-matrix least-squares on $F^2$ | |
| Data / restraints / parameters | 12824 / 0 / 266 | |
| Goodness-of-fit on $F^2$ | 1.031 | |
| Final R indices [I>2sigma(I)] | $R_1$ = 0.0310, $wR_2$ = 0.0771 | |
| R indices (all data) | $R_1$ = 0.0409, $wR_2$ = 0.0810 | |
| Extinction coefficient | n/a | |
| Largest diff. peak and hole | 1.325 and -1.316 e.$Å^{-3}$ | |

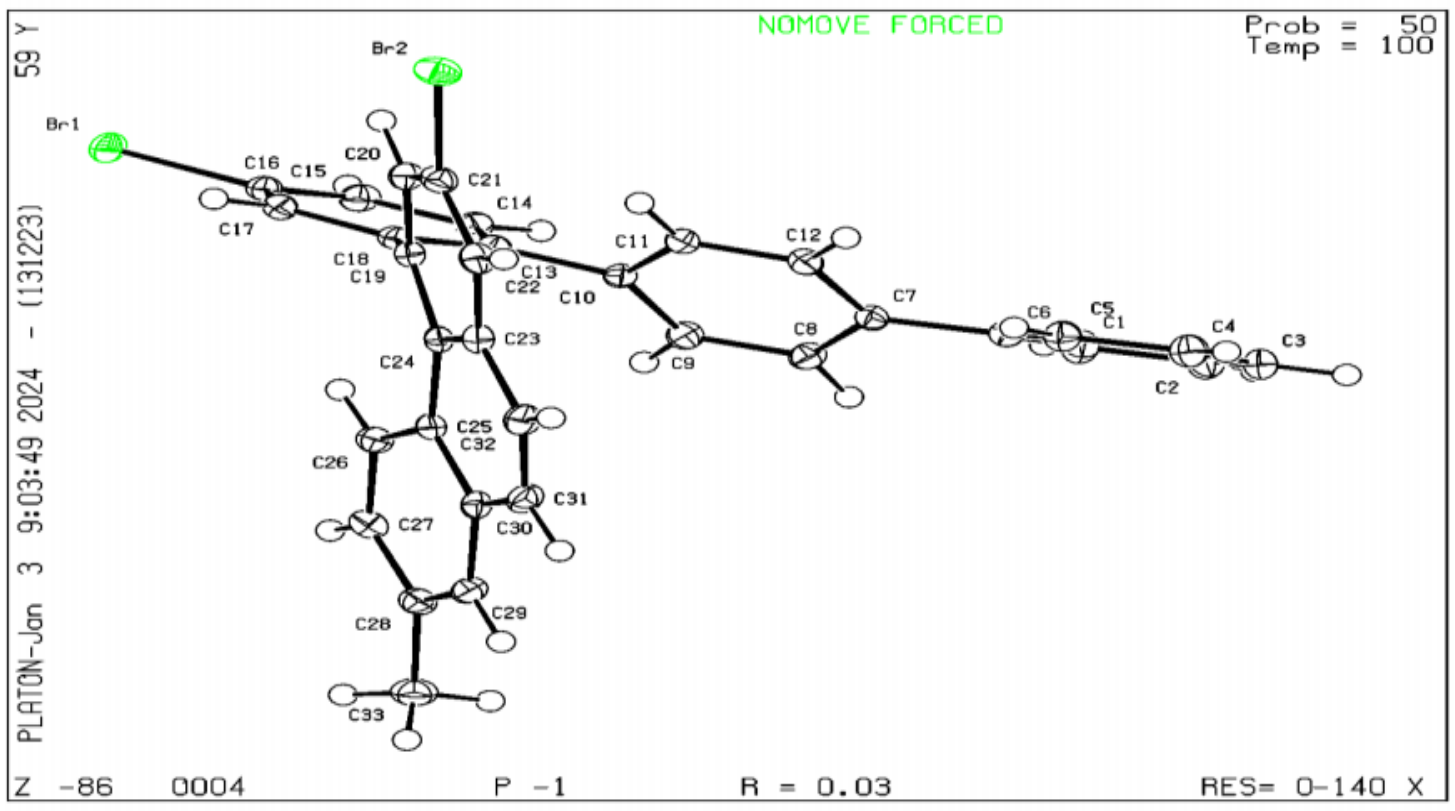

Table S2. Crystal data and structure refinement for precursor **2**.

| | | |
|---|---|---|
| Identification code | O004 | |
| Empirical formula | $C_{33} H_{22} Br_2$ | |
| Formula weight | 578.32 | |
| Temperature | 100(2) K | |
| Wavelength | 0.71073 Å | |
| Crystal system | Triclinic | |
| Space group | P-1 | |
| Unit cell dimensions | a = 9.91993(8) Å | a= 86.831(3)°. |
| | b = 11.332(1) Å | b= 87.026(3)°. |
| | c = 11.3728(8) Å | g = 72.729(3)°. |
| Volume | 1218.09(17) $Å^3$ | |
| Z | 2 | |
| Density (calculated) | 1.577 $Mg/m^3$ | |
| Absorption coefficient | 3.348 $mm^{-1}$ | |
| F(000) | 580 | |
| Crystal size | 0.213 x 0.212 x 0.138 $mm^3$ | |
| Theta range for data collection | 2.406 to 29.593°. | |
| Index ranges | -13<=h<=13, -15<=k<=15, -15<=l<=15 | |
| Reflections collected | 43032 | |
| Independent reflections | 6824 [R(int) = 0.0459] | |
| Completeness to theta = 29.593° | 99.7 % | |
| Absorption correction | Semi-empirical from equivalents | |
| Max. and min. transmission | 0.7459 and 0.6363 | |
| Refinement method | Full-matrix least-squares on $F^2$ | |
| Data / restraints / parameters | 6824 / 0 / 317 | |
| Goodness-of-fit on $F^2$ | 1.041 | |
| Final R indices [I>2sigma(I)] | $R_1$ = 0.0279, $wR_2$ = 0.0685 | |
| R indices (all data) | $R_1$ = 0.0332, $wR_2$ = 0.0710 | |
| Extinction coefficient | n/a | |
| Largest diff. peak and hole | 1.582 and -0.472 e.$Å^{-3}$ | |

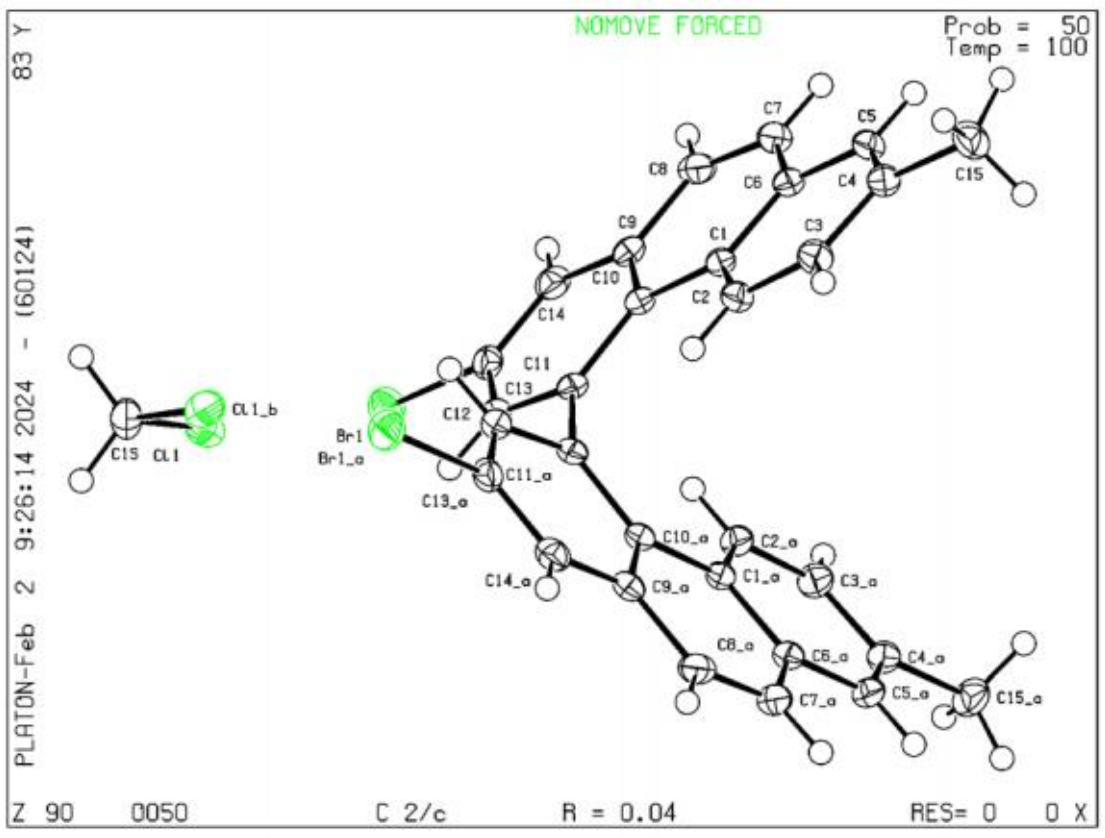

Table S3. Crystal data and structure refinement for precursor **3** with a DCM solvent molecule.

| | | |
|---|---|---|
| Identification code | O050 | |
| Empirical formula | C31 H22 Br2 Cl2 | |
| Formula weight | 625.20 | |
| Temperature | 100(2) K | |
| Wavelength | 0.71073 Å | |
| Crystal system | Monoclinic | |
| Space group | C2/c | |
| Unit cell dimensions | a = 17.5679(7) Å | a= 90°. |
| | b = 10.7387(4) Å | b= 108.0360(10)°. |
| | c = 14.0186(5) Å | g = 90°. |
| Volume | 2514.74(16) Å3 | |
| Z | 4 | |
| Density (calculated) | 1.651 Mg/m$^3$ | |
| Absorption coefficient | 3.456 mm$^{-1}$ | |
| F(000) | 1248 | |
| Crystal size | 0.075 x 0.118 x 0.133 mm$^3$ | |
| Theta range for data collection | 2.255 to 28.277°. | |
| Index ranges | -23<=h<=23, -13<=k<=14, -18<=l<=18 | |
| Reflections collected | 39252 | |
| Independent reflections | 3125 [R(int) = 0.0544] | |
| Completeness to theta = 25.242° | 100.0 % | |
| Absorption correction | Semi-empirical from equivalents | |
| Max. and min. transmission | 0.7461 and 0.6733 | |
| Refinement method | Full-matrix least-squares on F$^2$ | |
| Data / restraints / parameters | 3125 / 0 / 160 | |
| Goodness-of-fit on F$^2$ | 1.068 | |
| Final R indices [I>2sigma(I)] | $R_1$ = 0.0361, $wR_2$ = 0.0907 | |
| R indices (all data) | $R_1$ = 0.0425, $wR_2$ = 0.0941 | |
| Extinction coefficient | n/a | |
| Largest diff. peak and hole | 1.928 and -0.505 e.Å$^{-3}$ | |

$^1$H and $^{13}$C nuclear magnetic resonance spectra and mass spectra for all new compounds.

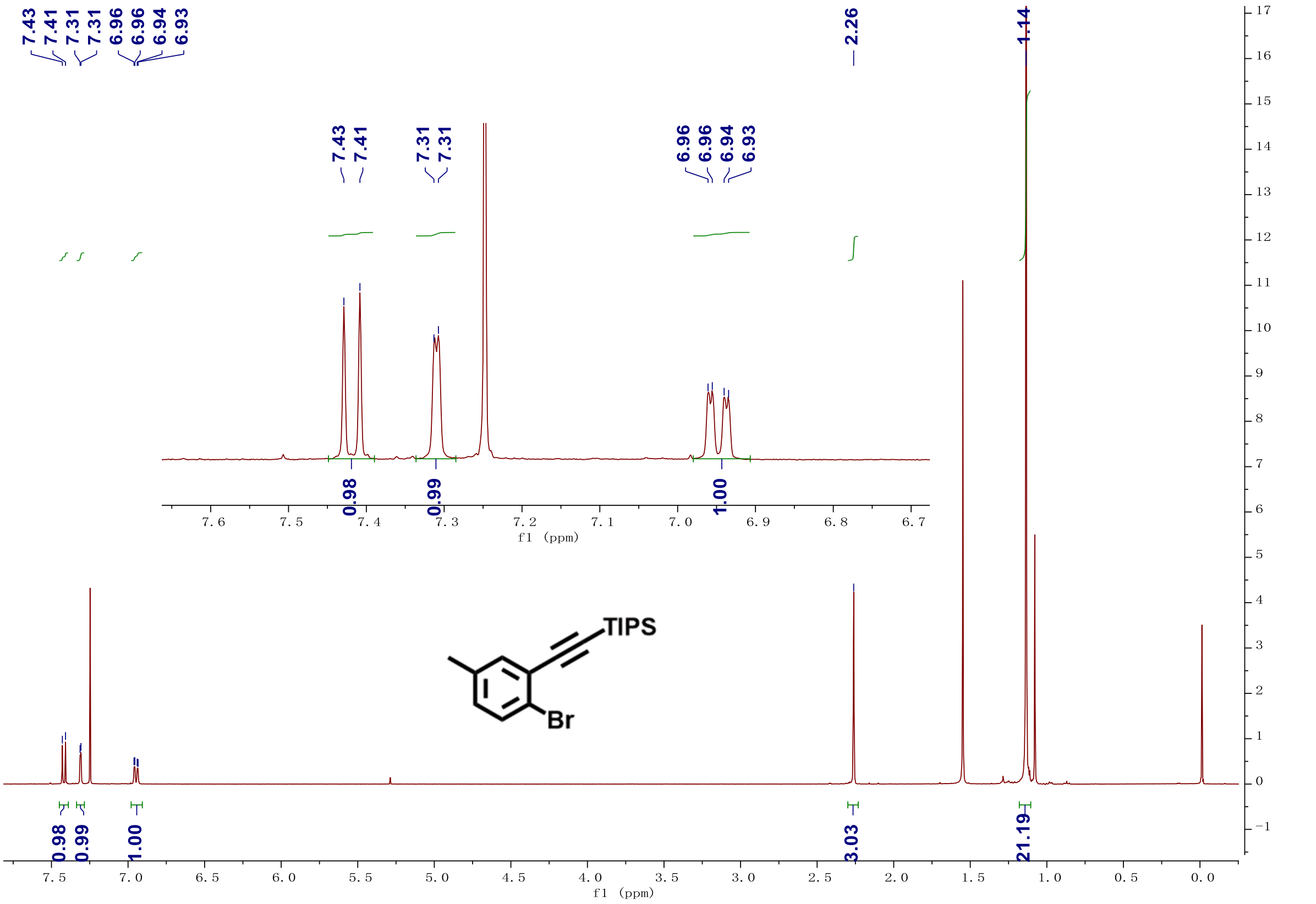

$^1$H NMR spectrum of compound **8** (400 MHz, $CDCl_3$, rt).

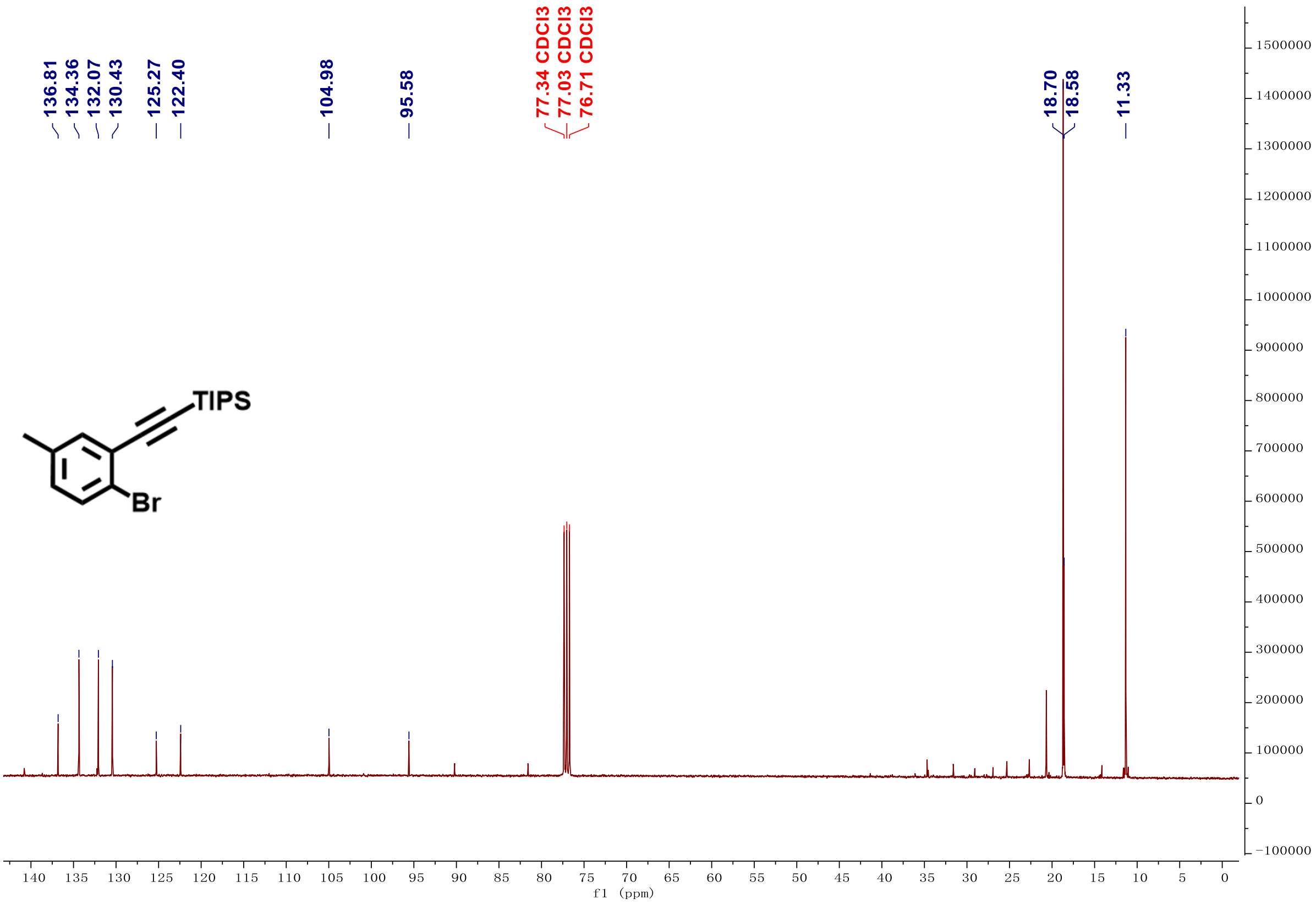

$^{13}$C NMR spectrum of compound **8** (101 MHz, $CDCl_3$, rt).

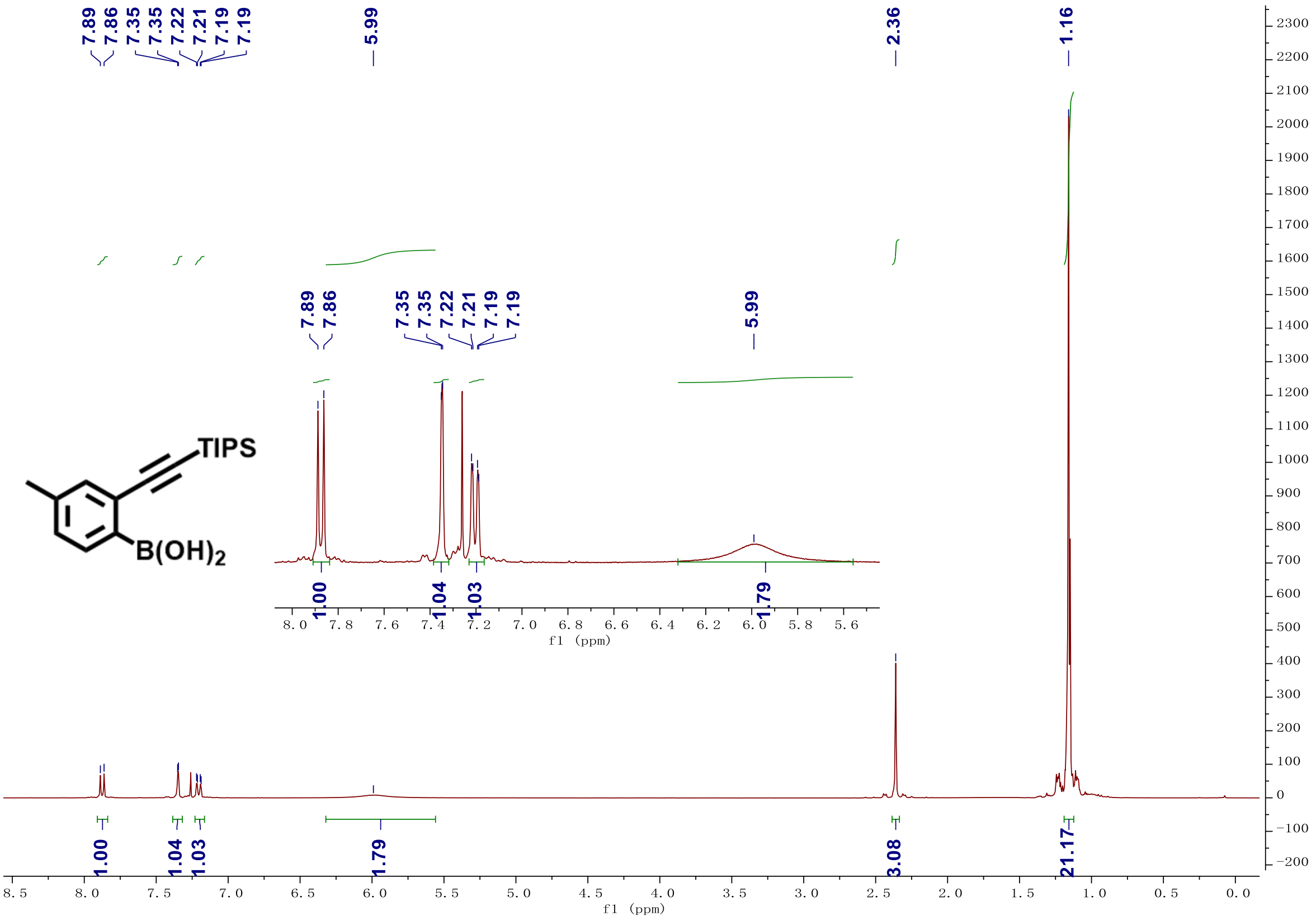

$^{1}$H NMR spectrum of compound **7** (300 MHz, $CDCl_3$, rt).

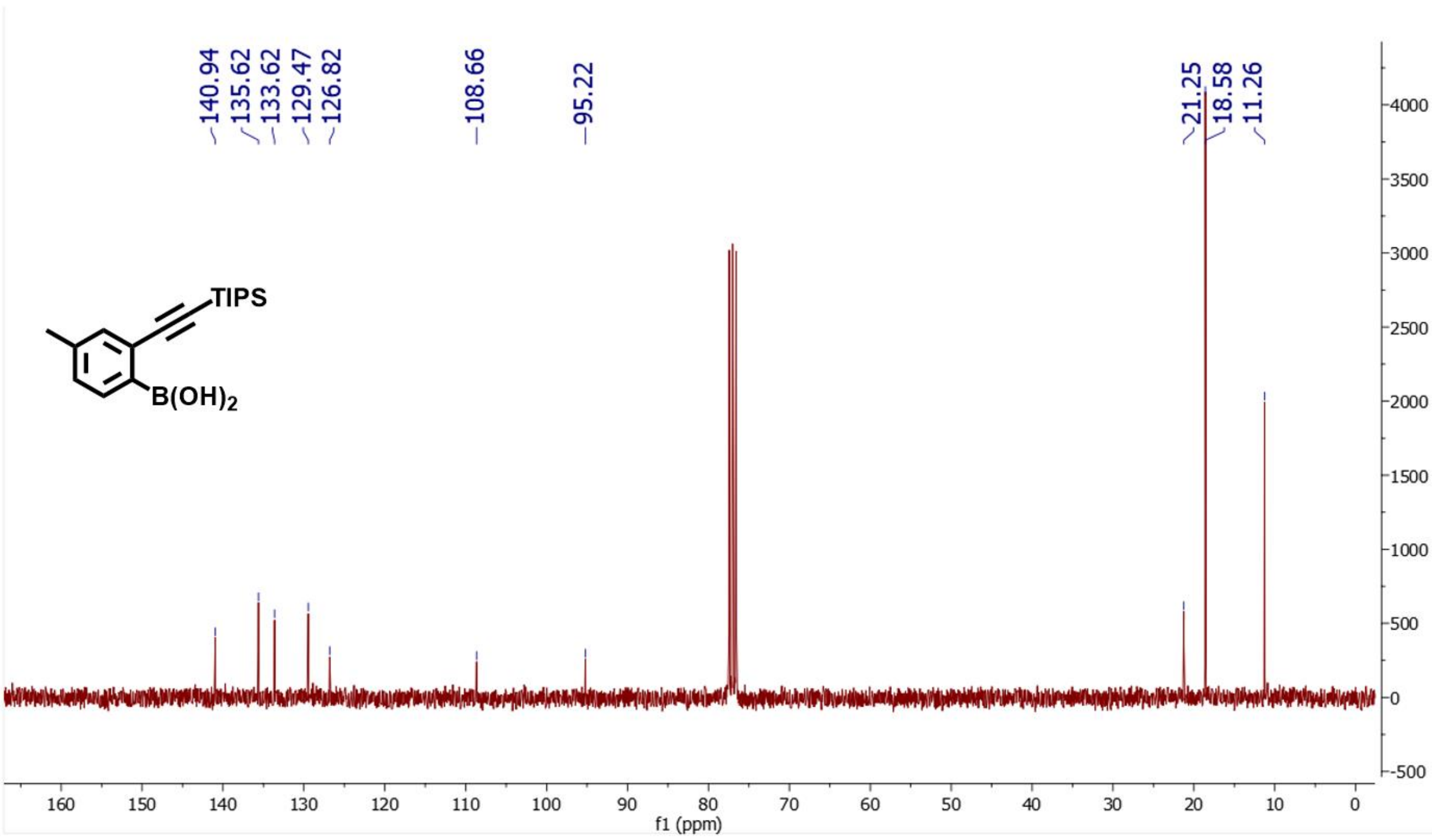

$^{13}$C NMR spectrum of compound **7** (75 MHz, $CDCl_3$, rt).

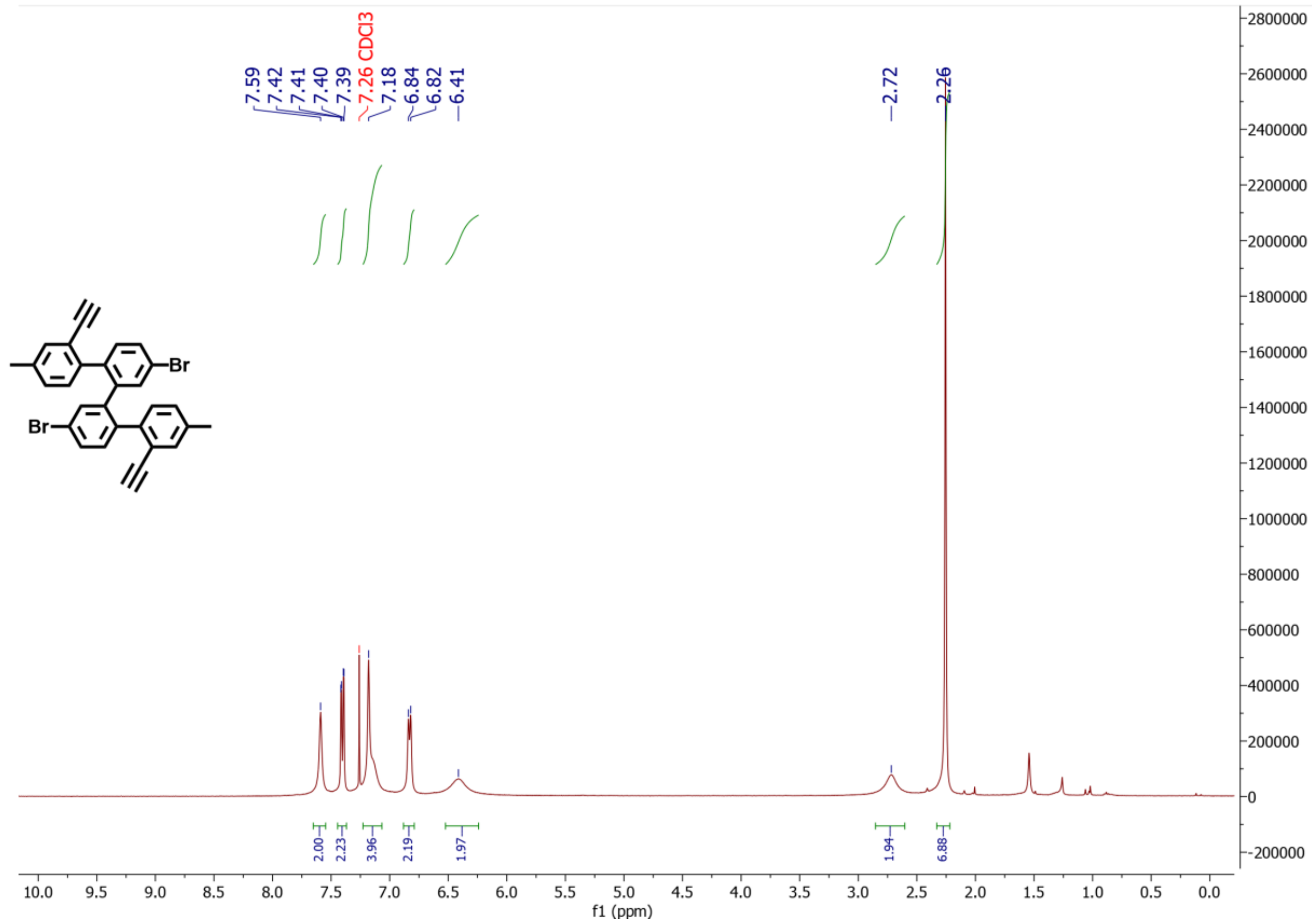

[1]H NMR spectrum of compound **4** (400 MHz, $CDCl_3$, rt).

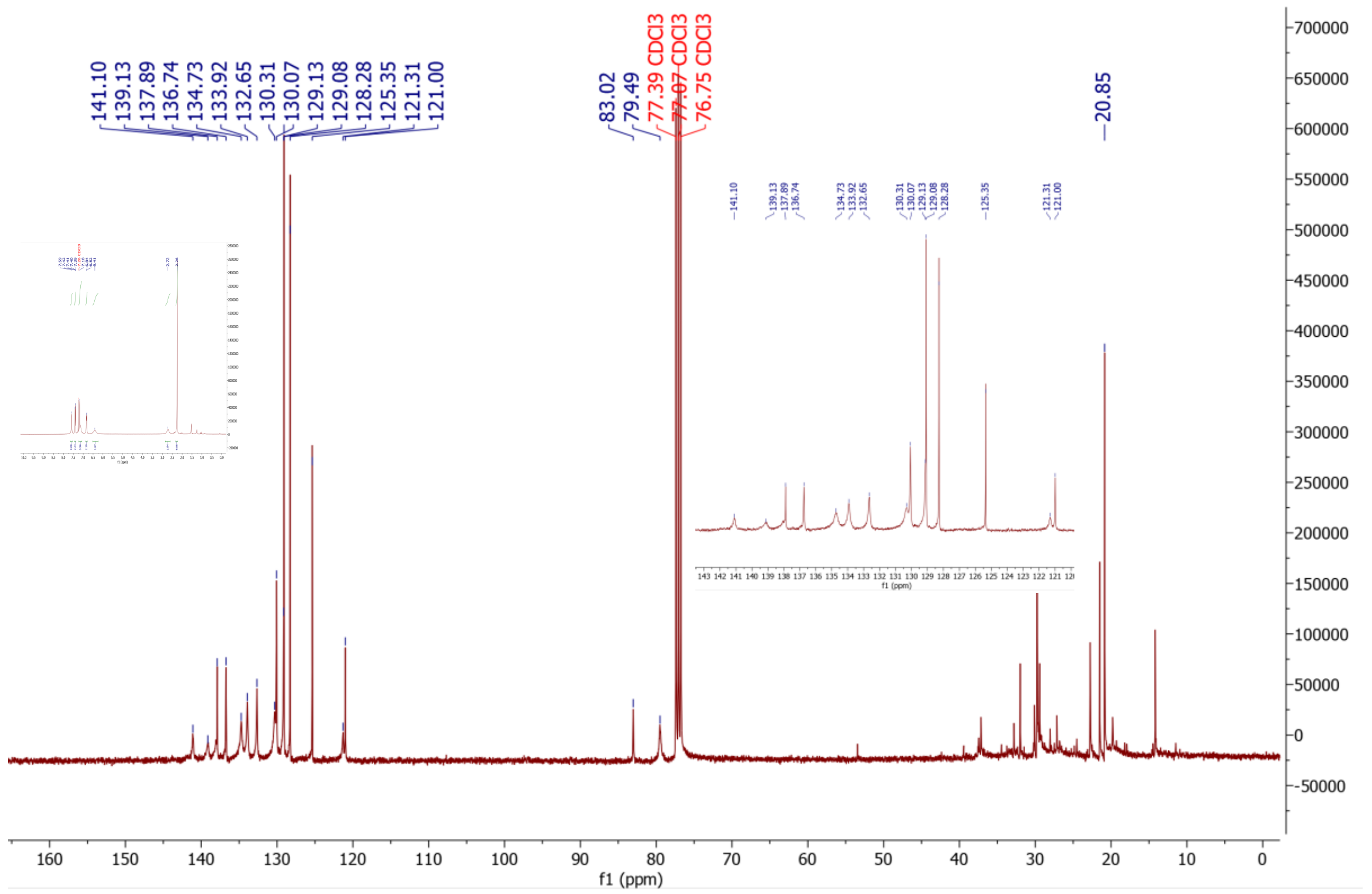

[13]C NMR spectrum of compound **4** (101 MHz, $CDCl_3$, rt).

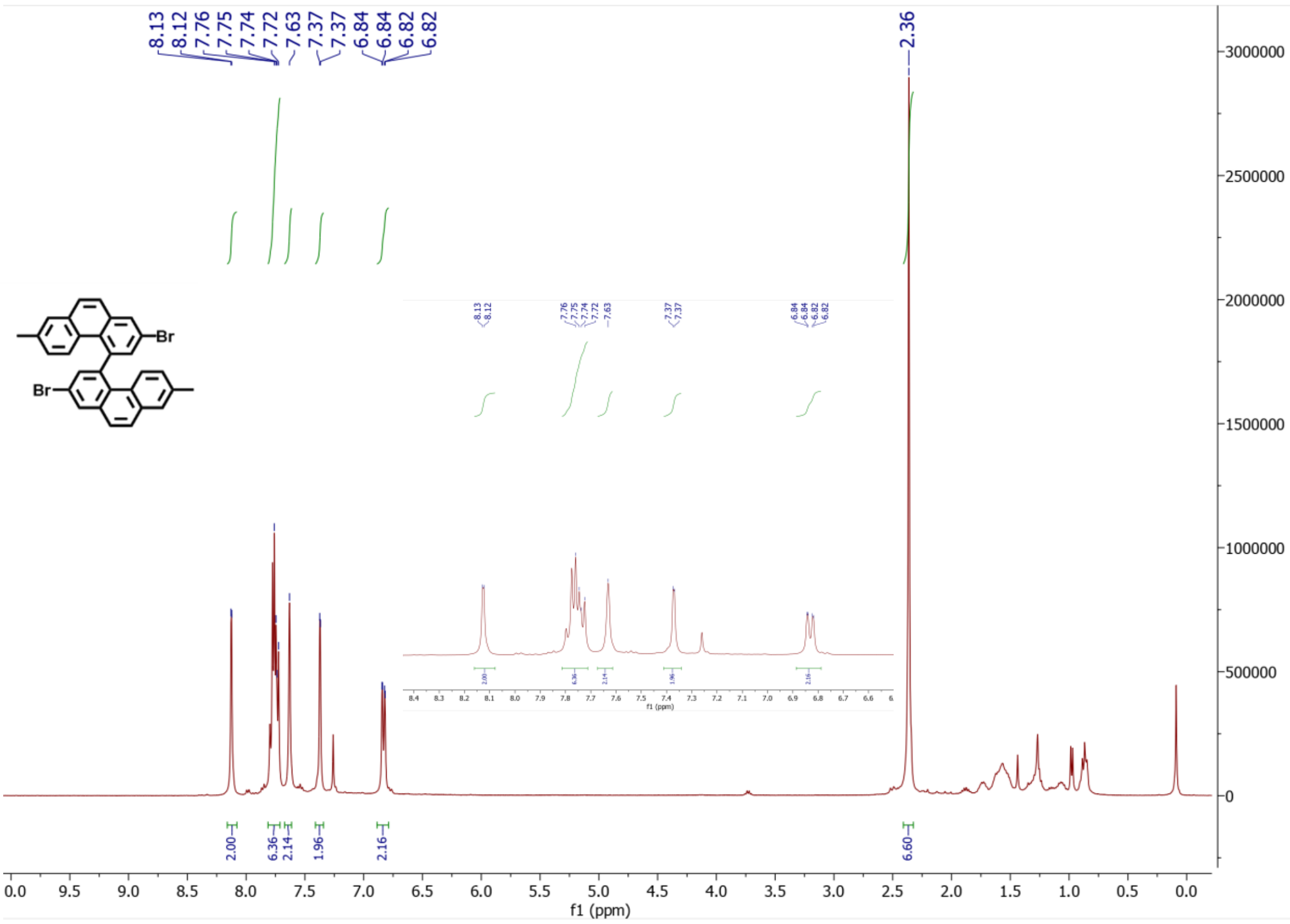

[1]H NMR spectrum of compound **3** (400 MHz, $CDCl_3$, rt).

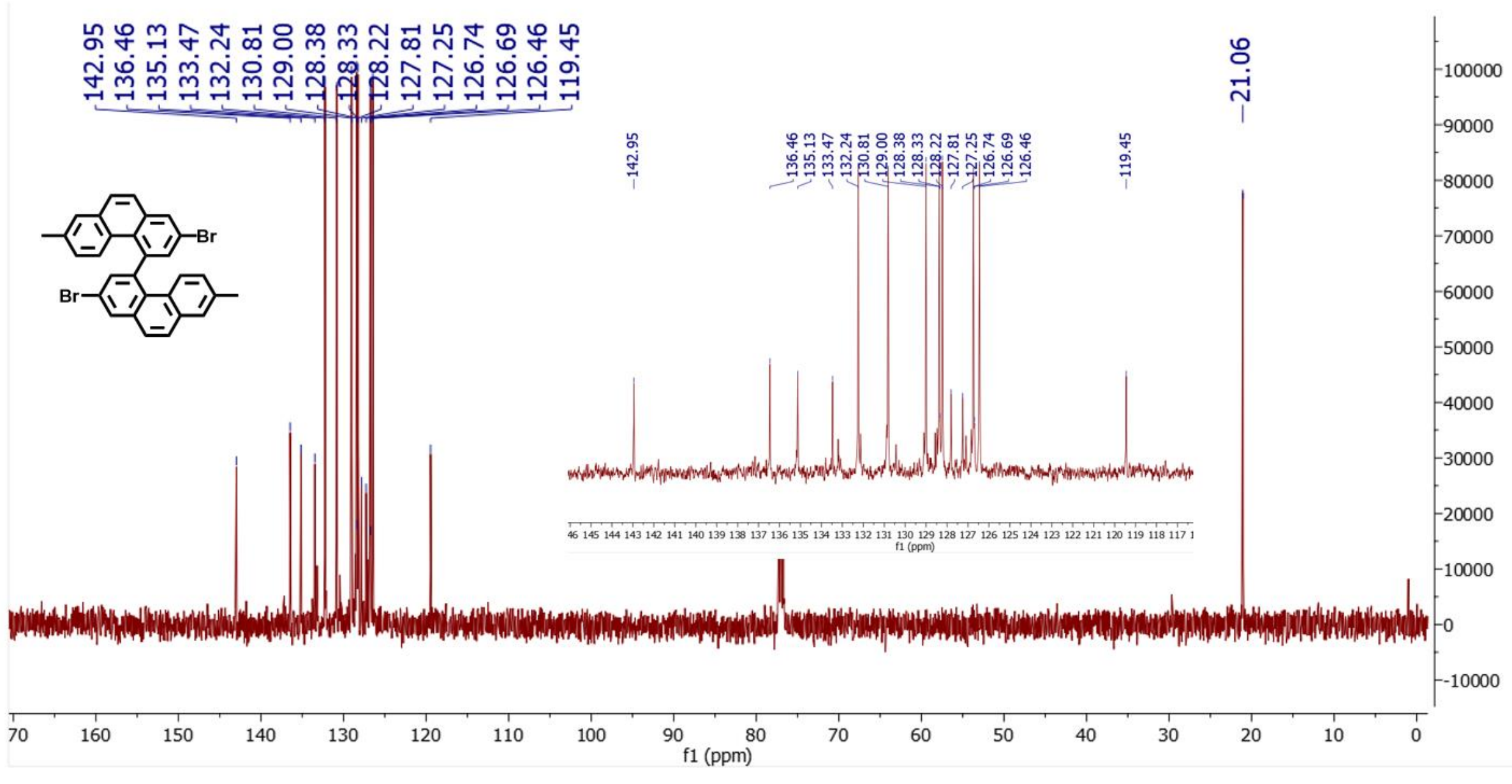

[13]C NMR spectrum of compound **3** (126 MHz, $CDCl_3$, rt).

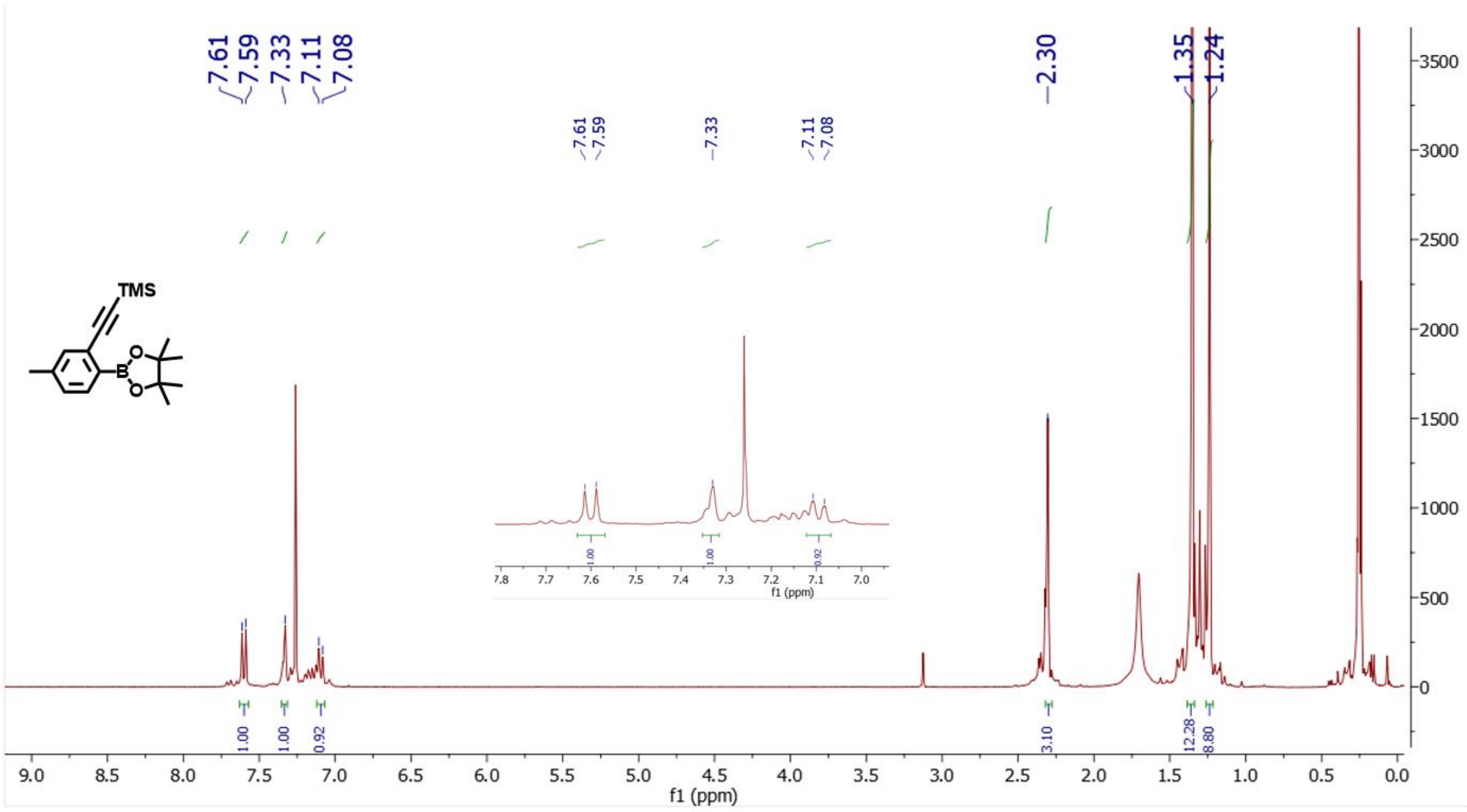

$^{1}H$ NMR spectrum of compound **13** (300 MHz, $CDCl_3$, rt).

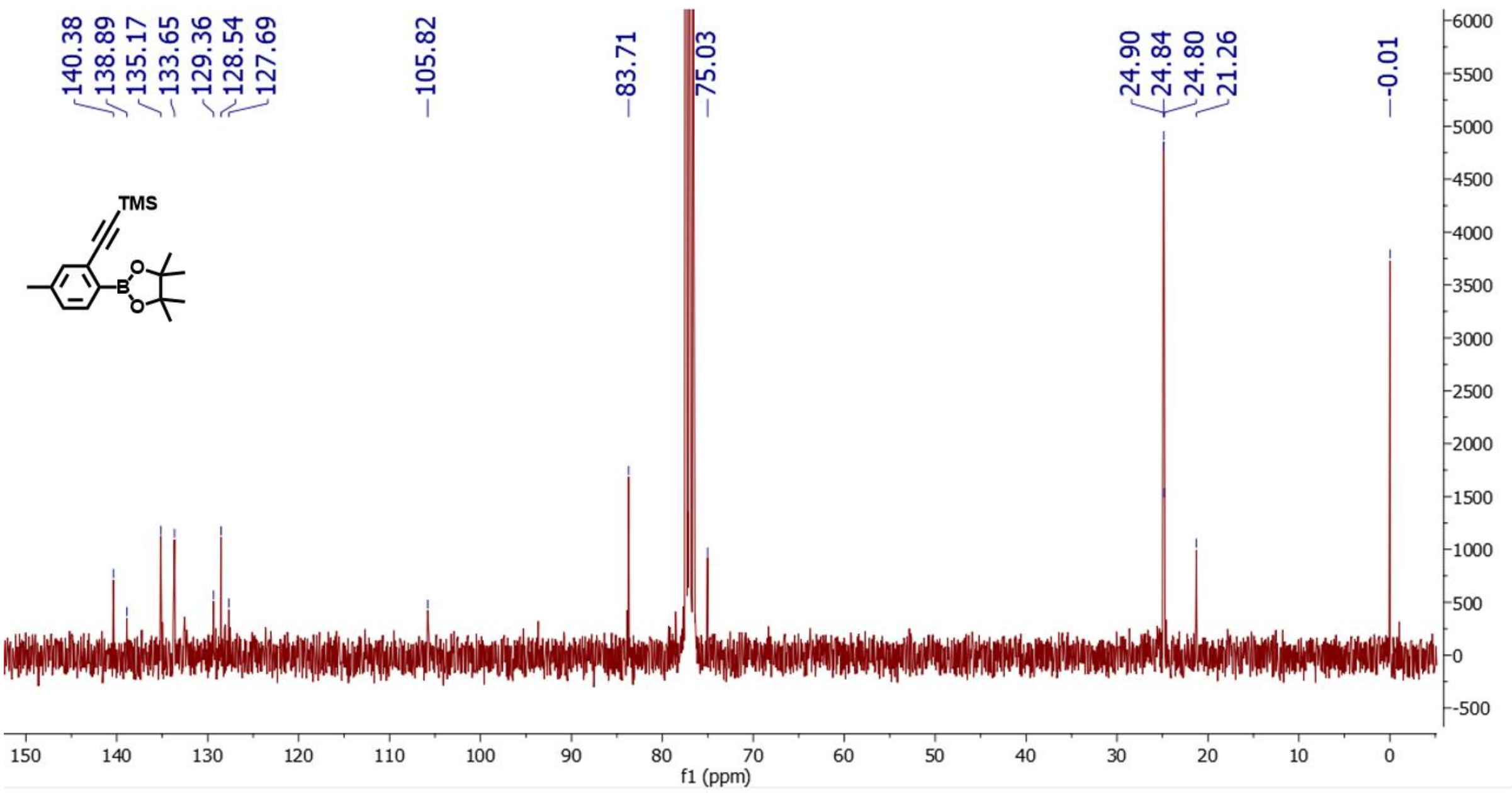

$^{13}C$ NMR spectrum of compound **13** (75 MHz, $CDCl_3$, rt).

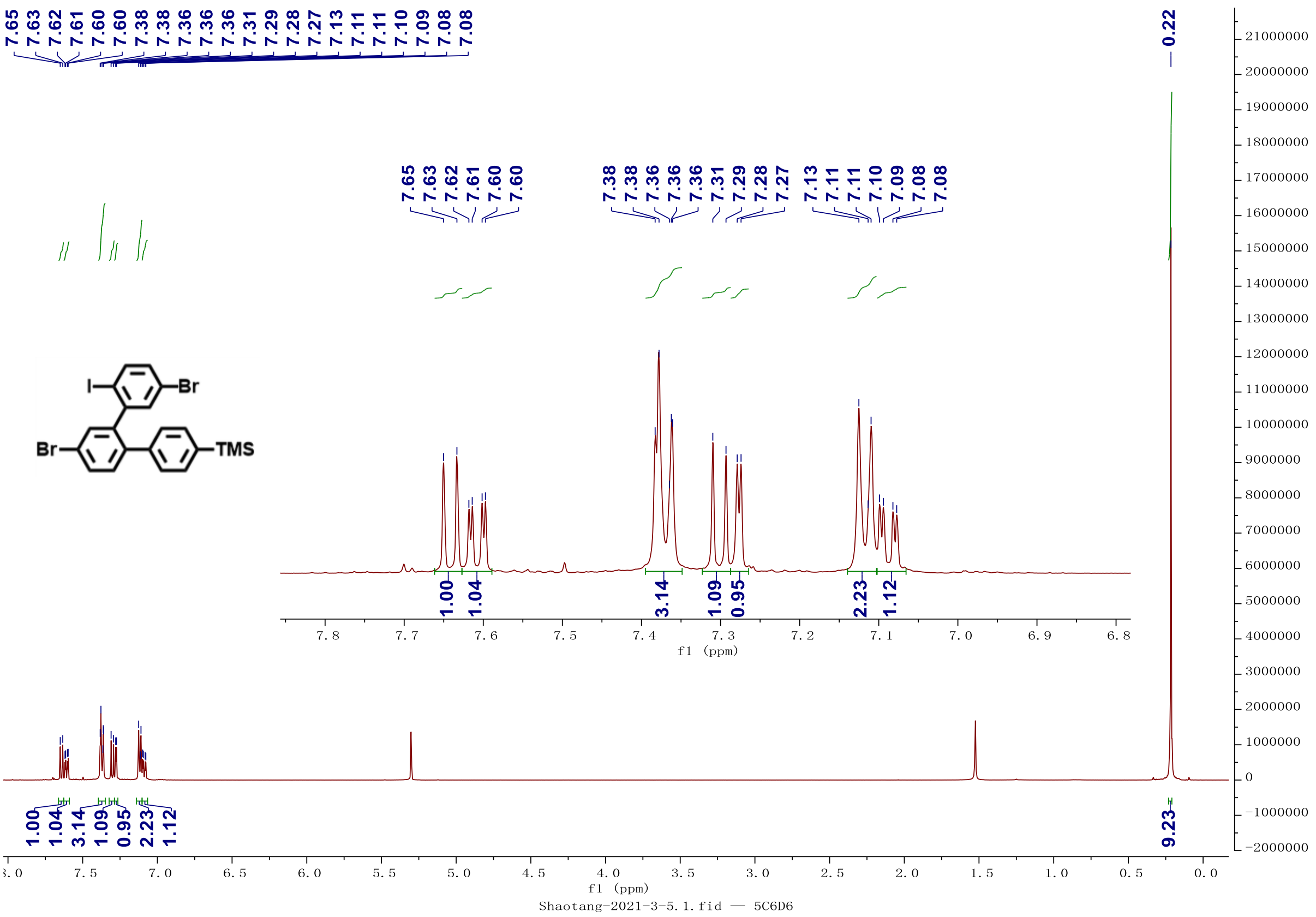

[1]H NMR spectrum of compound **12** (500 MHz, $CD_2Cl_2$, rt).

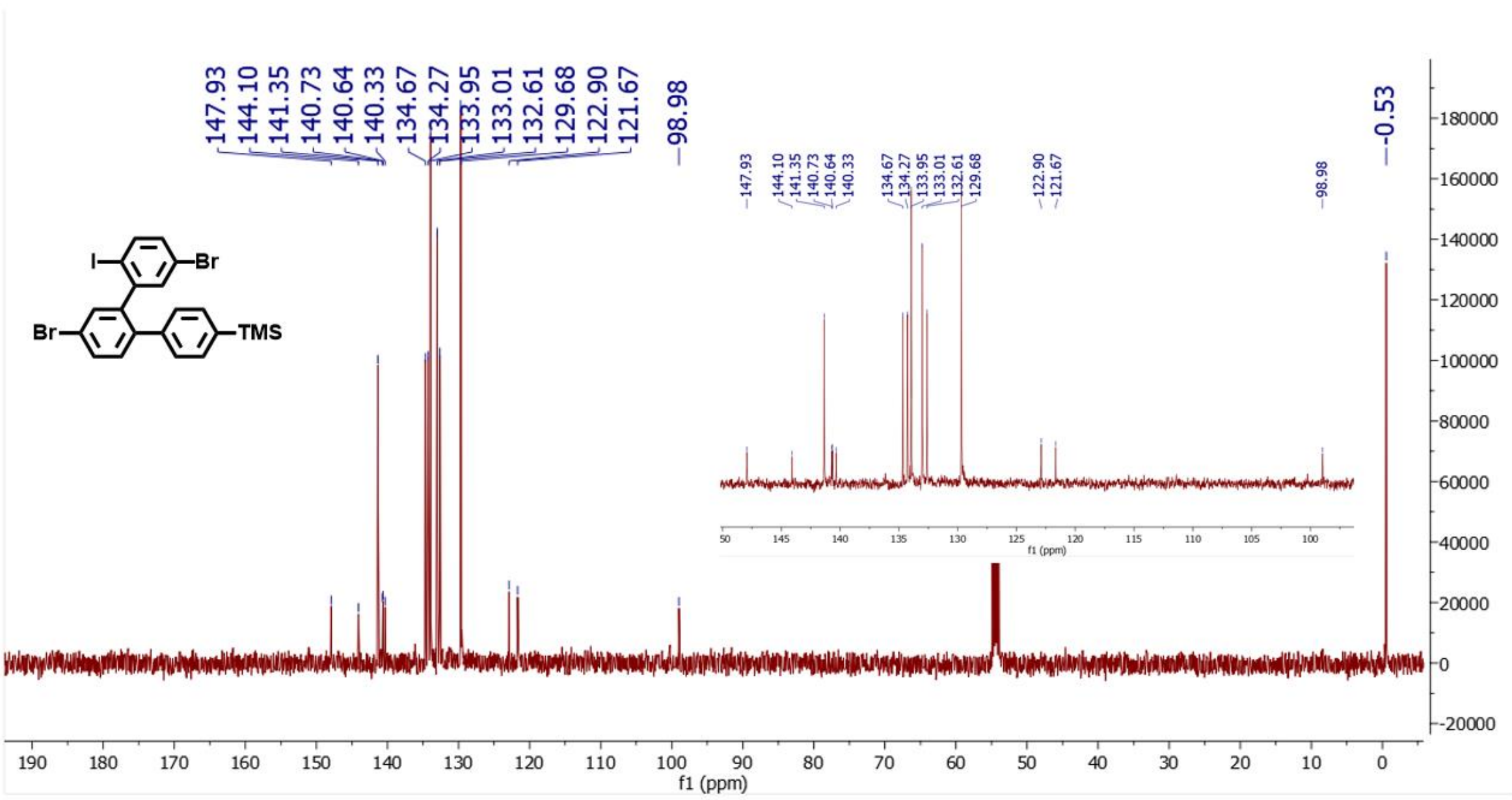

[13]C NMR spectrum of compound **12** (126 MHz, $CD_2Cl_2$, rt).

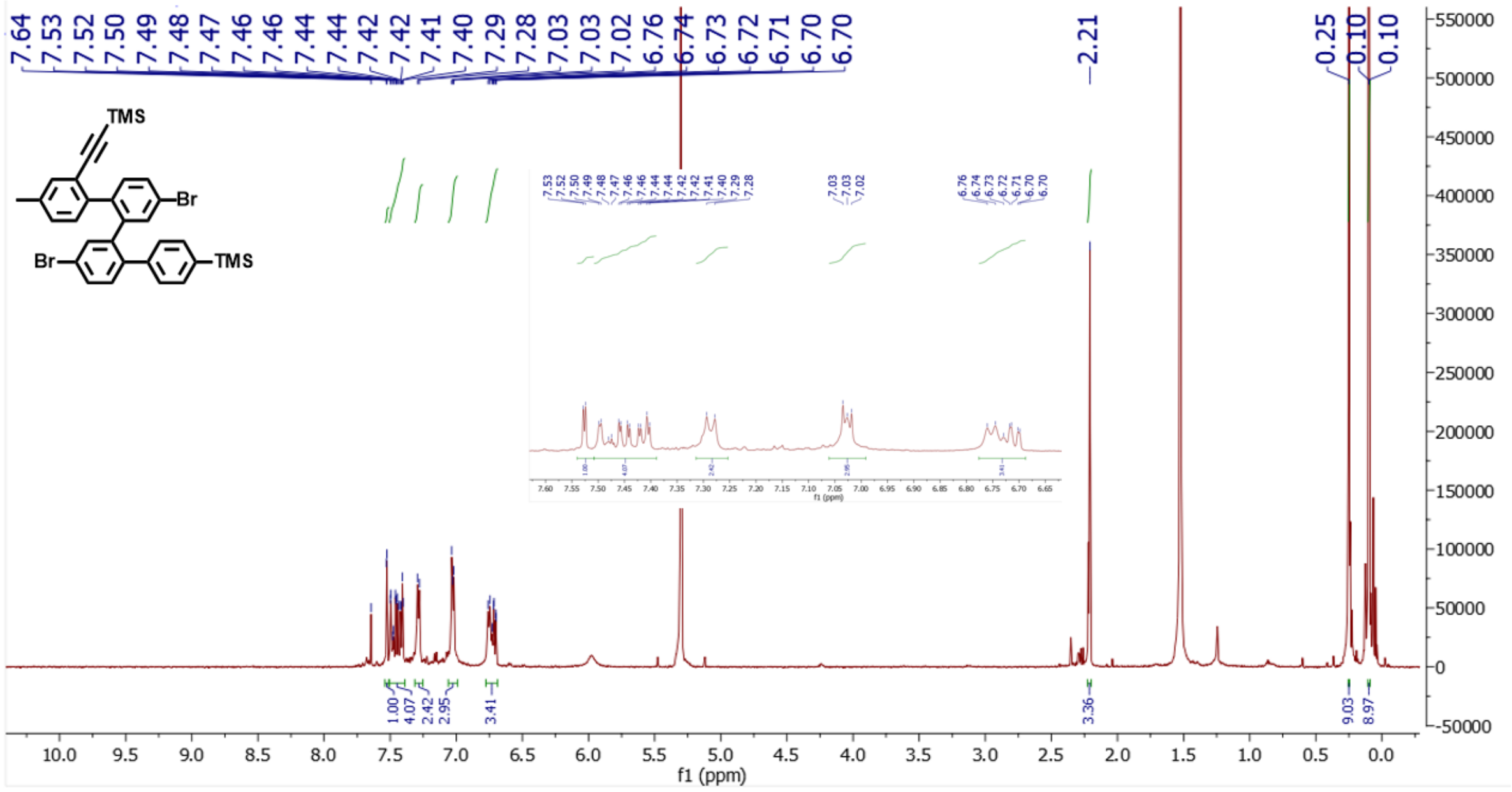

[1]H NMR spectrum of compound **11** (500 MHz, $CD_2Cl_2$, rt).

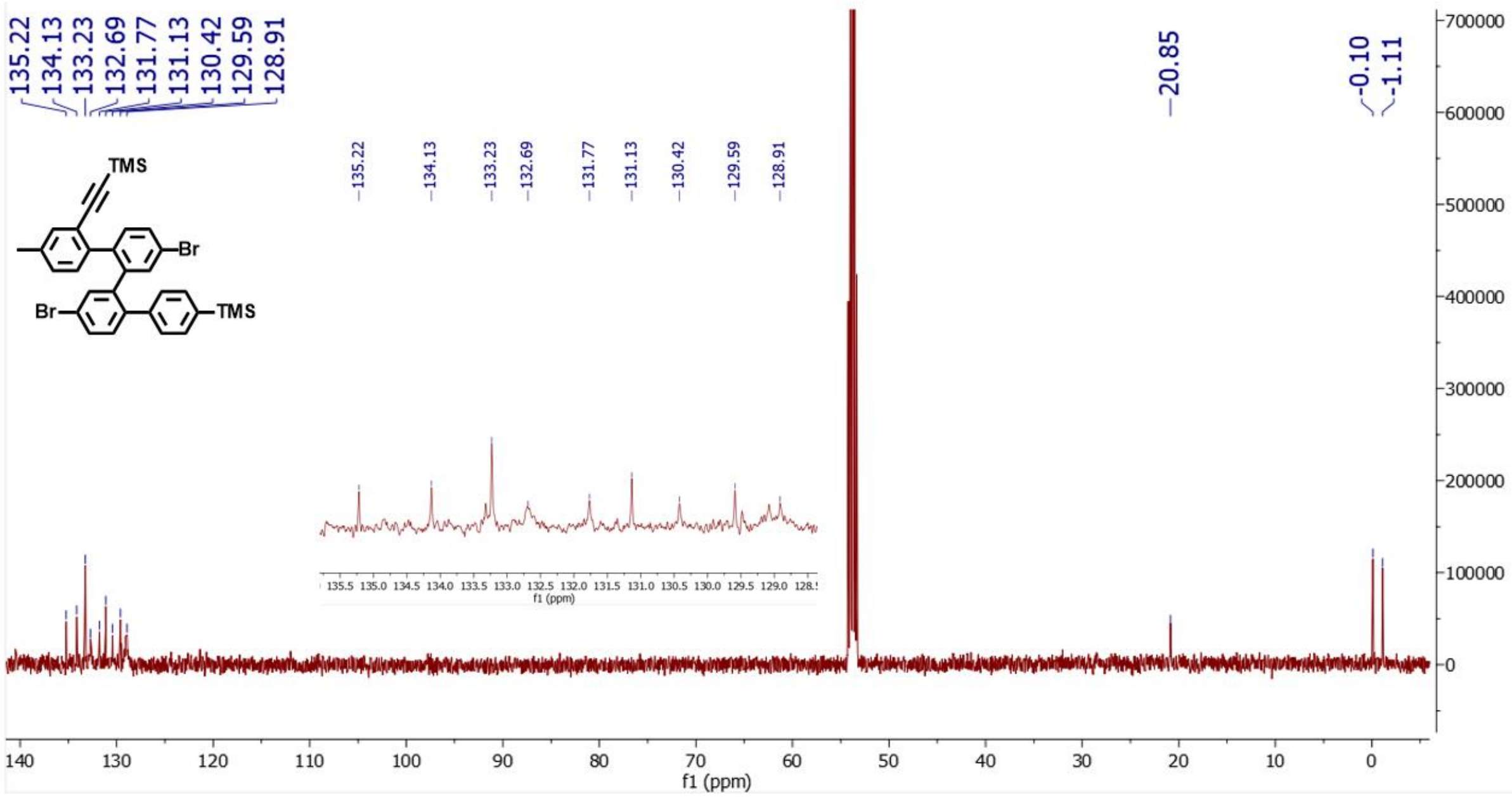

[13]C NMR spectrum of compound **11** (126 MHz, $CD_2Cl_2$, rt).

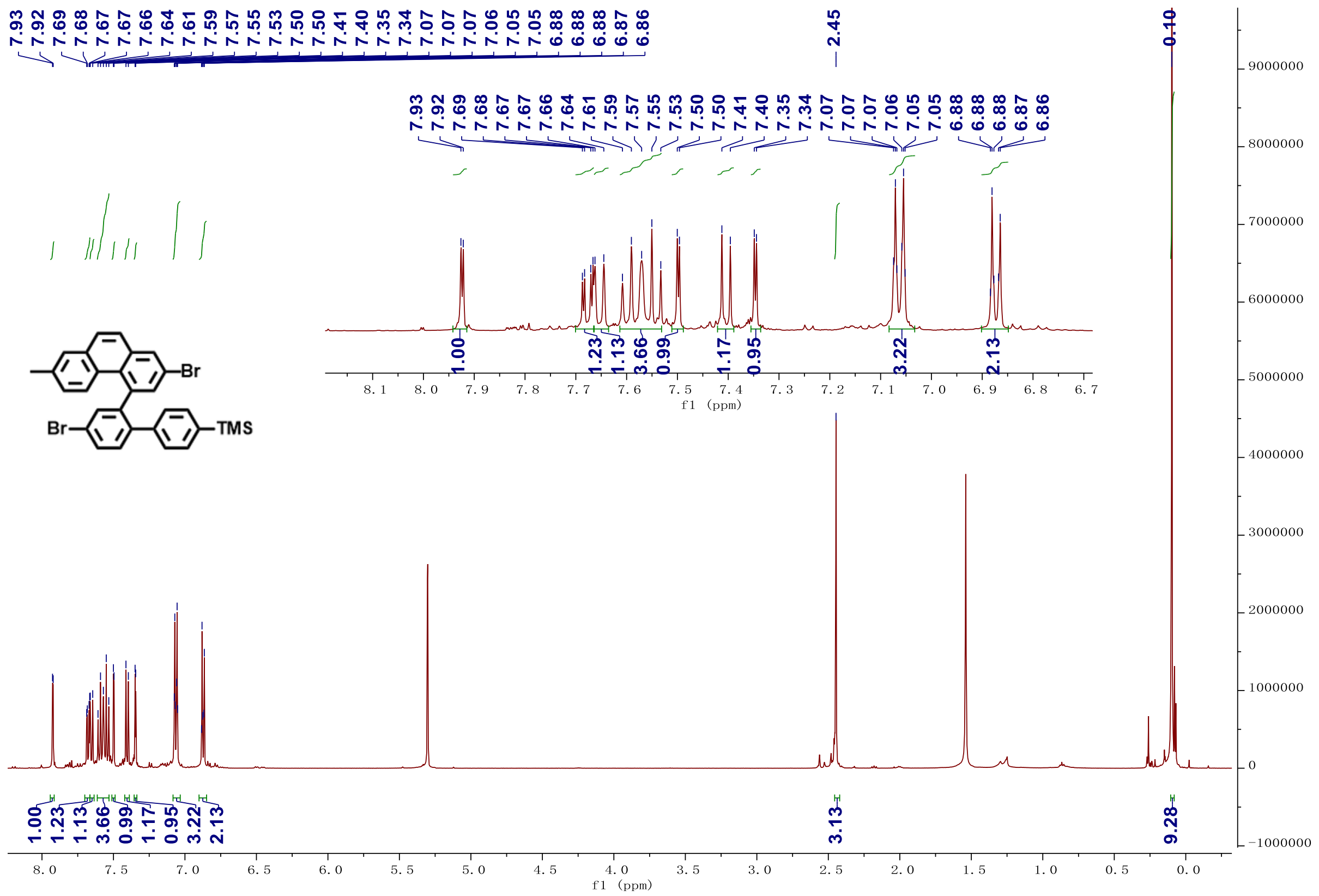

[1]H NMR spectrum of compound **9** (500 MHz, $CD_2Cl_2$, rt).

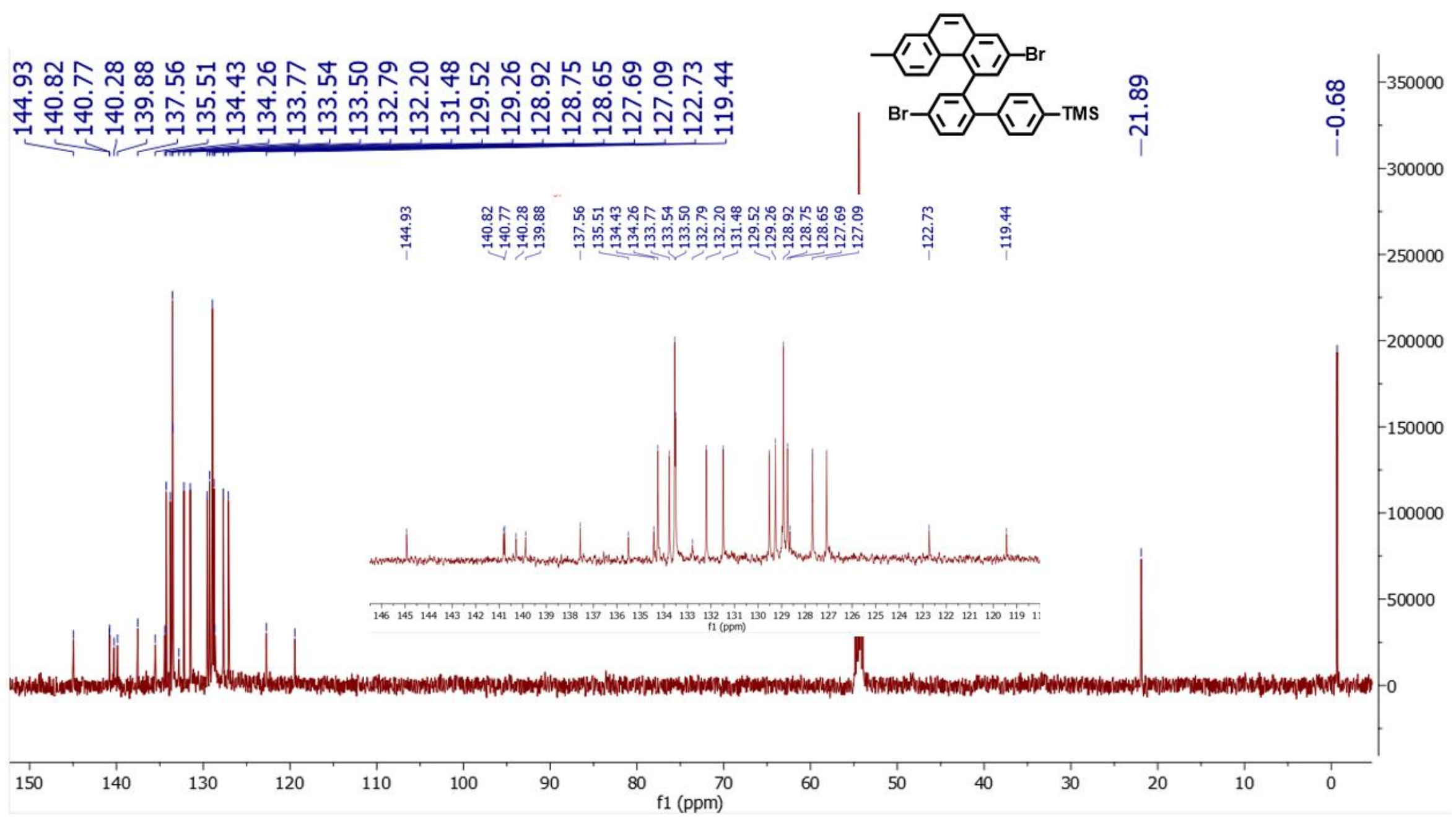

[13]C NMR spectrum of compound **9** (126 MHz, $CD_2Cl_2$, rt).

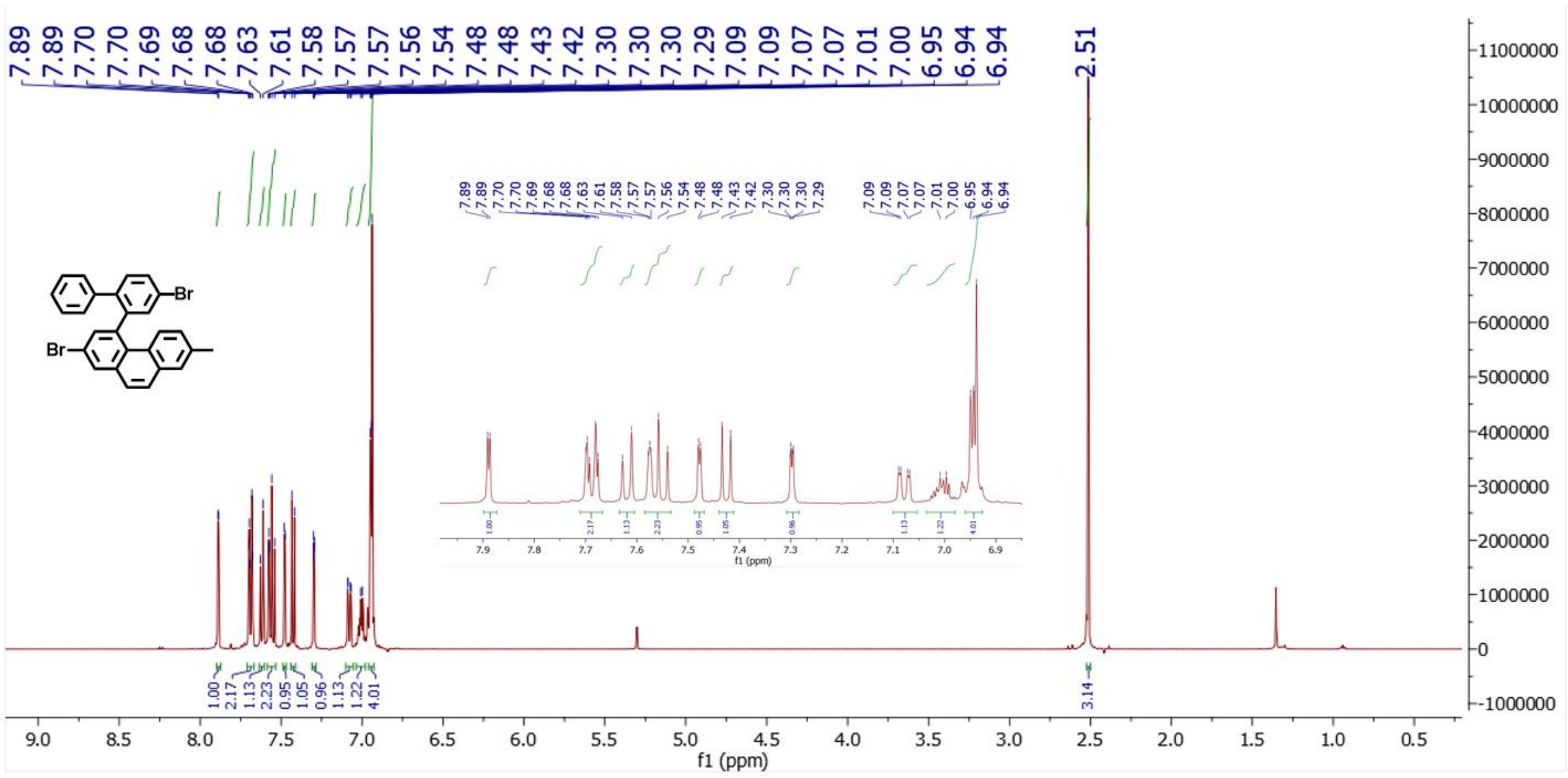

[1]H NMR spectrum of compound **1** (500 MHz, $CD_2Cl_2$, rt).

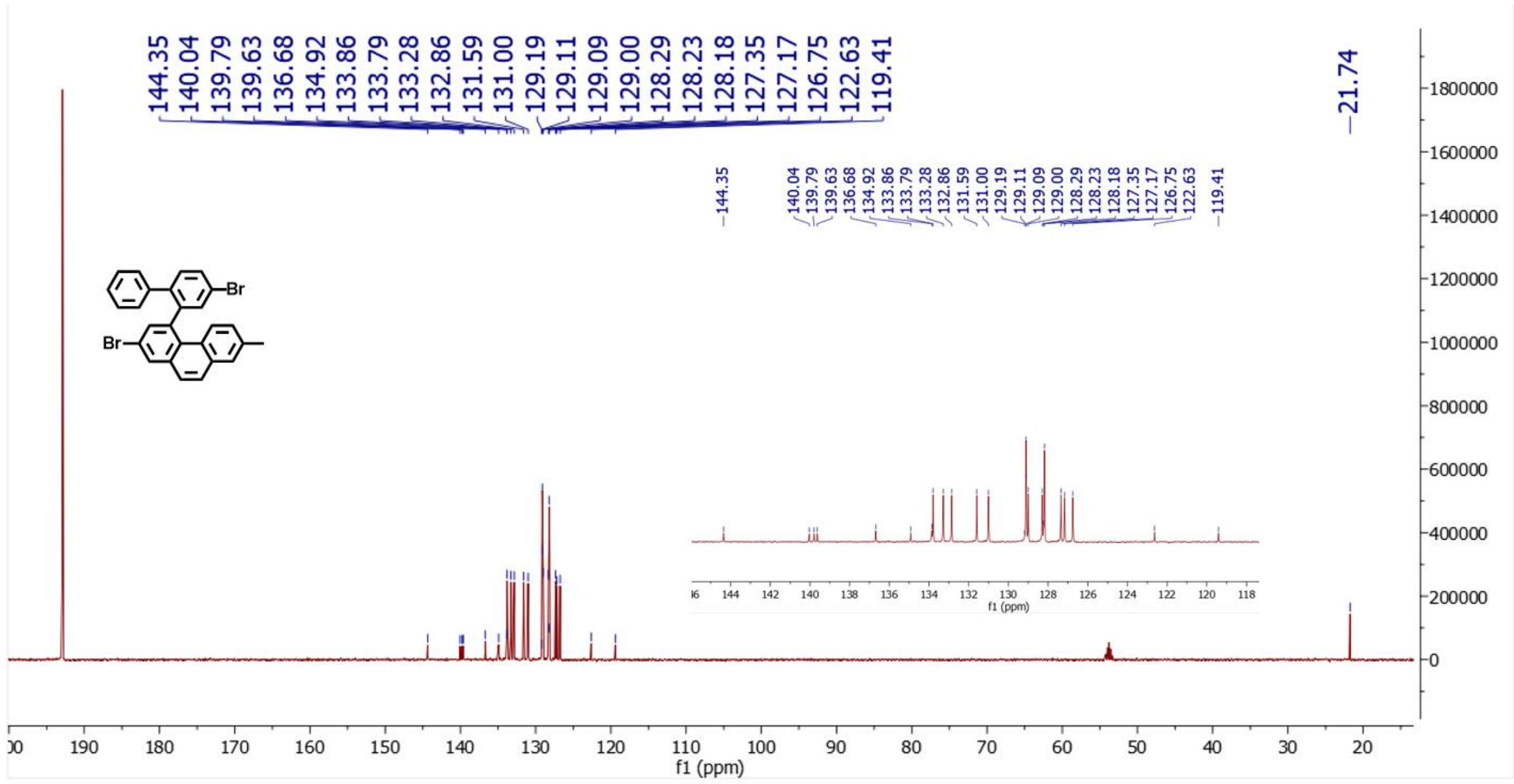

[13]C NMR spectrum of compound **1** (126 MHz, $CD_2Cl_2$, rt).

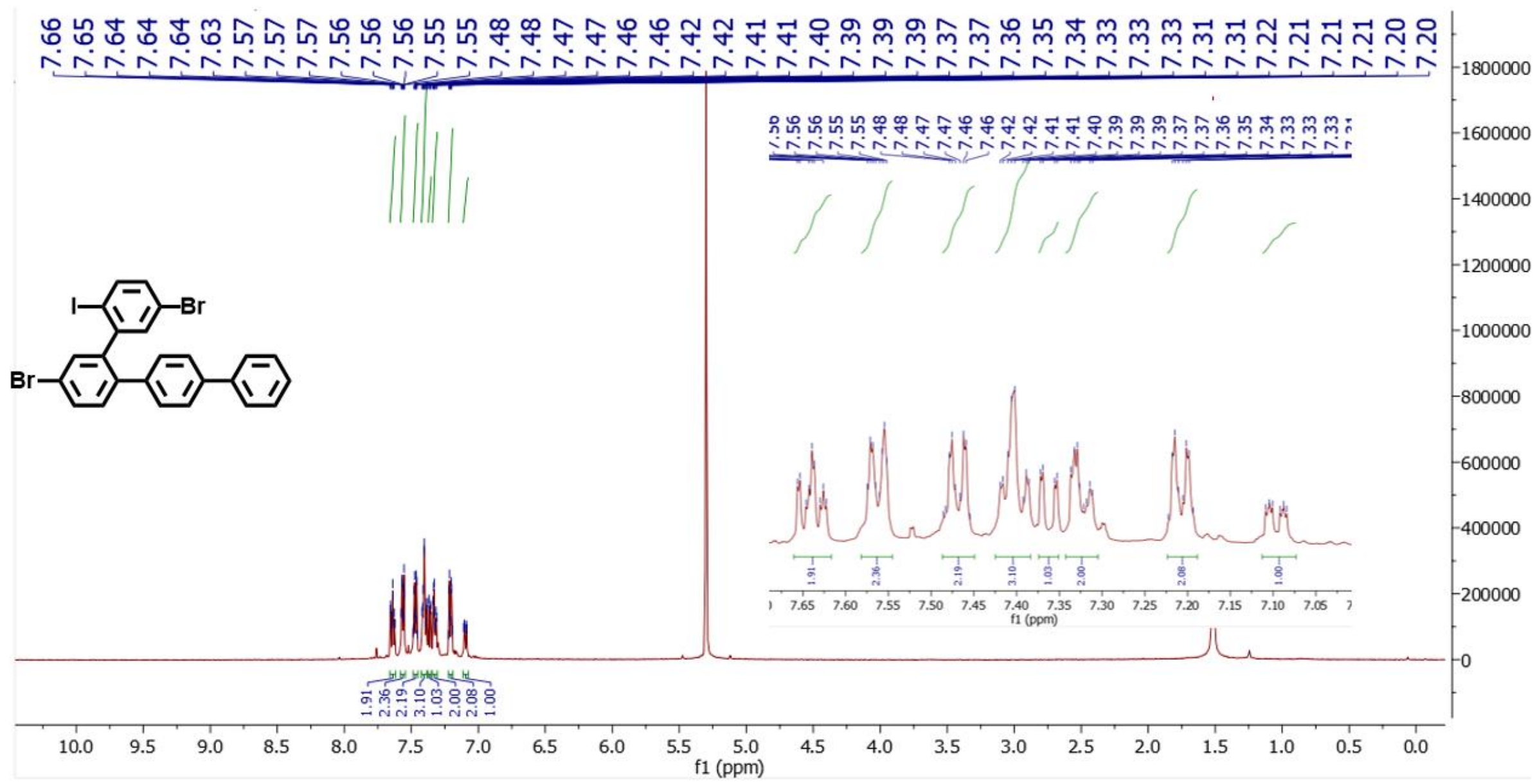

[1]H NMR spectrum of compound **17** (500 MHz, $CDCl_3$, rt).

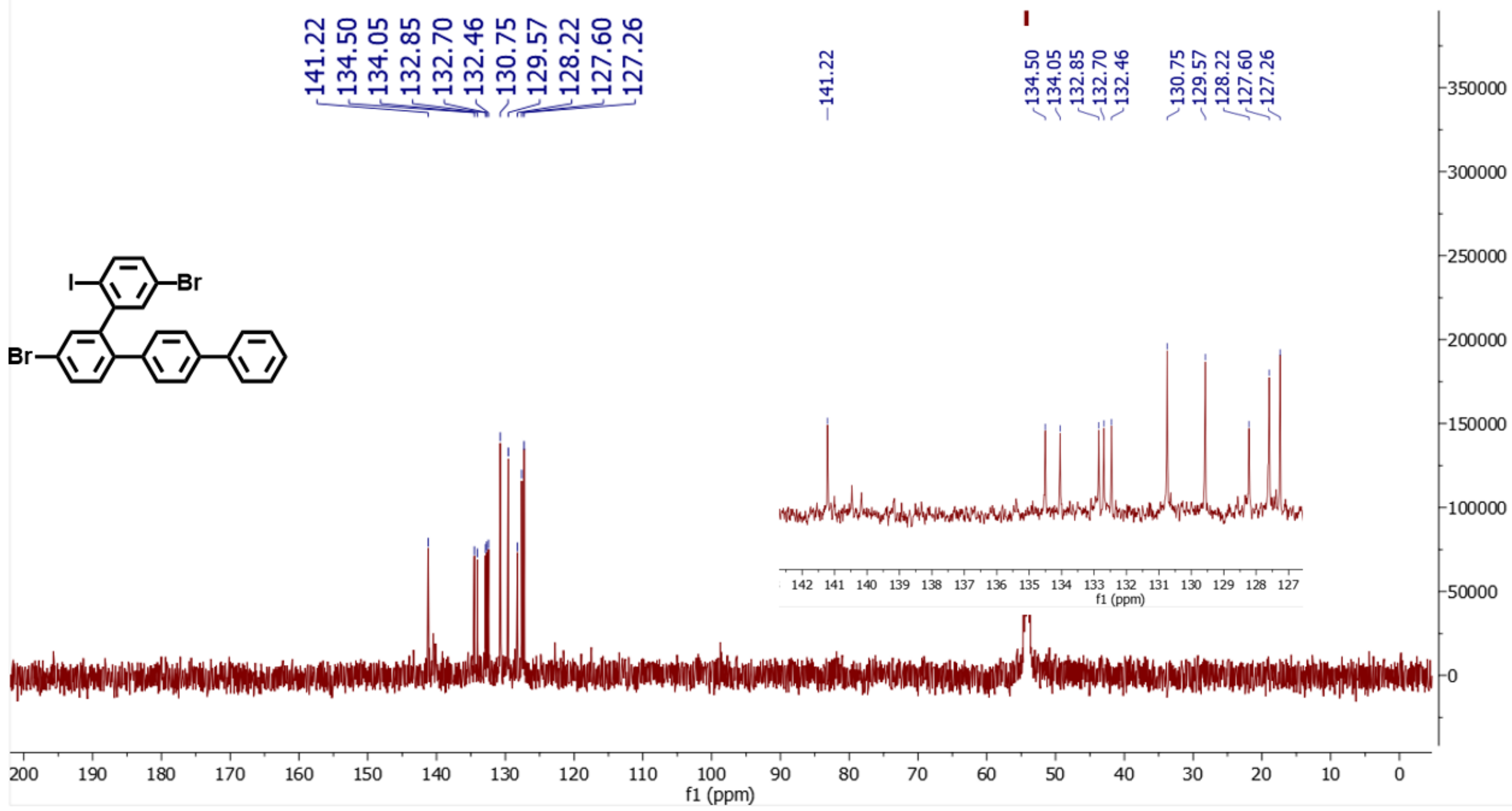

[13]C NMR spectrum of compound **17** (126 MHz, $CDCl_3$, rt).

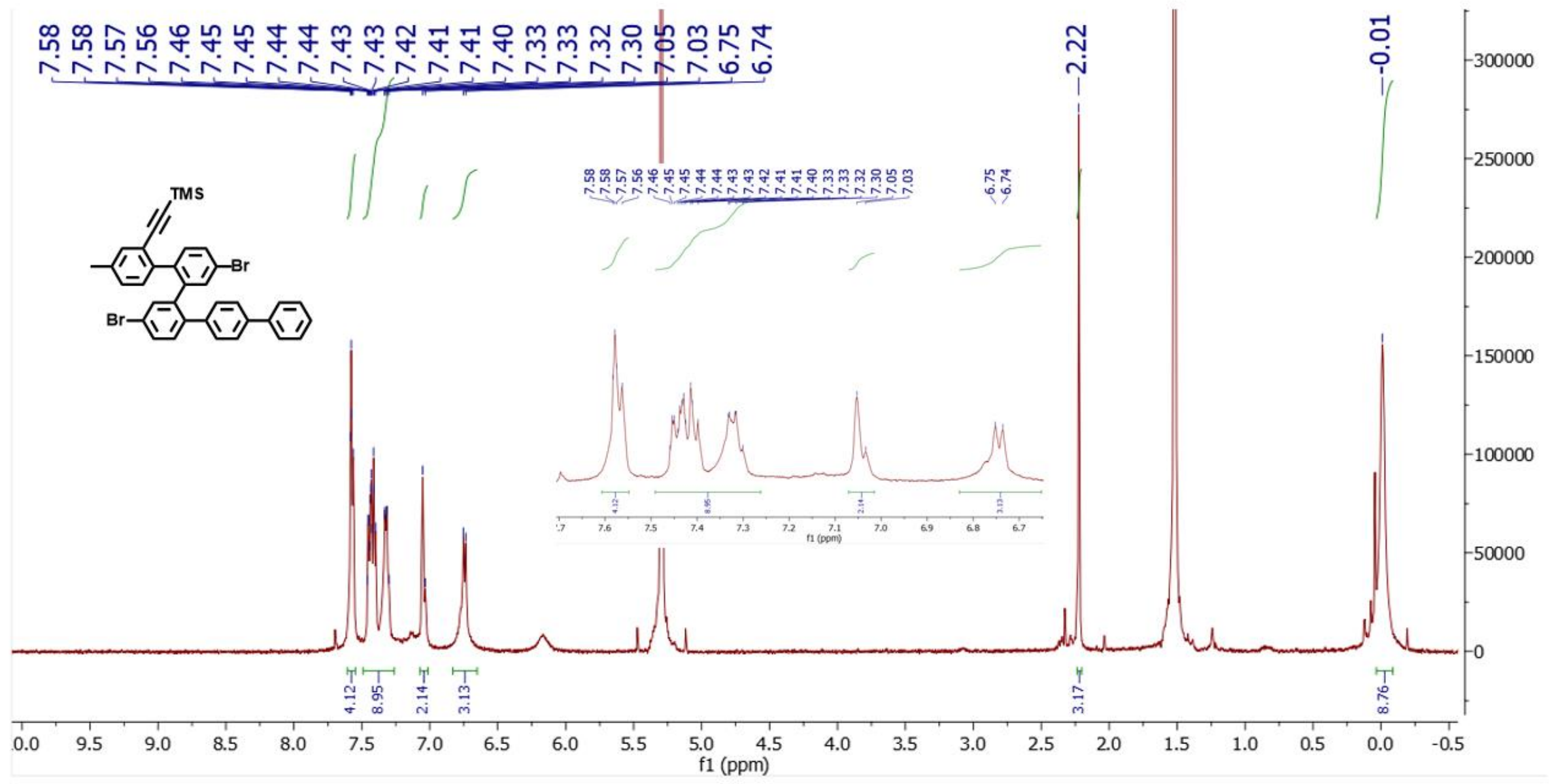

$^{1}H$ NMR spectrum of compound **16** (500 MHz, $CD_2Cl_2$, rt).

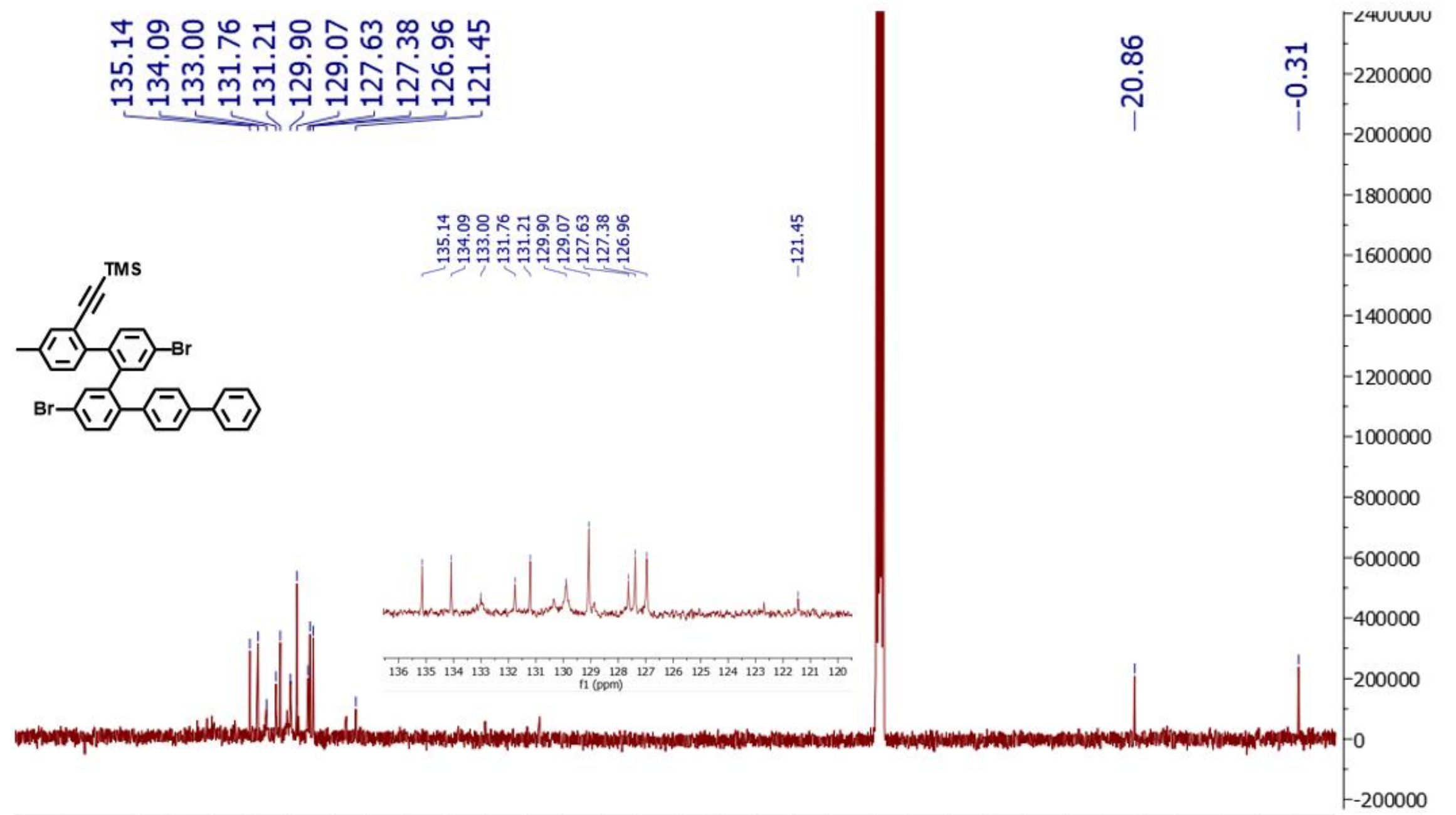

$^{13}C$ NMR spectrum of compound **16** (126 MHz, $CD_2Cl_2$, rt).

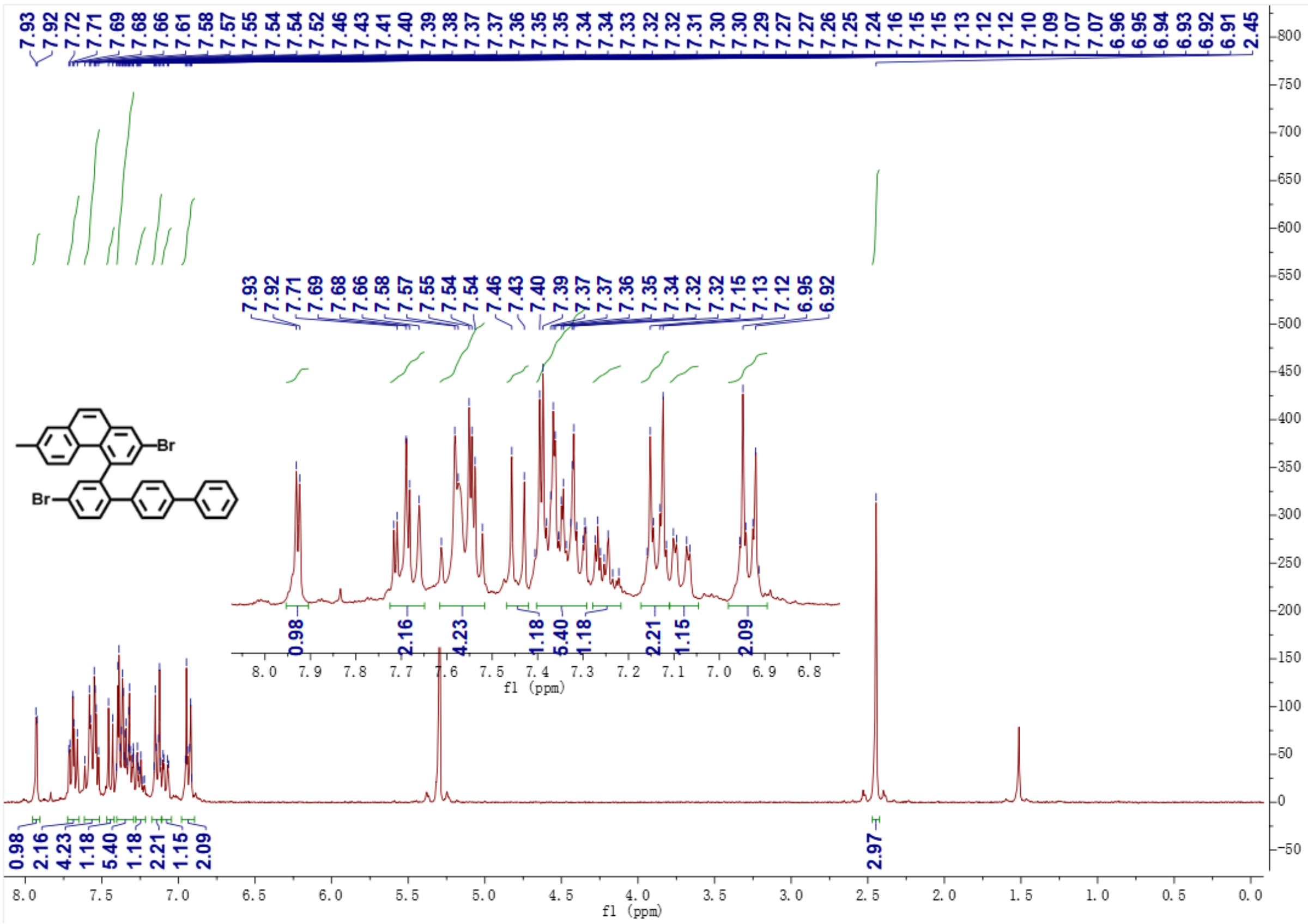

[1]H NMR spectrum of compound **2** (300 MHz, $CD_2Cl_2$, rt).

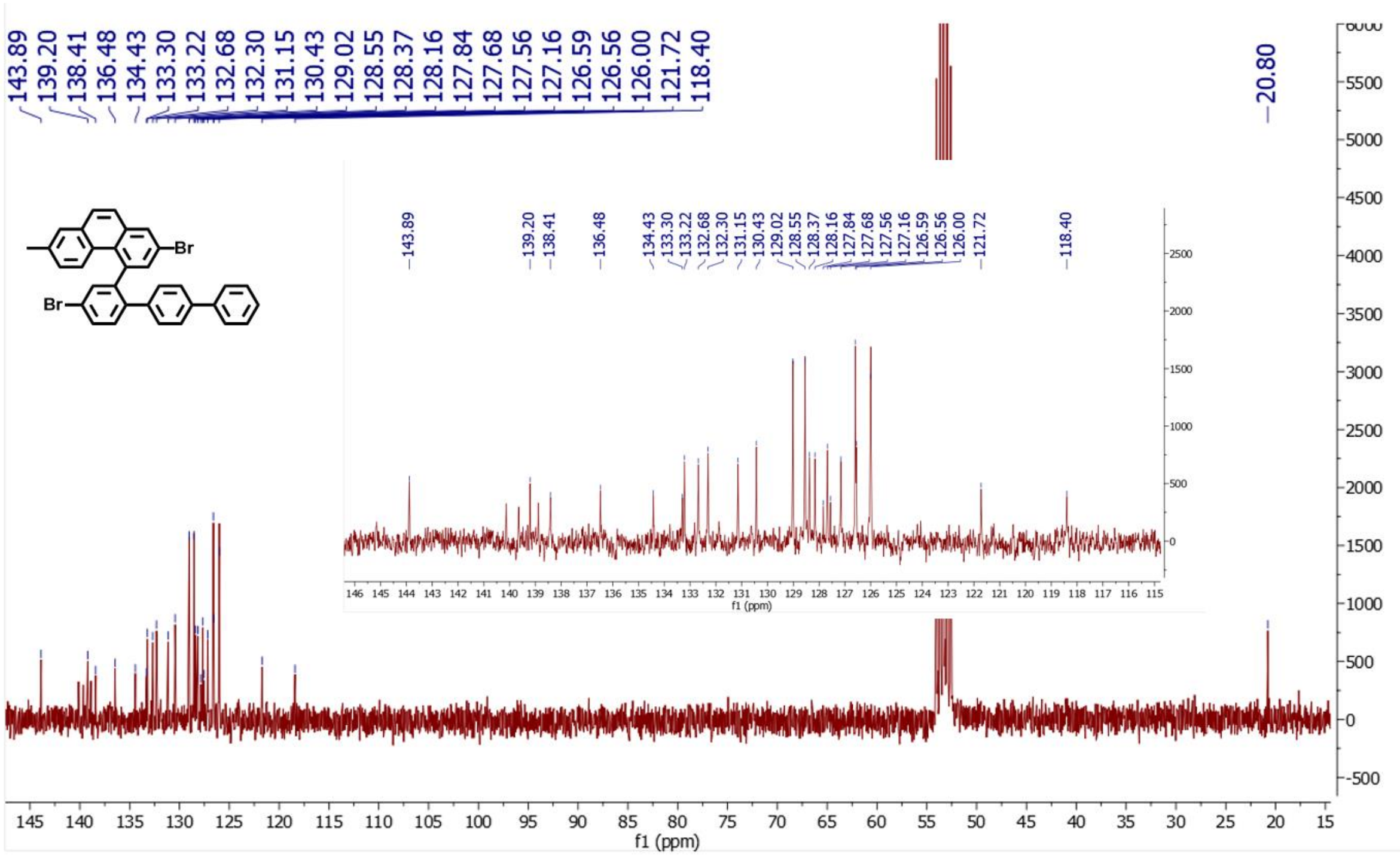

[13]C NMR spectrum of compound **2** (75 MHz, $CD_2Cl_2$, rt).